\documentclass[a4paper,11pt]{article}
\pdfoutput=1 

\usepackage{jheppub}

\usepackage{graphicx}
\usepackage{amsmath,amssymb}
\usepackage{bm}
\usepackage{color}

\usepackage{hyperref}
\hypersetup{ 
 setpagesize=false,
 bookmarksnumbered=true,
 colorlinks=true,
 linkcolor=blue,
 citecolor=red,
}

\title{\boldmath Flavor constraints on the Two Higgs Doublet Models of $Z_2$ symmetric and aligned types}

\author[a]{Tetsuya ENOMOTO}
\author[b]{Ryoutaro WATANABE}

\affiliation[a]{Department of Physics, Graduate School of Science, Osaka University, Toyonaka, Osaka 560-0043, Japan}
\affiliation[b]{Center for Theoretical Physics of the Universe, Institute for Basic Science (IBS), Daejeon, 34051, Republic of Korea}
\emailAdd{tetsuya@het.phys.sci.osaka-u.ac.jp}
\emailAdd{wryou1985@ibs.re.kr}

\abstract{
We give a comprehensive study from flavor observables of $\pi$, $K$, $D_{(s)}$, and $B_{(s)}$ mesons for limiting the Two Higgs Doublet Models (2HDMs) with natural flavor conservation, 
namely, $Z_2$ symmetric (type~I, II, X, Y) and aligned types of models. 
With use of updated theoretical predictions and experimental analyses of $B\to\tau\nu$, $D\to\mu\nu$, $D_s\to\tau\nu$, $D_s\to\mu\nu$, $K\to\mu\nu$, $\pi\to\mu\nu$, $B^0_s \to \mu^+ \mu^-$, $B^0_d \to \mu^+ \mu^-$, $\tau \to K\nu$, $\tau \to \pi\nu$, $\bar B \to X_s \gamma$, $K$-$\bar{K}$ mixing, $B^0_d$-$\bar{B}^0_d$ mixing, and $B^0_s$-$\bar{B}^0_s$ mixing, we obtain constraints on the parameters in the 2HDMs. 
To calculate the constraints, we pay attention to a determination of CKM matrix elements and re-fit them to experimental data so that new contributions from additional Higgs bosons do not affect the determination.  
As a result, we find that the charged Higgs boson mass less than around 490~GeV is ruled out from $\bar B \to X_s \gamma$ in the type~II and Y models, 
whereas large $\tan\beta$ is excluded from $B^0_s \to \mu^+ \mu^-$ in the type~II. 
We also see that severe constraints on the mass and couplings are put from $\bar B \to X_s \gamma$, $B^0_s \to \mu^+ \mu^-$, and $B^0_s$-$\bar{B}^0_s$ in the aligned model. 
In addition, we discuss excesses of observables in the muon anomalous magnetic moment and the semi-tauonic $B$ meson decays in the context of the 2HDM, 
and find that the aligned model can explain part of the excesses, compatible with the other constraints.} 

\begin{document}

\maketitle
\flushbottom

%%%%%%%%%%%%%%%%%%%%%%%%%%%%%%%%%%%%%%%%%%%%%%%%%%
\section{Introduction}
%%%%%%%%%%%%%%%%%%%%%%%%%%%%%%%%%%%%%%%%%%%%%%%%%%
Many experiments have investigated the validity of the standard model (SM). 
It has turned out that the SM can mostly accommodate the present low-energy experimental data. 
One of the most impressive features of the SM is the flavor structure. 
It is described by the Cabibbo-Kobayashi-Maskawa (CKM) matrix~\cite{Cabibbo:1963yz,Kobayashi:1973fv} in the Yukawa sector, which in turn requires only one Higgs doublet.  
Charged currents and $CP$-violations in the quark sector are controlled by the CKM matrix elements, 
whereas flavor changing neutral currents (FCNCs) are naturally suppressed due to the Glashow-Iliopoulos-Maiani (GIM) mechanism~\cite{Glashow:1970gm}.  
Extensions of the SM, especially in the Yukawa sector, easily and/or drastically change such a structure. 
Thus, flavor observables are good tools to test new physics involving such an extension.

Two-Higgs-Doublet Models (2HDMs) are minimal extensions of the SM with a limited number of parameters~\cite{Gunion:1989we}. 
In the general 2HDM, there exist tree-level FCNCs in Higgs boson interactions with fermions. 
A condition of {\it natural flavor conservation}~\cite{Glashow:1976nt} is usually imposed on the Yukawa sector to avoid such dangerous FCNCs.  
In principle, there are two ways to realize {\it natural flavor conservation} in the 2HDM. 
The first one is to introduce a $Z_2$ symmetry in the Lagrangian so that each fermion doublet couples only to one of the two Higgs doublets. 
Under this condition, there are four distinct types of $Z_2$ symmetric model, usually referred to as type~I, II, X (``lepton specific''), and Y (``flipped'') models (see, {\it e.g.}, Ref.~\cite{Aoki:2009ha}). 
A more general method is to impose an alignment condition on the flavor space of the Yukawa matrices~\cite{Pich:2009sp}, which we call as an aligned model in this paper. 
%If {\it natural flavor conservation} is imposed, the flavor structure is composed of the CKM matrix as well as the SM.  
From the viewpoint of low-energy phenomenology, the differences between these five models appear only in the parametrization of the Yukawa interaction terms that are sources of flavor transitions. 
The 2HDMs have been under detailed investigation for a long time, in particular the type~II model, which corresponds to the Yukawa interaction in the minimal supersymmetric SM. 
In Ref.~\cite{Deschamps:2009rh}, a constraint on the 2HDM of type~II from flavor observables have been calculated as well as the global fit of the CKM matrix elements, taking the charged Higgs effect into account. 
In Ref.~\cite{Jung:2010ik}, a comprehensive study for flavor phenomenologies in the aligned model has been done. 
A study for the $Z_2$ symmetric models is also given in Ref.~\cite{Cheng:2014ova}. 
As for studies focusing on individual processes and collider phenomenology in the 2HDMs, there are numerous papers these days (see Ref.~\cite{Branco:2011iw} and its references and citations for a review).

In recent years, theoretical calculations of higher order corrections on several flavor observables have been developed. 
The next-to-leading order (NLO) electroweak (EW) and next-to-next-to-leading order (NNLO) QCD corrections on $B_{s,d}^0 \to \ell^+\ell^-$ 
in the SM have been evaluated in Refs.~\cite{Bobeth:2013uxa,Hermann:2013kca,Bobeth:2013tba}. 
The contribution of the extra Higgs bosons in the 2HDM is also known~\cite{Li:2014fea,Cheng:2015yfu}. 
In Refs.~\cite{Misiak:2015xwa,Czakon:2015exa}, the updated NNLO QCD prediction of the branching ratio for $B \to X_s \gamma$ in the SM, considering all the available non-perturbative effects, has been reported, 
and the NNLO contribution in the 2HDM was also obtained in Ref.~\cite{Hermann:2012fc}. 
The complete one-loop calculation of the EW correction on the $B_s^0$-$\bar{B}_s^0$ mixing in the 2HDM is given in Ref.~\cite{Chang:2015rva}. 
Moreover, lattice studies on non-perturbative QCD quantities such as the meson decay constants and the bag parameters of neutral meson mixings have also been updated and collected recently~\cite{Aoki:2013ldr}.

Taking all the recent updates both for the theoretical calculations and the experimental results along with the other significant processes, we perform a comprehensive study of the type~I, II, X, Y, and aligned models 
and obtain the current status of the flavor constraints on the parameters in these models. 
In our study, we consider: the leptonic meson decays $B\to\tau\nu$, $D\to\mu\nu$, $D_s\to\tau\nu$, $D_s\to\mu\nu$, $K\to\mu\nu$, $\pi\to\mu\nu$, $B^0_s \to \mu^+ \mu^-$, and $B^0_d \to \mu^+ \mu^-$; 
the hadronic tau lepton decays $\tau \to K\nu$ and $\tau \to \pi\nu$; the radiative $B$ meson decay $B \to X_s \gamma$; and the neutral meson mixings $\Delta M_s$, $\Delta M_d$, and $| \epsilon_K|$. 
In addition, we discuss several observables in which deviations between SM predictions and experimental results have been reported, 
such as the semi-tauonic $B$ meson decays $R(D^{(*)})$ and the muon anomalous magnetic moment $\Delta a_\mu$, in the context of the 2HDMs. 
We also summarize formulae for the observables of these processes in the 2HDM and utilize them to calculate constraints on the parameters.   
To obtain the constraints, we underline all the uncertainties taken into account in our evaluation and, in particular, pay close attention to the determination of the CKM matrix elements since this is one of the dominant uncertainties. 
In this paper we consider a CP conserved Higgs potential, which means that the CP-odd Higgs boson is not mixed with the CP-even Higgs bosons.

Our paper is organized as follows. 
In Sec.~\ref{Sec:Yukawa}, we review the Yukawa sector and the parametrization of the 2HDM with the hypothesis of {\it natural flavor conservation}. 
The flavor observables used in our analysis are summarized in Sec.~\ref{Sec:FO}, and we show useful formulae for them. 
In Sec.~\ref{Sec:DetCKM} we obtain values of the CKM matrix elements by re-fitting CKM parameters to experimental data, so as to avoid effects from extra Higgs bosons in the 2HDM. 
In Sec.~\ref{Sec:Cons} we show the inputs and experimental data, and then we obtain the constraints on the 2HDMs. 
We also discuss the anomalies in $R(D^{(*)})$ and $\Delta a_\mu$. 
A summary is given in Sec.~\ref{Sec:Summary}.

%%%%%%%%%%%%%%%%%%%%%%%%%%%%%%%%%%%%%%%%%%%%%%%%%%
\section{Yukawa sector in 2HDM}
\label{Sec:Yukawa}
%%%%%%%%%%%%%%%%%%%%%%%%%%%%%%%%%%%%%%%%%%%%%%%%%%
When we consider two Higgs doublet fields $\Phi_1$ and $\Phi_2$ in a model with the SM fermion field contents, the most general Yukawa Lagrangian is given by
\begin{align}
\label{Eq:generalY}
-\mathcal{L}_Y= \sum_{a=1,2}
\left[ 
\bar Q_L \,\mathcal Y_d^a\, \Phi_a d_R +\bar Q_L \,\mathcal Y_u^a\, \tilde{\Phi}_a u_R +\bar L_L \,\mathcal Y_\ell^a\, \Phi_a \ell_R +\text{h.c.} 
\right] \,, 
\end{align}
where $\mathcal Y_f^a$ are flavor mixing complex matrices and $\tilde \Phi_a \equiv i\sigma_2 \Phi^*_a$. 
The vacuum expectation values (VEVs) are defined as $\langle \Phi_a \rangle = (0 \quad v_a)^T$. 
In general, this term immediately induces FCNCs via neutral Higgs bosons even at the tree level. 
To see it clearly, we can change the basis of the Higgs fields $\Phi_i$ into $\Psi_i$ so that 
\begin{align}
 \begin{pmatrix} \Phi_1 \\ \Phi_2 \end{pmatrix} 
 = 
 \mathcal R (\beta)
 \begin{pmatrix} \Psi_1 \\ \Psi_2 \end{pmatrix} \,,\quad
 \mathcal R(\theta)
 =
 \begin{pmatrix} \cos\theta & -\sin\theta \\ \sin\theta & \cos\theta \end{pmatrix} \,,
\end{align}
with  
\begin{align}
 \Psi_1 =
 \begin{pmatrix} G^+ \\ (v + h_1 +iG^0)/\sqrt 2 \end{pmatrix} \,,\quad
 \Psi_2 = 
 \begin{pmatrix} H^+ \\ (h_2 +i A)/\sqrt 2 \end{pmatrix} \,,
\end{align}
where $\tan\beta = v_2/v_1$ and $H^\pm (A)$ is a charged (CP odd) Higgs boson. 
The neutral Higgs fields are indicated as $h_1$ and $h_2$, which are not yet in the mass eigen basis. 
In this basis, the SM Higgs VEV $(v =\sqrt{v_1^2 +v_2^2}\,)$ and the NG bosons $(G^\pm, G^0)$ are contained in $\Psi_1$. 
Thus, we can rewrite the Yukawa Lagrangian in (\ref{Eq:generalY}) as 
\begin{align}
\label{Eq:higgsbasisY}
 -\mathcal{L}_Y= 
 \bar Q_L  \left( \hat Y_d \Psi_1 + \rho_d \Psi_2 \right) d_R 
 +\bar Q_L \left( \hat Y_u \tilde\Psi_1 + \rho_u \tilde\Psi_2 \right) u_R  
 +\bar L_L \left( \hat Y_\ell \Psi_1 + \rho_\ell \Psi_2 \right) \ell_R +\text{h.c.} \,,
\end{align}
where $\hat Y_f^{ij}$ are the SM Yukawa matrices which derive fermion mass matrices, and $\rho_f^{ij}$ are new couplings which are in general not diagonalized in the mass eigen basis. 
Note that $\rho_f^{ij}$ do not contribute to the fermion masses. 
The forms of $\hat Y_f$ and $\rho_f$ are described in terms of the original matrices $\mathcal Y_f^a$ as
\begin{align}
\label{Eq:MandY}
\hat Y_f = \mathcal Y_f^1\cos\beta + \mathcal Y_f^2\sin\beta \,,\quad 
\rho_f = -\mathcal Y_f^1\sin\beta + \mathcal Y_f^2\cos\beta \,. 
\end{align} 
We can see that the off-diagonal elements of $\rho_f$ cause FCNCs in the neutral Higgs sector at the tree level. 
So, it is required to impose a natural condition so that such FCNCs are suppressed, namely {\it natural flavor conservation}~\cite{Glashow:1976nt}, or directly constrain parameters inducing FCNCs by experiments. 
It is known that there are two ways to archive the flavor conservation in the neutral current.

\subsection{$Z_2$ symmetry}
%%%%%%%%%%%%%%%%%%%%%%%%
%%%%%%%%%%%%%%%%%%%%%%%%

%Table%
\begin{table}[t]\begin{center}
{\renewcommand\arraystretch{1.5}
\begin{tabular}{c|ccccccccc|c}
\hline\hline
&$\xi_u^h$ &$\xi_d^h$&$\xi_\ell^h$&$\xi_u^H$&$\xi_d^H$&$\xi_\ell^H$ &$\xi_u^A$&$\xi_d^A$&$\xi_\ell^A$ & condition\\ 
\hline
Type-I &$\frac{\cos\alpha}{\sin\beta}$&$\frac{\cos\alpha}{\sin\beta}$&$\frac{\cos\alpha}{\sin\beta}$&$\frac{\sin\alpha}{\sin\beta}$&$\frac{\sin\alpha}{\sin\beta}$&$\frac{\sin\alpha}{\sin\beta}$&$\cot\beta$&$-\cot\beta$&$-\cot\beta$&$\mathcal Y_{u,d,\ell}^1=0$\\ 
\hline
Type-II &$\frac{\cos\alpha}{\sin\beta}$&$-\frac{\sin\alpha}{\cos\beta}$&$-\frac{\sin\alpha}{\cos\beta}$&$\frac{\sin\alpha}{\sin\beta}$&$\frac{\cos\alpha}{\cos\beta}$&$\frac{\cos\alpha}{\cos\beta}$&$\cot\beta$&$\tan\beta$&$\tan\beta$&$\mathcal Y_{u}^1=\mathcal Y_{d,\ell}^2=0$\\
\hline
Type-X &$\frac{\cos\alpha}{\sin\beta}$&$\frac{\cos\alpha}{\sin\beta}$&$-\frac{\sin\alpha}{\cos\beta}$&$\frac{\sin\alpha}{\sin\beta}$&$\frac{\sin\alpha}{\sin\beta}$&$\frac{\cos\alpha}{\cos\beta}$&$\cot\beta$&$-\cot\beta$&$\tan\beta$&$\mathcal Y_{u,d}^1=\mathcal Y_{\ell}^2=0$\\
\hline
Type-Y &$\frac{\cos\alpha}{\sin\beta}$&$-\frac{\sin\alpha}{\cos\beta}$&$\frac{\cos\alpha}{\sin\beta}$&$\frac{\sin\alpha}{\sin\beta}$&$\frac{\cos\alpha}{\cos\beta}$&$\frac{\sin\alpha}{\sin\beta}$&$\cot\beta$&$\tan\beta$&$-\cot\beta$&$\mathcal Y_{u,\ell}^1=\mathcal Y_{d}^2=0$\\
\hline\hline
\end{tabular}}
\caption{Relations of the scaling factors defined in (\ref{Eq:Z2interaction}) for each type of $Z_2$ symmetric models.}
\label{Tab:Z2factor}
\end{center}\end{table}
%Table%
It is well-known that the FCNC can be prohibited by imposing a $Z_2$ symmetry on the fields in the Yukawa sector~\cite{Gunion:1989we}. 
This is realized so that two Higgs doublets $\Phi_1$ and $\Phi_2$ are assigned to be $Z_2$-even and -odd respectively such as $\Phi_1 \to + \Phi_1$ and $\Phi_2 \to -\Phi_2$. 
Due to this assignment, each field $f (=u,d,\ell)$ cannot have one of two original Yukawa matrices $\mathcal Y_f^a$, which immediately leads to the relation $\rho_f^{ij} \propto \hat Y_f^{ij}$ 
and thus the FCNC term does not appear in the Lagrangian. 
The protection against FCNCs is valid at any scale in the $Z_2$ symmetric models~\cite{Li:2014fea}, as can be seen later in the formula for $B_q^0 \to\ell^+\ell^-$.  
This procedure leads to four distinct types of Yukawa interactions.
In the mass eigen basis, it is summarized as  
\begin{align}
\label{Eq:Z2interaction}
\mathcal L_Y^\text{$Z_2$ symmetric} =
&- \sum_f^{u,d,\ell} \frac{m_f}{v} \Big[  (\xi_f^h\, h+ \xi_f^H\, H)\, {\bar f}\, P_R\, f - i \xi_f^A A {\bar f} \, P_R \, f \Big] \notag \\
&+ \frac{\sqrt{2}}{v} \Big[ V_{ud}\, \bar{u} \left( \xi_u^A\, m_u P_L+ \xi_d^A\, m_d P_R \right) d\, H^+ + \xi_\ell^A\, m_\ell\, \bar\nu\, P_R\, \ell \, H^+ \Big] +\text{h.c.}, 
\end{align}
where $V_{ud}$ is the CKM matrix element, $P_{\,^R_L} =(1\pm \gamma^5)/2$, 
$h$ and $H$ are the neutral CP even Higgs bosons in their mass eigen basis obtained by $(h_1 \quad h_2)^T = \mathcal R(\alpha -\beta) (H \quad h)^T$, 
and $\xi_f^\phi$ are scaling factors of the Yukawa couplings, dependent on the type of the model. 
The explicit expression for $\xi_f^\phi$ is listed in Table~\ref{Tab:Z2factor}. 
We see that the Yukawa sector is controlled by $\tan\beta$ and $\sin(\alpha-\beta)$ in this model.

\subsection{Alignment}
%%%%%%%%%%%%%%%%%%%%%%%%
%%%%%%%%%%%%%%%%%%%%%%%%

%Table%
\begin{table}[t]\begin{center}
\begin{tabular}{ccccc}
\hline\hline
 Aligned\,\,	&	\,\,Type I\,\,	&	\,\,Type II\,\,	&	\,\,Type X	\,\,	&	\,\,Type Y \\
\hline
\,\,$\zeta_u$	&	$\cot\beta$	&	$\cot\beta$	&	$\cot\beta$	&	$\cot\beta$ \\
\hline
\,\,$\zeta_d$	&	$\cot\beta$	&	$-\tan\beta$	&	$\cot\beta$	&	$-\tan\beta$ \\
\hline
\,\,$\zeta_\ell$	&	$\cot\beta$	&	$-\tan\beta$	&	$-\tan\beta$	&	$\cot\beta$ \\
\hline\hline
\end{tabular}
\caption{The relation between the $Z_2$ models and the aligned model.}
\label{Tab:Alinedfactor}
\end{center}\end{table}
%Table%
Another way to naturally forbid the tree level FCNC is worked out by taking the two Yukawa matrices to be aligned such as $\mathcal Y_f^2 \propto \mathcal Y_f^1$, or equivalently 
\begin{align}
 \label{Eq:Acondition}
 \rho_f^{ij} \equiv \zeta_f \hat Y_f^{ij} \,, 
\end{align}
where $\zeta_f$ for $f=u,d$, and $\ell$ are family universal scaling factors\footnote{
We can also define the alignment condition such as $\mathcal Y_f^2 = \bar\zeta_f \mathcal Y_f^1$. 
Under this condition, $\rho_f$ is also proportional to $\hat Y_f$. 
The relation between $\bar\zeta_f$ and $\zeta_f$ is written by 
\begin{align*}
 \zeta_f = {\bar\zeta_f - \tan\beta \over 1+ \bar\zeta_f \tan\beta} \,.
\end{align*}
}. 
The parameters $\zeta_f$ can be complex values, in principle. 
If $\zeta_{u,d}$ are complex values, determinations of the CP phase in the CKM matrix are affected. 
In this paper, we take $\zeta_f$ to be real. 
In this case, the scaling factors of the Yukawa couplings in (\ref{Eq:Z2interaction}) are written as 
\begin{align}
\label{Eq:ALinteraction}
 & \xi_f^H = \cos(\alpha-\beta) + \zeta_f \sin(\alpha-\beta) \,, \\
 & \xi_f^h = -\sin(\alpha-\beta) + \zeta_f \cos(\alpha-\beta) \,, \\
 & \xi^A_f = \begin{cases} + \zeta_f & \quad f=u \\ - \zeta_f & \quad f=d,\ell \end{cases}  \,. 
\end{align}
We can see from Table~\ref{Tab:Alinedfactor} that the $Z_2$ symmetric types can be considered as the particular cases of the aligned model. 
In this model, the alignment condition as in (\ref{Eq:Acondition}) is only guaranteed at the scale where the model is defined. 
Namely, non-zero contributions from running effects can be generated at a low energy scale, see {\it e.g.}, Ref.~\cite{Li:2014fea}. 
We see such effects in $B_q^0 \to\ell^+\ell^-$ later. 
We have practically $\sin(\alpha-\beta)$ and $\zeta_f$ as the model parameters in the Yukawa sector and $\tan\beta$ is irrelevant to this sector.  
A typical difference of this model from the $Z_2$ symmetric models is that $\zeta_f$ is nothing to do with the fermion mass and the VEVs. 
Hence it can be arbitrary.

\subsection{Type III}
%%%%%%%%%%%%%%%%%%%%%%%%
%%%%%%%%%%%%%%%%%%%%%%%%
Indeed, we can consider the case that the FCNC interactions appear at the tree level. 
This class of model is called as a type III model and directly obtained from (\ref{Eq:higgsbasisY}) without taking any condition. 
In this model, there are a lot of couplings which can induce the FCNC transitions and thus, they are severely constrained by experiments. 
For an overview of this model and flavor constraints, see Ref.~\cite{Crivellin:2013wna}. 
Recent studies for the top quark FCNC in the type~III are given in Refs.~\cite{Kao:2011aa,Craig:2012vj,Chen:2013qta,Atwood:2013ica,CMS:2014qxa,Altunkaynak:2015twa,Gaitan-Lozano:2014nka,Gaitan:2015hga}. 
As for the lepton flavor violating decay of the Higgs, there are several investigations in Refs.~\cite{Harnik:2012pb,Sierra:2014nqa,deLima:2015pqa,Bressler:2014jta}. 
The decay can be related to a muon anomalous magnetic moment in this class of model~\cite{Omura:2015nja}. 
Some flavor structures can be derived if a global symmetry is imposed on the fields. 
The BGL model has been proposed by considering such a symmetry~\cite{Branco:1996bq}. 
This model prohibits FCNCs in the up-type quarks. 
On the other hand, FCNCs in the down-type quarks are controlled by the CKM matrix elements. 
This class of models has been recently analyzed in Refs.~\cite{Botella:2014ska,Botella:2015hoa}.

%%%%%%%%%%%%%%%%%%%%%%%%%%%%%%%%%%%%%%%%%%%%%%%%%%
\section{Flavor observables}
\label{Sec:FO}
%%%%%%%%%%%%%%%%%%%%%%%%%%%%%%%%%%%%%%%%%%%%%%%%%%
In the 2HDM, the charged Higgs boson $H^\pm$ can contribute to flavor observables via the charged current. 
By considering the $Z_2$ symmetry or the alignment condition, interesting features appear in the Yukawa interaction term as follows. 
%(1) 
The tree-level interaction term of $H^\pm$ with quarks has the same CKM structure with that of $W^\pm$. 
%(2) 
Higher oder corrections due to loop diagrams through extra Higgs bosons provide additional contributions to FCNC processes in the quark sector.  
%even though the tree level contribution from neutral Higgs bosons $h,H,A$ are forbidden.  
%(3) 
On the other hand,  the lepton flavor violation is quite suppressed due to a tiny neutrino mass contribution.

A sensitivity of an observable to a new physics model, from the viewpoint of limiting a new physics model, depends on the situation 
such that how much precisely the observable is measured and how much the parameters involved in the new physics model can enhance the effect on the observable. 
To clarify which observables are important to constrain the 2HDM, we classify flavor observables as follows. 
\begin{itemize}
 \item[(a)] Potentially sensitive to the extra Higgs bosons
 \item[(b)] Insensitive to the extra Higgs bosons due to \vspace{-0.5em}
 \item[] \begin{itemize} \item[ -1. ] no enhancement in the contributions \item[ -2. ] low precision and/or large uncertainty \end{itemize}
 \item[(c)] Anomalies 
\end{itemize}
Obviously, observables in the category (a) are significant in order to limit the 2HDM. 
The category (b) plays an important role when we fit the CKM matrix elements in the 2HDM.  
In the SM, they can be determined by a fit to experimental data for each element, or we can also perform a global fit to a specific parametrization. 
In the 2HDM, however, some of observables are affected by the extra Higgs bosons and thus the CKM matrix elements must be re-fitted by taking such an effect into account. 
In particular, the CKM matrix elements should be determined by the observables classified as (b-1) and not (b-2).  
The category (c) means that a discrepancy between a SM prediction and an experimental result exists in an observable. 
Such an observable is also important for an indirect evidence of the additional scalar bosons.

In the following subsection, we  summarize theoretical formulae of flavor observables classified as (a) and (c), which are useful to constrain the 2HDM. 
After that in the next section, we discuss the way to obtain the CKM matrix elements by the global fit  with use of observables in (b).

\subsection{$M^\pm \to \ell^\pm \nu$ and $\tau^\pm \to M^\pm \nu$}
%%%%%%%%%%%%%%%%%%%%%%%%
%%%%%%%%%%%%%%%%%%%%%%%% 
In the 2HDM, decay processes $M^\pm \to \ell^\pm \nu$ and $\tau^\pm \to M^\pm \nu$ occur at the tree level and their branching ratios are given by 
\begin{align}
 &\mathcal B (M^\pm \to \ell^\pm \nu) =
 \frac{\tau_M G_F^2 m_M m_\ell^2}{8\pi} 
 \left( 1-\frac{m_\ell^2}{m_M^2} \right)^2
 \left | V_{ud} \right |^2 f_M^2 |1 + \mathcal C_H|^2\,, \\
 &\mathcal B (\tau^\pm \to M^\pm \nu) = 
  \frac{\tau_\tau G_F^2 m_\tau^3}{16\pi} 
 \left( 1-\frac{m_M^2}{m_\tau^2} \right)^2
 \left | V_{ud} \right |^2 f_M^2 |1 + \mathcal C_H|^2\,,
\end{align}
where $M^+$ is the meson which consists of $(u \bar d)$, $f_M$ is its decay constant, $\tau_M$ is its lifetime, and $V_{ud}$ is the relevant CKM matrix element in the process.  
In the following part, we omit the notation of charge assignment unless otherwise stated. 
The contribution from the charged Higgs boson is encoded in $\mathcal C_H$, written by 
\begin{align}
  \label{Eq:MtoEllNu_CH}
  \mathcal C_H = \frac{ \xi_\ell^{A} \xi_u^A\, m_u - \xi_\ell^{A} \xi_d^A\, m_d }{m_u + m_d} \cdot \frac{m_M^2}{m_{H^\pm}^2}\,,
\end{align}
where $\xi_f^A$ is defined in (\ref{Eq:Z2interaction}) and applied to each specific model. 
The main sources of non-negligible uncertainties for theoretical evaluations are $f_M$ and $V_{ud}$ as shown later.

The branching ratios of $B \to\tau\nu$, $D\to\mu\nu$, $D_s\to\tau\nu$, and $D_s\to\mu\nu$ are notable observables in the 2HDM. 
The processes $B \to\mu\nu$ and $D\to\tau\nu$ are less sensitive to new physics since only upper limits of the branching ratios are given by experiments for now. 
For the $K$, $\pi$, and $\tau$ decays, the ratio
\begin{align}
 \frac{\mathcal B (K \to \mu \nu)}{\mathcal B (\pi \to \mu \nu)} \,, \quad \frac{\mathcal B (\tau \to K \nu)}{\mathcal B (\tau \to \pi \nu)} \,,
 \label{Eq:Bratios}
\end{align}
can be significant thanks to smaller theoretical uncertainties and precise experimental data. 
In this case, the long distance electromagnetic corrections enter as $(1 + \delta_\text{EM}^{K/\pi})$ and $(1 + \delta_\text{EM}^{K/\pi,\tau})$ in (\ref{Eq:Bratios}), respectively. 
These corrections are very small, but include uncertainties taken into account. 
Input values are shown in Sec.~\ref{Sec:Cons}.

\subsection{$M^0\to\ell^+\ell^-$}
%%%%%%%%%%%%%%%%%%%%%%%%
%%%%%%%%%%%%%%%%%%%%%%%%
Pure leptonic decays of neutral meson $M^0\to\ell^+\ell^-$ can probe the effect of the 2HDM via loop contributions. 
For instance, the branching ratio of $B^0_{q}\to\ell^+\ell^-$ can be written by
\begin{align}
 & \mathcal B (B^0_q \to \ell^+ \ell^-) 
 = \mathcal B (B^0_q \to \ell^+ \ell^-)_\text{SM} \left( |P|^2 +|S|^2 \right) \,, \\
 & \mathcal B (B^0_q \to \ell^+ \ell^-)_\text{SM} 
 = \frac{\tau_{B^0_q} G_F^4 m_W^4}{8\pi^5} \left| V_{tb} V_{tq}^*\, \mathcal C_{10}^\text{SM} \right|^2 f_{B_q^0}^2 m_\ell^2 m_{B_q^0} \sqrt{1 -\frac{4m_\ell^2}{m_{B_q^0}^2} } \,,
\end{align}
with   
\begin{align}
  \label{Eq:PSfunction}
  P = \frac{\mathcal C_{10}}{\mathcal C_{10}^\text{SM}} + \frac{m_{B_q^0}^2}{2m_W^2} \left( \frac{m_b}{ m_b + m_q } \right) \frac{\mathcal C_P }{\mathcal C_{10}^\text{SM}}, \quad\quad 
  S = \sqrt{1 - \frac{4m_\ell^2}{m_{B_q^0}^2} } \frac{m_{B_q^0}^2}{2m_W^2} \left( \frac{m_b}{ m_b + m_q } \right) \frac{\mathcal C_S }{\mathcal C_{10}^\text{SM}} \,, 
\end{align}
where the Wilson coefficients $\mathcal C_{10}$, $\mathcal C_{P}$, and $\mathcal C_{S}$ show the contributions from the effective operators 
$\mathcal O_{10}=(\bar b\gamma_\mu P_Ls)(\bar\ell\gamma^\mu\gamma_5 \ell)$, $\mathcal O_{S}=\frac{m_\ell m_b}{m_W^2} (\bar b P_R s)(\bar\ell  \ell)$, 
and $\mathcal O_{P}=\frac{m_\ell m_b}{m_W^2} (\bar b P_R s)(\bar\ell \gamma_5 \ell)$, respectively. 
Since the branching ratio is quite small, the effect of neutral meson mixing is not negligible. 
To involve such an effect, the averaged time-integrated branching ratio is defined~\cite{DeBruyn:2012wj,DeBruyn:2012wk,Buras:2013uqa} as 
\begin{align} 
 & \overline{\mathcal B} (B^0_q \to \ell^+ \ell^-) = \overline{\mathcal B} (B^0_q \to \ell^+ \ell^-)_\text{SM} \left[ |P|^2 + \left( 1 - \Delta\Gamma_q\, \tau_{B^0_q}^L \right) |S|^2 \right] \,, \label{Eq:barBrNewPhysics} \\
 & \overline{\mathcal B} (B^0_q \to \ell^+ \ell^-)_\text{SM} =  \frac{\tau_{B^0_q}^H}{\tau_{B^0_q}} \mathcal B (B^0_q \to \ell^+ \ell^-)_\text{SM}  \,, \label{Eq:barBrStandardModel}
\end{align} 
where $\tau_{B^0_q}^{H(L)}$ is the life time corresponding to the heavier (lighter) eigenstate of $B^0_q$, and $\Delta\Gamma_q$ is the decay width difference.  
This is the actual observable which can be compared with experimental data. 
A brief review of this observable is shown in Appendix~\ref{App:form_Bsmumu}.

The coefficient $\mathcal C_{10}^\text{SM}$ includes the dominant SM contribution from $\mathcal O_{10}$. 
At the leading order, it is evaluated as $\mathcal C_{10}^\text{SM-LO} = - \frac{x_t}{8} [ \frac{x_t-4}{x_t-1} + \frac{3x_t}{(x_t-1)^2} \ln x_t ]$, where we define $x_q$ as 
%\begin{align}
% \mathcal C_{10}^\text{SM-LO} = - \frac{x_t}{8} \left[ \frac{x_t-4}{x_t-1} + \frac{3x_t}{(x_t-1)^2} \ln x_t \right] \,,
%\end{align}
%where we define the mass ratio as 
\begin{align}
 x_q = \frac{m_q^2}{m_W^2} \,. 
\end{align}
From recent studies in Refs.~\cite{Bobeth:2013uxa,Hermann:2013kca,Bobeth:2013tba} evaluating $\mathcal C_{10}^\text{SM}$ up to the NNLO QCD and NLO EW corrections, we obtain the following fit formula: 
\begin{align}
 \mathcal C_{10}^\text{SM} = -0.9380 \left( \frac{M_t}{173.1\,\text{GeV}} \right)^{1.53} \left( \frac{\alpha_s(m_Z)}{0.1184} \right)^{-0.09}  \,,
 \label{Eq:C10smNNLO}
\end{align}
where $\alpha_s(m_Z)$ is the QCD running coupling at the $m_Z$ scale and $M_t$ shows the pole mass of the top quark. 
In our analysis, we use (\ref{Eq:C10smNNLO}) as the SM prediction. 
The scalar and pseudo-scalar type effects in $\mathcal C_S$ and $\mathcal C_P$ are suppressed in the SM.

The explicit one-loop order calculation for $B^0_{q}\to\ell^+\ell^-$ in the aligned 2HDM has been done in Ref.~\cite{Li:2014fea}. 
The similar calculation in the $Z_2$ symmetric 2HDMs is also given in Ref.~\cite{Cheng:2015yfu}. 
In the 2HDM, the charged Higgs boson contributes to $\mathcal C_{10}$ via $Z$-penguin diagrams. 
It is described as  
\begin{align}
 \mathcal C_{10} = \mathcal C_{10}^\text{SM} + ( \xi_u^A )^2\, \frac{x_t^2}{8}\, \left[\frac{1}{x_{H^+}-x_t} + \frac{x_{H^+}}{(x_{H^+}-x_t)^2}\,\left(\ln x_t - \ln x_{H^+}\right)\right] \,. 
\end{align}
The neutral Higgs bosons give contributions in $\mathcal C_{S}$ and $\mathcal C_{P}$. 
The contributions are divided by two parts such as 
 \begin{align}
 \mathcal C_{S,P} = \mathcal C_{S,P}^\text{c} + \mathcal C_{S,P}^\text{n} \,.
\end{align}
The first terms $\mathcal C_{S}^c$ and $\mathcal C_{P}^c$ come from box diagrams in the Unitary gauge (box, $Z$-penguin, and Goldstone-penguin diagrams in the Feynman gauge) and are given by 
\begin{align}
 \mathcal C_{S}^\text{c} = 
%Feynman Gauge:
 & \,\,\mathcal C_{S}^\text{c,\,SM} + \frac{x_t}{8 (x_{H^+}-x_t)}\,\Bigg\{ 2\xi_d^A \xi_\ell^{A} \left[\frac{1}{x_{H^+}-1}\,\ln x_{H^+} - \frac{1}{x_t-1}\,\ln x_t\right] \nonumber\\[0.2cm]
 & +\xi_u^A \xi_\ell^{A} \left[\frac{1}{x_{H^+}-1} + \frac{ x_{H^+}}{(x_{H^+}-x_t)(x_t-1)}\,\ln x_t - \frac{x_{H^+}(2x_{H^+}-x_t-1)}{(x_{H^+}-x_t)(x_{H^+}-1)^2}\,\ln x_{H^+}\right] \nonumber\\[0.2cm]
 & -\xi_\ell^A \xi_u^{A}\, \left[\frac{x_t-x_{H^+}}{(x_{H^+}-1)(x_t-1)} + \frac{x_t}{(x_t-1)^2}\,\ln x_t - \frac{x_{H^+}}{(x_{H^+}-1)^2}\,\ln x_{H^+}\right] \Bigg\}\,, 
 \label{Eq:CSc}
\end{align}
\begin{align}
 \mathcal C_{P}^\text{c} = 
 & \,\, \mathcal C_{P}^\text{c,\,SM} - \frac{x_t}{8 (x_{H^+}-x_t)}\,\Bigg\{ 2\xi_d^A \xi_\ell^{A} \left[\frac{1}{x_{H^+}-1}\,\ln x_{H^+} - \frac{1}{x_t-1}\,\ln x_t\right] \notag \\[0.2cm]
 & +\xi_u^A \xi_\ell^{A} \left[\frac{1}{x_{H^+}-1} + \frac{ x_{H^+}}{(x_{H^+}-x_t)(x_t-1)}\,\ln x_t - \frac{x_{H^+}(2x_{H^+}-x_t-1)}{(x_{H^+}-x_t)(x_{H^+}-1)^2}\,\ln x_{H^+}\right] \notag \\[0.2cm]
 & +\xi_\ell^A \xi_u^{A}\, \left[\frac{x_t-x_{H^+}}{(x_{H^+}-1)(x_t-1)} + \frac{x_t}{(x_t-1)^2}\,\ln x_t - \frac{x_{H^+}}{(x_{H^+}-1)^2}\,\ln x_{H^+}\right] \Bigg\} \notag \\[0.2em] 
 & +\frac{x_t}{4 (x_{H^+}-x_t)^2}\, \Bigg\{ -\xi_d^A \xi_u^{A} \left[ -\frac{x_t + x_{H^+}}{2} + \frac{x_t x_{H^+}}{x_{H^+} - x_t}  \ln \frac{x_{H^+}}{x_t} \right] \notag \\[0.2em] 
 & + \frac{( \xi_u^A )^2}{ 6(x_{H^+}-x_t) } \left[  \frac{ x_{H^+}^2 -8x_{H^+} x_t -17x_t^2 }{6} +\frac{x_t^2 (3x_{H^+} +x_t)}{x_{H^+}-x_t} \ln \frac{x_{H^+}}{x_t} \right] \Bigg\} \notag \\[0.2em] 
 & + s_W^2 \frac{x_t}{ 6(x_{H^+}-x_t)^2 } \Bigg\{ -\xi_d^A \xi_u^{A} \left[ \frac{5x_t -3x_{H^+}}{2} +\frac{x_{H^+} (2x_{H^+} -3x_t)}{x_{H^+}-x_t} \ln \frac{x_{H^+}}{x_t} \right] \notag \\[0.2em]
 & +\frac{( \xi_u^A )^2}{ 6(x_{H^+}-x_t) } \left[ \frac{4x_{H^+}^3 \!-\!12x_{H^+}^2x_t \!+\! x_{H^+}x_t^2 \!+\! 3x_t^3}{x_{H^+}-x_t} \ln \frac{x_{H^+}}{x_t}
     -\frac{17x_{H^+}^2 \!-\! 64x_{H^+}x_t \!+\! 71x_t^2}{6}  \right] \Bigg\} \notag \\[0.2em]
 & + c_W^2 ( \xi_u^A )^2 \frac{x_t^2}{4(x_{H^+}-x_t)^2} \left[ x_{H^+} \ln \frac{x_{H^+}}{x_t} + x_t - x_{H^+} \right]  \,\,,
 \label{Eq:CPc}
\end{align}
where $\mathcal C_{S,P}^\text{c,\,SM}$ show the SM contributions that are given in Appendix~\ref{App:form_Bsmumu}.  
The second terms $\mathcal C_{S}^\text{n}$ and $\mathcal C_{P}^\text{n}$ indicate the contribution from scalar boson exchange diagrams. 
They can be expressed as  
\begin{align}
 \mathcal C_{S}^\text{n} =
 & \, x_t \Big[ F_0 + \xi_\ell^{A} \left( \xi_d^A F_1 +\xi_u^A F_2 \right) +\xi_\ell^{A}\xi_u^{A} F_3  \Big] \notag \\[0.2em] 
 & +\frac{x_t}{2 x_h} \left( s_{\alpha-\beta} + c_{\alpha-\beta}\, \xi_\ell^{A} \right)  
 \Big[ 
 s_{\alpha-\beta}\, G_1 (\xi_u^{A},\xi_d^{A},x_{H^+},x_t) + c_{\alpha-\beta}\, G_2 (\xi_u^{A},\xi_d^{A},x_{H^+},x_t)
 \Big] \notag \\[0.2em]
 & +\frac{x_t}{2 x_H} \left( c_{\alpha-\beta} - s_{\alpha-\beta}\,\xi_\ell^{A} \right) 
 \Big[ 
 c_{\alpha-\beta}\, G_1 (\xi_u^{A},\xi_d^{A},x_{H^+},x_t) - s_{\alpha-\beta}\, G_2 (\xi_u^{A},\xi_d^{A},x_{H^+},x_t)
 \Big] \,, 
 \label{Eq:CSn}
\end{align}
\begin{align}
 \mathcal C_{P}^\text{n} = 
  \, x_t \Big[ - \xi_\ell^{A} \left( \xi_d^A F_1 +\xi_u^A F_2 \right) +\xi_\ell^{A}\xi_u^{A} F_3  \Big] +\frac{x_t}{2 x_A} \xi_\ell^{A} G_3 (\xi_u^{A},\xi_d^{A},x_{H^+},x_t)  \,, 
 \label{Eq:CPn}
\end{align}
where $c_\theta = \cos\theta$, $s_\theta = \sin\theta$, and $F_i \equiv F_i (x_t,x_{H^+})$ are functions in terms of $x_t$ and $x_{H^+}$. 
The analytic expressions for $G_i$ and $F_i$ are given in Appendix~\ref{App:form_Bsmumu}. 
Note that the effects from $\mathcal C_{P}$ and $\mathcal C_{S}$ on the branching ratio is suppressed by the factor $m_{B_q^0}^2 /m_W^2 \sim 0.004$ as seen in (\ref{Eq:PSfunction}). 
In the $Z_2$ symmetric model, however, $\mathcal C_{P,S}$ can be enhanced for large $\tan\beta$, relative to $\mathcal C_{10}$. 
In the aligned model, we can obtain dominant constraints on $\zeta_d$ and $\zeta_\ell$ from this observable. 
Further details for this analytic formula are found in Appendix~\ref{App:form_Bsmumu}. 
The dominant uncertainties for this process are the decay constant $f_{B_q^0}$ and the product of CKM matrix elements $V_{tb} V_{tq}^*$ as well as the $M \to\ell\nu$ case.

The formulae for $D^0 \to\ell^+\ell^-$ and $K_L \to\ell^+\ell^-$ are easily given by replacing masses and CKM components as appropriate. 
In these two cases, however, there exist non-negligible long distance contributions and thus we must concern that effect in addition to the short distance contributions given above. 
Then they are less significant to constrain the 2HDMs.

\subsection{Neutral meson mixing}
%%%%%%%%%%%%%%%%%%%%%%%%
%%%%%%%%%%%%%%%%%%%%%%%%
The 2HDM gives additional contributions to neutral meson mixings and affects measurements of the CKM matrix elements from mixing observables. 
For the $B_q^0$-$\bar B_q^0$ mixing, the mass difference between two CP eigenstates of $B_q^0$ and $\bar B_q^0$ defined as 
\begin{align}
  \Delta M_q = 2 \left| \langle B_q^0 | H^{\Delta B=2} | \bar B_q^0 \rangle \right| \,,
\end{align}
is used to determine the parameters in the CKM matrix in the case of the SM, since this is less sensitive to a long distance effect. 
In the 2HDM, the exchanges of the charged Higgs boson in the box diagrams contribute to $\Delta M_q$~\cite{Buras:1989ui,Barger:1989fj}. 
A complete one-loop calculation without neglecting the term proportional to $x_b$ is given in Ref.~\cite{Chang:2015rva}. 
We express the analytic formula as    
\begin{align}
  \Delta M_q 
  = \frac{G_F^2 m_W^2 m_{B_q} }{24\pi^2} |V_{tq} V_{tb}^* |^2 f_{B_q}^2 \left[ \hat B_{B_q} \eta_{B_q}\, \mathcal C_{V} +\hat B_{B_q}^{ST} \eta_{B_q}^{ST}\, \mathcal C_{ST} \right] \,, 
  \label{Eq:deltaM}
\end{align}
where $\mathcal C_{V}$ comes from the effective operator $\mathcal O_{VLL} = \bar s^\alpha \gamma_\mu (1-\gamma_5) b^\alpha \bar s^\beta \gamma^\mu (1-\gamma_5) b^\beta$; 
$\mathcal C_{ST}$ shows the combined contribution from $\mathcal O_{SRR} = \bar s^\alpha (1+\gamma_5) b^\alpha \bar s^\beta (1+\gamma_5) b^\beta$ 
and $\mathcal O_{TRR} = \bar s^\alpha \sigma_{\mu\nu} (1+\gamma_5) b^\alpha \bar s^\beta \sigma^{\mu\nu} (1+\gamma_5) b^\beta$; 
$\hat B_{B_q}$ ($\hat B_{B_q}^{ST}$) is the bag parameter for $\mathcal O_{VLL}$ ($\mathcal O_{SRR}$ and $\mathcal O_{TRR}$); 
%the SM (scalar and tensor) operator, usually referred to as ``$VLL$'' (``$SRR$'' and ``$TRR$''), 
and $\eta_{B_q}^{(ST)}$ involves a running effect of the QCD correction from the matching scale of $\mathcal C_{V(ST)}$ to the low energy scale. 
The Wilson coefficients are then written by 
\begin{align}
 &\mathcal C_{V} = x_t \Big( A_{WW}(x_t) + 2x_t\, A_{WH}(x_t,x_b) +x_t\, A_{HH}(x_t,x_b) \Big) \,, \\
 &\mathcal C_{ST} = 4\,x_b\,x_t^2 \Big(  A_{WH}^{ST}(x_t) + A_{HH}^{ST}(x_t) \Big) \,,
\end{align}
where $A_{WW}(x_t)$ contains the SM contribution and the others $A_{VV'}$ are from the charged Higgs boson. 
The explicit forms are given in Appendix~\ref{App:form_mixing}. 
Note that there exist SM contributions in $\mathcal C_{ST}$, which is usually neglected due to a large suppression by $x_b$. 
In the 2HDM, the term proportional to $x_b$ can be enhanced by additional factors such as $\xi_u^A$ and $\xi_d^A$. 
The formulae of $A_{WH}^{ST}$ and $A_{HH}^{ST}$, as given in (\ref{Eq:WHst}) and (\ref{Eq:HHst}), have been obtained in Ref.~\cite{Chang:2015rva} 
by taking nonzero external momenta into account and thus they are different from those given in Ref.~\cite{Barger:1989fj}. 
We independently obtained the same result for (\ref{Eq:WHst}) and (\ref{Eq:HHst}). 
In addition, we found the $x_b$ terms in $A_{WH}$ and $A_{HH}$ as shown in (\ref{Eq:WH}) and (\ref{Eq:HH}). 
We stress that they are new contributions which are not described in Ref.~\cite{Chang:2015rva}.

As for the $K^0$-$\bar K^0$ mixing, the mass difference $\Delta M_K$ is not a good observable due to a non-negligible long distance effect. 
Instead, the $\epsilon$ parameter which is defined as 
\begin{align}
  \epsilon_K \equiv \frac{1}{ \sqrt{\Delta M_K^2 +\Delta \Gamma_K^2/4} } \, \text{Im} \left( \langle K^0 | H^{\Delta K=2} | \bar K^0 \rangle \right) \,,
\end{align}
is the measurement of the CP violation in the $K^0$-$\bar K^0$ system. 
In the 2HDM, the analytic formula can be written\cite{Barger:1989fj} as 
\begin{align}
 \epsilon_K = 
 &\frac{G_F^2 m_W^2 m_K}{48\sqrt 2 \pi^2 \Delta M_K} f_K^2 \hat B_K \notag \\
 &\times \Big\{
 \eta_{tt} \left( V_{ts}V_{td}^* \right)^2 x_t \big[ A_{WW}(x_t) + 2x_t A_{WH}(x_t,0) +x_t A_{HH}(x_t,0) \big] \notag  \\
 &\hspace{2em} +
 \eta_{cc} \left( V_{cs}V_{cd}^* \right)^2 x_c \big[ A_{WW}(x_c) + 2x_c A_{WH}(x_c,0) +x_c A_{HH}(x_c,0) \big]  \notag \\
 &\hspace{2em} +
 2\eta_{ct} \left( V_{cs}V_{cd}^* V_{ts}V_{td}^* \right) x_t x_c \big[ B_{WW}(x_t,x_c) + 6B_{WH}(x_t,x_c) + B_{HH}(x_t,x_c)  \big] 
 \Big\} \,, \label{Eq:epsilonK}
\end{align}
where $A_{VV'}$ are the same forms with those for $\Delta M_q$, $B_{VV'}$ are additional functions expressed in Appendix~\ref{App:formulae}, and $\eta_{qq'}$ indicates the QCD corrections for each pair of the internal quarks. 
In this formula, $x_s=0$ is assumed and thus there is no contribution from the $SRR$ and $TRR$ operators\footnote{
It is often stated that the matrix element for the scalar type operator can give a larger contribution than that for the SM vector operator due to a chiral enhancement. 
But, the loop diagram for the coefficient of the matrix element provides the suppression by $x_s$ in the 2HDM and then in total the contribution from the scalar sector is suppressed by $\sim m_K^2/m_W^2$. 
}.   
Note that we use the relation $\Delta\Gamma_K \simeq 2\Delta M_K$ in (\ref{Eq:epsilonK}) and the experimental data of $\Delta M_K$ as the input value in our numerical analysis.

For $\epsilon_K$ and $\Delta M_{q}$, the uncertainty from the bag parameter $\hat B_M^{(ST)}$ is taken into account as well as the decay constant and the CKM matrix element. 
Later we show and summarize the detail of input parameters. 
As for the QCD corrections in the $VLL$ operator, we use the following values\cite{Buchalla:1995vs}, 
\begin{align}
 \eta_{B_d} = \eta_{B_s} = 0.551 \,, 
 \quad 
 \eta_{cc} = 1.380 \,,
 \quad
 \eta_{ct} = 0.470 \,,
 \quad
 \eta_{tt} = 0.574 \,.
\end{align}
The QCD correction of $\eta_{B_q}$ from the extra Higgs bosons has been obtained in Ref.~\cite{Urban:1997gw}. 
In our study, we simply neglect that effect. 
Practical input values of the bag parameter and QCD correction $\hat B_{B_q}^{ST} \eta_{B_q}^{ST}$ in the $SRR$ and $TRR$ operators are shown in Sec.~\ref{Sec:Cons}.

In the $D^0$-$\bar D^0$ mixing, there is no feasible observable in which a long distance effect is sufficiently suppressed. 
In principle, we can give a constraint in the $D^0$-$\bar D^0$ mixing by taking such an unknown effect as one of uncertainties. 
This strategy can be applied to other observables such as $\epsilon'_K$, $\Delta M_K$ and so on, but in this paper we do not consider such a case.

\subsection{$\bar B \to X_q \gamma$}
%%%%%%%%%%%%%%%%%%%%%%%%
%%%%%%%%%%%%%%%%%%%%%%%%
Inclusive radiative decays of $B$ meson, $\bar B \to X_q \gamma$, are one of the most interesting FCNC processes, 
and hence they have been precisely evaluated to bound on several new physics models. 
Since the hadron transition occurs in this process, perturbative QCD corrections are much important and non-perturbative effects must be concerned as well. 
According to the recent summary that collects all the available and relevant contributions presented in Refs.~\cite{Misiak:2015xwa,Czakon:2015exa}, the SM predicts 
\begin{align}
 &\overline{\mathcal B} (b \to s \gamma)_{E_\gamma > E_0} = (3.36 \pm 0.23) \times 10^{-4} \,,  \label{Eq:bsgamma} \\
 &\overline{\mathcal B} (b \to d \gamma)_{E_\gamma > E_0} = (1.73^{+0.12}_{-0.22}) \times 10^{-5} \,, \label{Eq:bdgamma}
\end{align}
for $E_0 = 1.6\,\text{GeV}$, where $E_0$ indicates the photon cutoff energy and $\overline{\mathcal B} (b \to q \gamma)$ is the CP- and isospin-averaged branching ratio of $\bar B \to X_q \gamma$. 
For the evaluation of (\ref{Eq:bsgamma}), perturbative QCD corrections up to the NNLO\cite{Misiak:2006zs,Misiak:2006ab,Becher:2006pu,Misiak:2010sk,Huber:2014nna} 
and calculable long-distance effects, (see Ref.~\cite{Czakon:2015exa} and its references for more detail), are taken into account. 
The uncertainty in (\ref{Eq:bsgamma}) comes from non-perturbative part ($\pm 5\%$), input parameters ($\pm 2\%$), and others ($\sim\!\pm 4\%$). 
The product of the CKM matrix elements, defined as 
\begin{align}
 r_V = \left| \frac{V_{ts}^*V_{tb}}{V_{cb}} \right|^2 \,,
\end{align}
is contained in the input parameters. 
In Refs.~\cite{Misiak:2015xwa,Czakon:2015exa}, the latest result of the CKM fit in Ref.~\cite{Charles:2015gya} are applied. 
To separate the CKM product from the observable, we employ the following expression for the SM prediction: 
\begin{align}
 \overline{\mathcal B} (b \to s \gamma)^\text{SM} 
 = \left[ 3.36 \left( \frac{r_V^{\,}}{0.9626_{\,}} \right) \pm \sqrt{0.23^2+\delta r_V^2} \right] \times 10^{-4} \,,
\end{align}
where $\delta r_V$ denotes the uncertainty derived from $r_V$. 
We note that $\delta r_V$ obtained from Ref.~\cite{Charles:2015gya} is negligible. 
Differently from the other observables in the present paper, we utilize the uncertainty {\it already} evaluated by the other well-sophisticated study as shown above.

The contribution from the charged Higgs boson is provided through the effective operators 
$\mathcal O_7 = \frac{e}{16\pi^2}m_b(\bar s_L \sigma^{\mu\nu} b_R) F_{\mu\nu}$ and $\mathcal O_8 = \frac{g_s}{16\pi^2}m_b(\bar s_L \sigma^{\mu\nu} t^a b_R) G_{\mu\nu}^a$ in the 2HDM\footnote
{
There exists the NLO correction to the effective operator $(\bar s_L \gamma_\mu t^a b_R) \sum_q (\bar q \gamma^\mu t^a q)$. 
The analytical formula for this is given in Ref.~\cite{Ciuchini:1997xe}. 
In the present analysis, we simply neglect this contribution. 
}. 
The LO correction is gained by one loop EW and QCD penguin diagrams and the NLO correction is calculated as in Refs.~\cite{Ciuchini:1997xe,Ciafaloni:1997un,Borzumati:1998tg,Borzumati:1998nx,Bobeth:1999ww}.  
To incorporate these effects, we obtain the useful numerical formula in terms of the Wilson coefficients at the scale $\mu_t = 160\,\text{GeV}$ as follows, 
\begin{align}
 & \overline{\mathcal B}(b\to s\gamma)_{E_\gamma>E_0=1.6\,\text{GeV}} = \overline{\mathcal B} (b \to s \gamma)^\text{SM} + \delta \overline{\mathcal B}(b\to s\gamma) \\[0.5em]
 & \delta \overline{\mathcal B}(b\to s\gamma)  = 10^{-4} \times \left( \frac{r_V^{\,}}{0.9626_{\,}} \right) \text{Re} 
 \Bigg [ 
 -8.100\, \mathcal C_7^\text{LO} -2.509\, \mathcal C_8^\text{LO} +2.767\, \mathcal C_7^\text{LO} \mathcal C_8^{\text{LO}*}  \notag \\
 & \hspace{6em}  + 5.348 \left|\mathcal C_7^\text{LO} \right|^2 +0.890\, \left|\mathcal C_8^\text{LO}\right|^2 -0.085\, \mathcal C_7^\text{NLO} -0.025\, \mathcal C_8^\text{NLO}  \\
 & \hspace{6em}  +0.095\, \mathcal C_7^\text{LO} \mathcal C_7^\text{NLO*} +0.008\, \mathcal C_8^\text{LO} \mathcal C_8^\text{NLO*} 
    +0.028\, \left( \mathcal C_7^\text{LO} \mathcal C_8^\text{NLO*} +\mathcal C_7^\text{NLO} \mathcal C_8^\text{LO*} \right)
 \Bigg] \,, \notag
 \label{Eq:bsgammaTHDM}
\end{align} 
where $\mathcal C_i^\text{LO (NLO)}$ show the new physics contributions from $\mathcal O_i$ for $i=7,8$ at the LO (NLO) part. 
The SM contributions are separated in advance. 
This numerical formula is estimated based on Refs.~\cite{Kagan:1998ym,Hurth:2003dk,Lunghi:2006hc} 
and we confirmed that the $\mathcal C_7^\text{LO}$ term, which is the most dominant, is consistent with Ref.~\cite{Misiak:2015xwa}. 
The explicit form of the coefficients $\mathcal C_i^\text{LO}$ and $\mathcal C_i^\text{NLO}$ is given by 
\begin{equation}
 \mathcal C_i^\text{LO} = \frac{ ( \xi_u^A )^2 }{3} \, G_1^i (y_{H^+}^t) + \xi_d^A \xi_u^{A}\, G_2^i (y_{H^+}^t) \,, 
 \label{Eq:bsgLO}  
\end{equation} 
\begin{equation}
 \mathcal C_i^\text{NLO} =  ( \xi_u^A )^2\, C_1^i (y_{H^+}^t) + \xi_d^A \xi_u^{A}\, C_2^i (y_{H^+}^t) +\Big( ( \xi_u^A )^2 D_1^i (y_{H^+}^t) + \xi_d^A \xi_u^{A} D_2^i (y_{H^+}^t) \Big) \ln \frac{\mu_t^2}{m_{H^+}^2} \,, 
 \label{Eq:bsgNLO}
\end{equation} 
for $i=7,8$ and $\mu_t =160\,\text{GeV}$, with respect to the mass ratio, 
\begin{equation}
 y_\phi^f = \frac{m_f^2}{m_\phi^2} \,.  
 \label{Eq:RatioY} 
\end{equation}
The loop functions $G_a^i$, $C_a^i$, and $D_a^i$ are described as in Appendix~\ref{App:form_bsgamma}. 
The NNLO QCD correction to this process has been studied in Ref.~\cite{Hermann:2012fc}. 
Within the present status on the uncertainties in the experimental result and the theoretical prediction, ignoring such a correction does not much change our result. 
As for $\overline{\mathcal B}(b\to d\gamma)$, the uncertainty in the prediction is still large and thus not of importance in limiting the 2HDM yet.

\subsection{Anomalies}
%%%%%%%%%%%%%%%%%%%%%%%%
%%%%%%%%%%%%%%%%%%%%%%%%
In this subsection we focus on formulae of observables in which discrepancies between the SM prediction and the experimental result have been reported, categorized as (c). 
The current status on the discrepancies are summarized in Sec.~\ref{Sec:Cons}.

\subsubsection{$\bar B \to D^{(*)} \tau\bar\nu$}
Semi-tauonic $B$ meson decays $\bar B \to D^{(*)} \tau\bar\nu$ are sensitive to the effect of the charged Higgs boson 
since its contribution is proportional to $\xi_d^A \xi_\ell^{A} m_b m_\tau/m_{H^+}^2$ and $ \xi_u^A \xi_\ell^{A} m_c m_\tau/m_{H^+}^2$ at the tree level\cite{Hou:1992sy,Tanaka:1994ay}.  
The results from the Belle, BaBar, and LHCb collaborations are nowadays available despite that it is difficult to identify the tau in these processes. 
Useful observables of these decays are given by 
\begin{align}
R(D^{(*)}) \equiv \frac{\mathcal{B}(\bar B\to D^{(*)}\tau^-\bar\nu_\tau)}{\mathcal{B}(\bar B\to D^{(*)}\ell^-\bar\nu_\ell)} \,, 
\end{align}
in which we can reduce uncertainties coming from input parameters. 
The SM predicts precise values of $R(D^{(*)})$ with the help of the heavy quark effective theory to evaluate form factors\cite{Caprini:1997mu,Amhis:2012bh}. 
The effect on $R(D^{(*)})$ in the context of the 2HDMs has been calculated as seen in Ref.~\cite{Nierste:2008qe,Kamenik:2008tj,Tanaka:2010se,Celis:2012dk,Celis:2013jha}.
Based on our previous study in Ref.~\cite{Tanaka:2012nw}, we give the numerical formulae of the branching ratios for the 2HDMs as follows.  
For the branching ratios of the light-leptonic modes,  
\begin{align}
&{\mathcal B}(\bar{B}\to D\ell\bar{\nu}) = \tau_{\bar{B}}G_F^2 |V_{cb}|^2 V_1(1)^2 \times 10^{-2} \left[ \Gamma_1^{D\ell} +\Gamma_2^{D\ell} \rho_1^2 +\Gamma_3^{D\ell} \rho_1^4 \right]\,,  \\
&{\mathcal B}(\bar{B}\to D^*\ell\bar{\nu}) = \tau_{\bar{B}}G_F^2 |V_{cb}|^2 A_1(1)^2 \times 10^{-2} \left[ \Gamma_1^{D^*\ell} +\Gamma_2^{D^*\ell} \rho_{A_1}^2 +\Gamma_3^{D^*\ell} \rho_{A_1}^4 \right] \,,
\end{align} 
where $\rho_1^2,\, \rho_{A_1}^2,\, R_1$, and $R_2$ are the form factor parameters fitted by energy distributions; 
$V_1(1)$ and $A_1(1)$ are overall normalizations of the form factors; 
and $\Gamma_i^{D^{(*)}\ell}$ are the coefficients of polynomial expansion with respect to $\rho_1^2$ ($\rho_{A_1}^2$).  
The explicit forms are given in Appendix~\ref{App:form_BDtaunu}.  
In this formula, we neglect the charged Higgs contribution to ${\mathcal B}(\bar{B}\to D\ell\bar{\nu})$ and ${\mathcal B}(\bar{B}\to D^*\ell\bar{\nu})$ since it is suppressed by the factor, $m_{q} m_\ell/m_{H^+}^2$. 
In the muonic mode, the contribution can be potentially a few \% and it can affect the determination of $|V_{cb}|$. 
We will discuss it in the next section. 
In the 2HDM, the numerical formulae for $\mathcal B (\bar{B}\to D^{(*)}\tau\bar{\nu})$ are written as 
\begin{align}
{\mathcal B}(\bar{B}\to D\tau\bar{\nu}) 
& = \tau_{\bar{B}}G_F^2 |V_{cb}|^2 V_1(1)^2 \times 10^{-2} \bigg[\Gamma_1^{D\tau} +\Gamma_2^{D\tau} \rho_1^2 +\Gamma_3^{D\tau} \rho_1^4 \notag \\
&~~~+(\Gamma_4^{D\tau} +\Gamma_5^{D\tau} \rho_1^2 +\Gamma_6^{D\tau} \rho_1^4) \left(\frac{3.45\,\text{GeV}}{m_b-m_c}\right) {\rm Re}( \mathcal C_{S_1} + \mathcal C_{S_2}) \notag \\
&~~~+(\Gamma_7^{D\tau} +\Gamma_8^{D\tau} \rho_1^2 +\Gamma_9^{D\tau} \rho_1^4) \left(\frac{3.45\,\text{GeV}}{m_b-m_c}\right)^2 | \mathcal C_{S_1}+ \mathcal C_{S_2}|^2 \bigg]\,, %\\[0.5cm]
\end{align}
\begin{align}
{\mathcal B}(\bar{B}\to D^*\tau\bar{\nu}) 
& = \tau_{\bar{B}}G_F^2 |V_{cb}|^2 A_1(1)^2 \times 10^{-2} \bigg[\Gamma_1^{D^*\tau} +\Gamma_2^{D^*\tau} \rho_{A_1}^2 +\Gamma_3^{D^*\tau} \rho_{A_1}^4 \notag \\
&~~~+(\Gamma_4^{D^*\tau} +\Gamma_5^{D^*\tau} \rho_{A_1}^2 +\Gamma_6^{D^*\tau} \rho_{A_1}^4) \left(\frac{6.2\,\text{GeV}}{m_b+m_c}\right) {\rm Re}( \mathcal C_{S_1}- \mathcal C_{S_2}) \notag \\
&~~~+(\Gamma_7^{D^*\tau} +\Gamma_8^{D^*\tau} \rho_{A_1}^2 +\Gamma_9^{D^*\tau} \rho_{A_1}^4) \left(\frac{6.2\,\text{GeV}}{m_b+m_c}\right)^2 | \mathcal C_{S_1}- \mathcal C_{S_2}|^2 \bigg]\,,
\end{align} 
where the charged Higgs contributions included in $\mathcal C_{S_1}$ and $\mathcal C_{S_2}$ are written as 
\begin{align}
 \label{Eq:BDtaunu_CS}
 \mathcal C_{S_1} = -\xi_d^A \xi_\ell^{A} \frac{m_b m_\tau}{m_{H^+}^2} \,, \quad \mathcal C_{S_2} = - \xi_u^A \xi_\ell^{A} \frac{m_c m_\tau}{m_{H^+}^2} \,.
\end{align}
As seen in the formulae, the overall normalizations $V_1(1)$, $A_1(1)$ and the CKM matrix element $|V_{cb}|$ are irrelevant to $R(D^{(*)})$. 
Later we show the fitted result for the input parameters.

\subsubsection{Muon anomalous magnetic moment}
The muon anomalous magnetic moment $a_\mu$ provides a sensitive test of quantum loop effects in the electroweak sector. 
The SM contributions are evaluated as in Refs.~\cite{Jegerlehner:2009ry,Davier:2010nc,Hagiwara:2011af} including several higher oder corrections\cite{Czarnecki:1995wq,Czarnecki:1995sz,Krause:1996rf,Prades:2009tw}.  
Recent studies for higher order corrections are also obtained in Refs.~\cite{Aoyama:2012wk,Gnendiger:2013pva,Kurz:2014wya,Colangelo:2014qya}. 
A discrepancy between the experimental result reported in Ref.~\cite{Bennett:2006fi} and the SM prediction, $a_\mu^\text{exp.} -a_\mu^\text{SM}$, can be compared with a new physics contribution.

In the 2HDM, the complete one-loop diagrams and the Barr-Zee type two-loop diagrams can be significant. 
The analytic formula for the one-loop diagrams is given\cite{Lautrup:1971jf,Leveille:1977rc,Dedes:2001nx} by 
\begin{align}
 a_\mu^\text{1loop} = \frac{G_F m_\mu^2}{4\sqrt 2\pi^2} \sum_{\phi=h,H,A,H^\pm}  ( \xi_\ell^\phi  )^2\, y_\phi^\mu\, F_\phi (y_\phi^\mu) \,,
\end{align}
where $y_\phi^f$ is defined in (\ref{Eq:RatioY}) and the loop functions $F_\phi$ are calculated as 
\begin{align}
 & F_h (a) = F_H (a) = \int^1_0 dz \frac{z^2(2-z)}{1 - z + a z^2} \simeq -\frac{7}{6} -\ln a - \left( \frac{13}{4} +3 \ln a \right) a + O(a^2) \,, \\
 & F_A (a) = \int^1_0 dz \frac{-z^3}{1 - z + a z^2} \simeq \frac{11}{6} +\ln a + \left( \frac{89}{12} +5 \ln a \right) a + O(a^2) \,, \\
 & F_{H^\pm} (a) = \int^1_0 dz \frac{-z (1-z)}{1-(1-z) a} \simeq -\frac{1}{6} - \frac{a}{12} +O(a^2) \,.
\end{align}
The result for the Barr-Zee type two-loop diagrams is given\cite{Czarnecki:1995wq,Chang:2000ii,Cheung:2001hz,Cheung:2003pw,Broggio:2014mna} by\footnote{
In the recent work of Ref.~\cite{Ilisie:2015tra}, new Barr-Zee type two-loop diagrams are calculated. 
It is stated that the contribution to $a_\mu$ can be drastically changed. 
However, our conclusion we will show later is not affected, although the values of the parameters are changed. 
} 
\begin{align}
 a_\mu^\text{2loop} 
 = \frac{G_F m_\mu^2}{4\sqrt 2\pi^2} \frac{\alpha}{\pi} \sum_{\phi=h,H,A} \sum_f N_f^c\, Q_f^2\, \xi_\ell^\phi\, \xi_f^{\phi}\, y_\phi^f\, G_\phi (y_\phi^f)  \,,
\end{align}
where the index $f$ represents the fermion in the loop, $Q_f$ and $N_f^c$ are the electric charge and color degrees of freedom of $f$. 
The functions $G_\phi$ are obtained by
\begin{align}
 & G_\phi (a) = \int^1_0 dz \frac{\widetilde g_\phi(z) }{z(1-z) -a} \ln \frac{z(1-z)}{a},  \\[0.2em]
 & \widetilde g_h(z)= \widetilde g_H(z) =2z(1-z)-1 \,, \quad\quad \widetilde g_A(z)=1 \,.
\end{align}
In the SM prediction, the contribution from the SM Higgs boson is already taken and thus we must care about this part when considering the 2HDM. 
Substituting the corresponding contribution, the 2HDM contribution which can be compared to $a_\mu^\text{exp.} -a_\mu^\text{SM}$ is represented as 
\begin{align}
 &\Delta a_\mu^\text{2HDM} = a_\mu^\text{1loop} + a_\mu^\text{2loop} - a_\mu^\text{SM Higgs} \,,  \\[0.2em]
 &a_\mu^\text{SM Higgs} = -1.4 \times 10^{-11} \,, \label{Eq:g2SMpart}
\end{align}
where the value in (\ref{Eq:g2SMpart}) is evaluated by fixing $\xi_f^h=1$ and $m_h=126\,\text{GeV}$ in the formulae relevant for $\phi=h$.

%%%%%%%%%%%%%%%%%%%%%%%%%%%%%%%%%%%%%%%%%%%%%%%%%%
\section{Determination of CKM in the 2HDM}
\label{Sec:DetCKM}
%%%%%%%%%%%%%%%%%%%%%%%%%%%%%%%%%%%%%%%%%%%%%%%%%%
As stressed in the previous section, it is necessary to concern the effect of the extra Higgs bosons when we determine the CKM matrix elements by fitting to experimental data, in the 2HDM. 
This is expected to be more crucial for the future flavor experiments at the SuperKEKB/Belle~II~\cite{Aushev:2010bq}. 
The global fit of the CKM matrix elements, together with the parameters of the 2HDM, to all the relevant experimental data is one of the approaches for the analysis\cite{Deschamps:2009rh}. 
In this paper, we employ a more visualized approach as follows.  
For the re-fit of the CKM matrix elements, we use the Wolfenstein parametrization which is defined as 
\begin{align}
 V_\text{CKM}=
\begin{pmatrix}
1 -\lambda^2/2 & \lambda & A \lambda^3 (\rho -i \eta) \\
-\lambda & 1 -\lambda^2/2 & A \lambda^2 \\
A \lambda^3 (1-\rho -i \eta) & -A \lambda^2 & 1
\end{pmatrix} \,, 
\end{align} 
where we neglect $O(\lambda^4 ) \sim O(0.001)$. 
Then, we obtain fitted values of $\lambda$, $A$, $\rho$, and $\eta$ by using observables in which contributions from the extra Higgs bosons are negligible. 
As for $r_V$ needed in the evaluation for $\bar B \to X_s \gamma$, we take $r_V \simeq 1 -\lambda^2 (1 -2 \rho)$.

\subsection{$\lambda$ and $A$}
%%%%%%%%%%%%%%%%%%%%%%%%
%%%%%%%%%%%%%%%%%%%%%%%%
The most precise value of the Cabibbo mixing parameter $\lambda$ is provided from the determination of $|V_{ud}|$ by the super allowed ($0^+\to 0^+$) nuclear beta decays. 
The experimental result $|V_{ud}| =0.97425\,(22)$~\cite{Hardy:2008gy} implies $\lambda = 0.2269 \pm 0.0010$. 
In the SM, $\lambda$ is also determined from leptonic $K$ decays such as $K \to (\pi) \ell \nu$ and $\tau \to K\nu$ for $\ell = e,\,\mu$. 
Among them, $K \to e \nu$ is usable to determine $\lambda$ in the 2HDM, since the effect of the extra Higgs bosons is safely negligible and its experimental data is available separately from the muonic mode.  
The experimental result $\mathcal B(K \to e\nu)=(1.581 \pm 0.008)\times 10^{-5}$~\cite{Agashe:2014kda} is translated into $\lambda = 0.2221 \pm 0.0014$, where the decay constant of $K$ we used is listed in Table~\ref{Tab:DecayConstant}. 
Therefore the combined result is given as
\begin{align}
 \lambda = 0.2253 \pm 0.0008 \,,
\end{align}
and we use this value for the following analysis in this paper.

\begin{figure}[t]
\begin{center}
\includegraphics[viewport=0 0 300 305, width=20em]{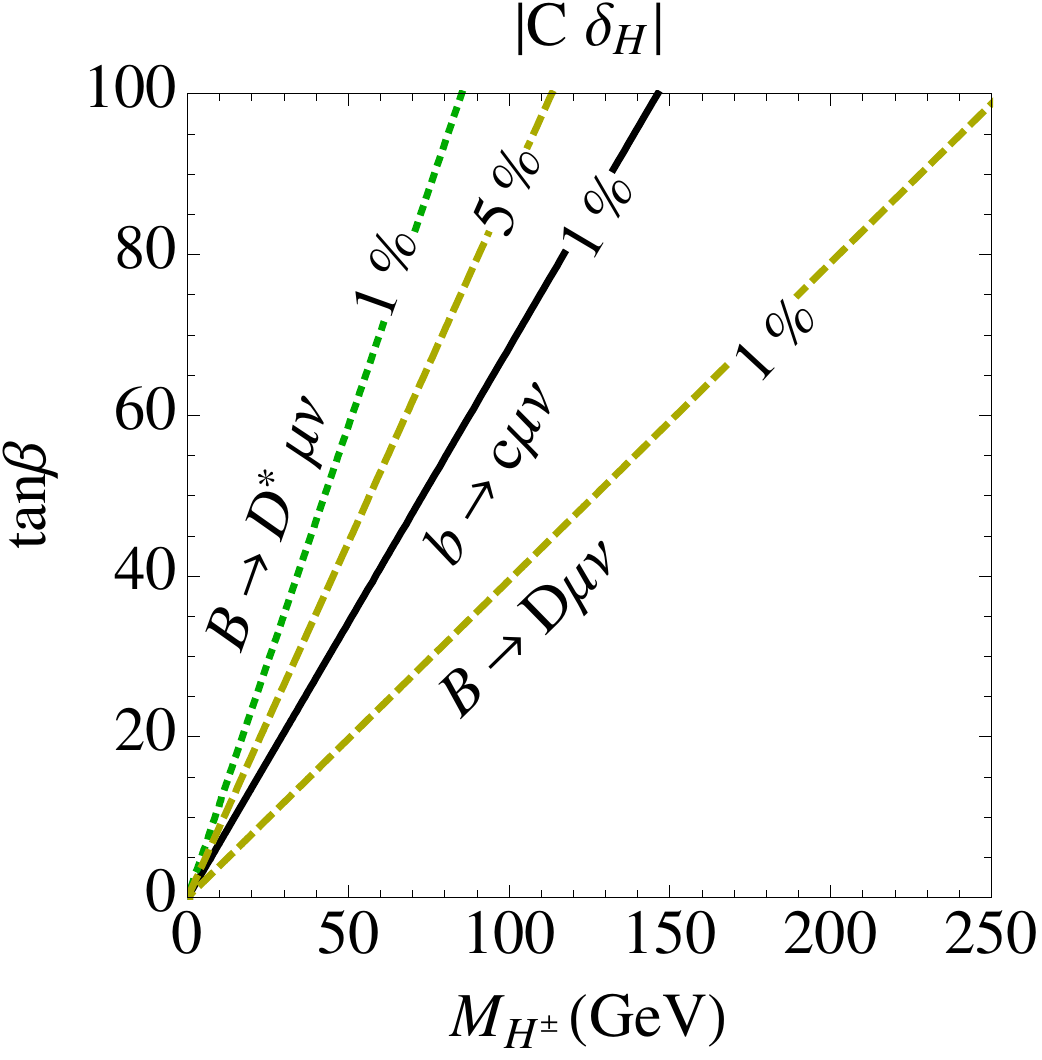}
\caption{
Corrections to the measurement of $|V_{cb}|$ caused by the charged Higgs contribution $C \delta_H$ on the plane of $(m_{H^+}, \tan\beta)$ in the type~II model. 
The green dotted, black solid, and yellow dashed lines are the results for $C=0.017$, $0.05$, and $0.15$ corresponding to $\bar B \to D^* \mu\bar\nu$, $b\to c\mu\nu$, and $\bar B \to D \mu\bar\nu$, respectively. 
}
\label{Fig:CHcont}
\end{center}
\end{figure}

The parameter $A$ is included in $V_{ub}$, $V_{cb}$, $V_{td}$, and $V_{ts}$, and usually obtained from the determination of $|V_{cb}|$. 
It is, however, known that the values of $|V_{cb}|$ obtained from inclusive ($\bar B \to X_c \ell\bar\nu$) and exclusive ($\bar B \to D^{(*)} \ell\bar\nu$) decay modes are not in good agreement~\cite{Agashe:2014kda,Amhis:2014hma}. 
In the 2HDM, although the charged Higgs boson affects the muonic modes ($\ell =\mu$), it is hard to compensate this discrepancy.  
In the present paper, we simply obtain a combined value of $|V_{cb}|$ considering the charged Higgs effect. 
For the determination from $\bar B \to X_c \ell\bar\nu$, a combined fit to moments of several variables (a hadronic-mass, a lepton-energy spectrum, and a photon-energy spectrum) are required. 
Calculating the charged Higgs effect on its distribution is beyond the scope of this paper. 
Instead, we roughly estimate such an effect by using the expression defined as 
\begin{align}
 & |V_{cb}|_\text{obs.} = |V_{cb}| \sqrt{1+C\, \delta_H +O(|\delta_H|^2) } \,, \label{Eq:RoughSketch} \\
 & \delta_H = - \xi_d^A \xi_\ell^{A} \frac{m_bm_\mu}{m_{H^+}^2} \,,
\end{align}
where $\delta_H$ indicates the contribution of the charged Higgs boson to the muonic decay mode and $|V_{cb}|_\text{obs.}$ is the experimental result of the fit by assuming the SM. 
The coefficient $C$ stems from the difference of the effective operator between the SM and the charged Higgs contributions. 
Considering the quark level process, which is the leading order contribution involved in $\bar B \to X_c \ell\bar\nu$, we obtain $C \simeq 0.05$ for $\ell =\mu$. 
Then we find that the correction from $C\, \delta_H$ is less than $1\%$ for $m_{H^+} > 150\, \text{GeV}$ and $\xi_d^A = \xi_\ell^{A} = \tan\beta < 100$ in the type~II and aligned models as shown in Fig~\ref{Fig:CHcont}. 
As for the other types, it is completely negligible. 
The similar estimation can be done for $\bar B \to D^{(*)} \ell\bar\nu$. 
In these exclusive processes, we obtain $C \simeq 0.15\, (0.017)$ with use of the formula for $\bar B \to D^{(*)} \tau\bar\nu$ replacing $m_\tau$ with $m_\ell$ in Ref.~\cite{Tanaka:2012nw}. 
In Fig~\ref{Fig:CHcont}, the correction to the measurement of $|V_{cb}|$ is shown in the type~II model. 
We can see that the charged Higgs effects in $\bar B \to D \mu\bar\nu$ are more important than that in the inclusive process but, in any case, for $m_{H^+} \gtrsim 300\, \text{GeV}$ they are not sizable. 
As for the combined experimental value of $|V_{cb}|$, we refer to the latest determination by the CKMfitter~\cite{Charles:2015gya} (not the result from the global fit in this reference). 
To conclude, we take  
\begin{align}
 A = 0.808 \pm 0.017 \,,
 \label{Eq:Aparameter}
\end{align}
in the case of the type~I, X, and Y models. 
We add an additional uncertainty to (\ref{Eq:Aparameter}) in accordance with $\delta_H$ in the case of the type~II and aligned models. 
%
%%%%%%%%%%%%FIG%%%%%%%%%%%%
%%%%%%%%%%%%FIG%%%%%%%%%%%%

\subsection{$\rho$ and $\eta$}
\begin{figure}[t]
\begin{center}
\includegraphics[viewport=0 0 799 327, width=40em]{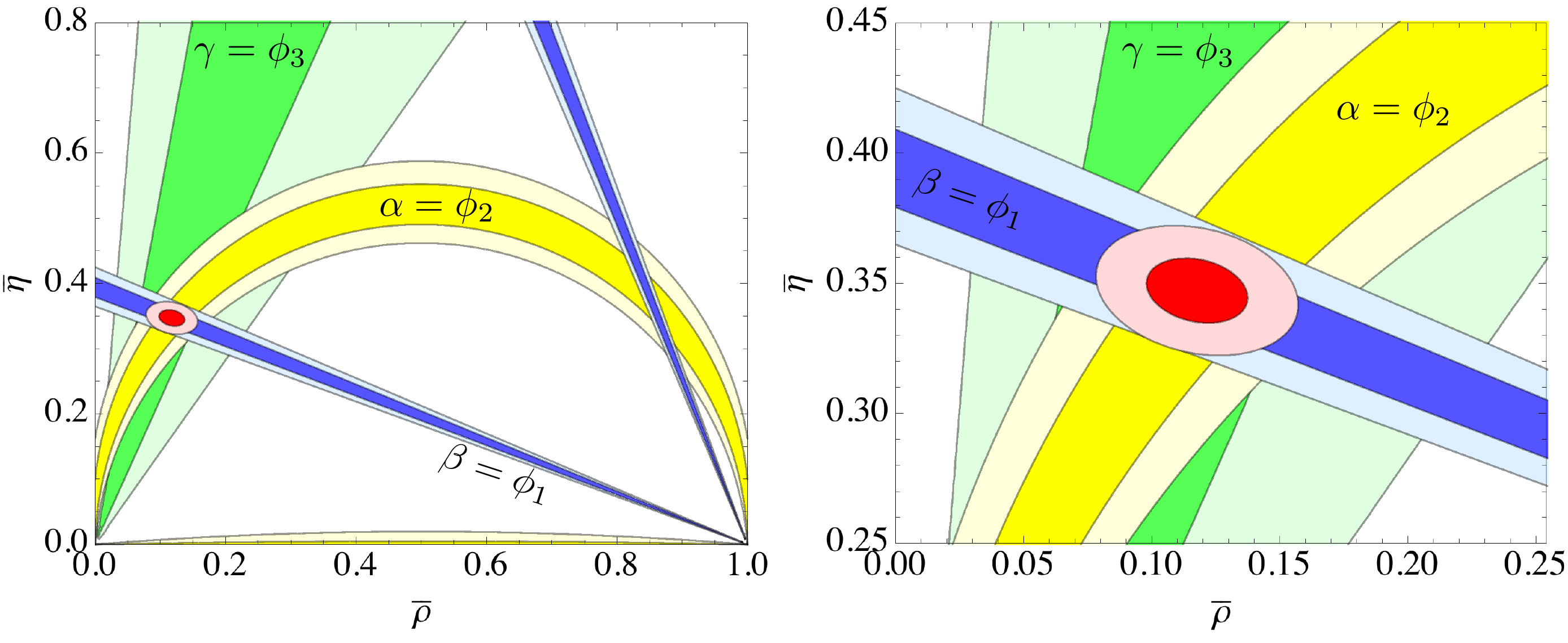}
\caption{
Allowed regions on the $(\bar\rho, \bar\eta)$ plane obtained from the measurements of $\alpha=\phi_2$ (yellow), $\beta=\phi_1$ (blue), and $\gamma=\phi_3$ (green). 
The red region indicates the combined result from these three measurements. 
The right panel is zoomed version of the left panel. 
}
\label{Fig:UTfitting}
\end{center}
\end{figure}

%%%%%%%%%%%%%%%%%%%%%%%%
%%%%%%%%%%%%%%%%%%%%%%%%
The CP phase in the CKM matrix is given by $\rho$ and $\eta$. 
For the actual observables, $\bar \rho + i \bar \eta = -(V_{ud} V_{ub}^*)/(V_{cd} V_{cb}^*)$ is defined and measured by experiments, where it is related as  
\begin{align}
 \rho + i  \eta = \frac{\bar \rho + i \bar \eta}{1 - A^2 \lambda^4 (\bar \rho + i \bar \eta)} \sqrt{ \frac{1-A^2 \lambda^4}{1- \lambda^2} } = (\bar \rho + i \bar \eta) (1-\lambda^2/2 +\cdots ) \,. 
 \label{Eq:rhoeta}
\end{align}
In the SM, $\bar \rho$ and $\bar \eta$ are fitted by several variables such as $\epsilon_K$, $\Delta M_{d}$, $\Delta M_{s}$, $|V_{ub}|$, and the angles of unitarity triangle, defined as 
\begin{align}
 \alpha \equiv \phi_2 = \arg \left( -\frac{V_{td} V_{tb}^* }{V_{ud} V_{ub}^*} \right) ,\,\,
 \beta \equiv \phi_1 = \arg \left( -\frac{V_{cd} V_{cb}^* }{V_{td} V_{tb}^*} \right) , \,\,
 \gamma \equiv \phi_3 = \arg \left( -\frac{V_{ud} V_{ub}^* }{V_{cd} V_{cb}^*} \right) .
\end{align}
In the 2HDM, we note that measuring these angles is not affected by the extra Higgs bosons as long as $\xi_f^A$ is real, whereas the others are potentially harmed. 
Thus we use only the unitary triangle to determine $\bar \rho$ and $\bar \eta$.  
The latest world averages of the angles are provided in Ref.~\cite{Charles:2015gya} and then related as  
\begin{align}
& \sin 2\phi_1 = \frac{2 \bar \eta (1 - \bar \rho)}{(1 - \bar \rho)^2 + \bar \eta^2} =0.682 \pm 0.019 \,, \label{Eq:beta} \\[0.2cm]
& \phi_2 = \frac{1}{2} \arcsin \left[ \frac{-2\eta( \bar \rho(1 - \bar \rho) - \bar \eta^2)}{(\bar \rho^2 + \bar \eta^2)((1 - \bar \rho)^2 + \bar \eta^2)} \right] = (87.7\pm 3.4)^\circ \,, \label{Eq:alpha}  \\[0.2cm]
& \phi_3 =  \frac{1}{2} \arcsin \left[ \frac{2\bar \rho \bar \eta}{\bar \rho^2 + \bar \eta^2} \right] = (73.2\pm 6.7)^\circ \,. \label{Eq:gamma}
\end{align} 
The fitted values from (\ref{Eq:beta})--(\ref{Eq:gamma}) are obtained as 
\begin{align}
 \bar \rho = 0.118 \pm 0.016 \,,\quad \bar \eta = 0.347 \pm 0.010 \,, \quad \text{correlation} = -0.22 \,, 
\end{align}
and we show the $(\bar\rho, \bar\eta)$ plot for the fit in Fig.~\ref{Fig:UTfitting}. 
%%%%%%%%%%%%FIG%%%%%%%%%%%%
%%%%%%%%%%%%FIG%%%%%%%%%%%%

We note that the SM global fit reported in Ref.~\cite{Charles:2015gya} results in 
$A=0.810^{+0.018}_{-0.024}$, $\lambda = 0.22548^{+0.00068}_{-0.00034}$, $\bar\rho = 0.145^{+0.013}_{-0.007}$, and $\bar\eta = 0.343^{+0.011}_{-0.012}$. 
Thus we can see that there is no significant difference in our determination of the CKM matrix elements.

%%%%%%%%%%%%%%%%%%%%%%%%%%%%%%%%%%%%%%%%%%%%%%%%%%
\section{Constraint}
\label{Sec:Cons}
%%%%%%%%%%%%%%%%%%%%%%%%%%%%%%%%%%%%%%%%%%%%%%%%%%
Here we give current constraints on the $Z_2$ symmetric and aligned types of the 2HDM with use of 
$\mathcal B (B\to\tau\nu)$, $\mathcal B (D\to\mu\nu)$, $\mathcal B (D_s\to\tau\nu)$, $\mathcal B (D_s\to\mu\nu)$, $\mathcal B (K\to\mu\nu) / \mathcal B (\pi\to\mu\nu)$, $\mathcal B (\tau \to K\nu) / \mathcal B (\tau \to \pi\nu)$, $\overline{\mathcal B} (B^0_s \to \mu^+ \mu^-)$, $\overline{\mathcal B} (B^0_d \to \mu^+ \mu^-)$, $\overline{\mathcal B}(b\to s\gamma)_{E_\gamma>1.6\,\text{GeV}}$, $\Delta M_s$, $\Delta M_d$, and $| \epsilon_K|$. 
The way to evaluate uncertainties and exclusion confidence levels (CLs) for the above observables is shown in Appendix~\ref{App:uncertainty}. 
To begin with, we exhibit input required to evaluate the observables and the experimental results. 
After that, we obtain the constraints and comment on them. 
We also discuss the anomalies of $R(D^{(*)})$ and $a_\mu$ in the context of the 2HDMs with the natural flavor conservation.

\subsection{Input and experimental data}
%%%%%%%%%%%%%%%%%%%%%%%%
%%%%%%%%%%%%%%%%%%%%%%%%
We apply our fit result for the Wolfenstein parametrization to the CKM matrix elements. 
Obtained from the previous section, we can express the result as 
\begin{align}
 \hspace{-2em}
 V_\text{CKM}=
 \begin{pmatrix}
 0.97462 \pm 0.00018			&  0.22530 \pm 0.00080 			& \begin{matrix} (0.00107 \pm 0.00014) \quad \\[-0.6em] \quad - i\, (0.00315 \pm 0.00012) \end{matrix} \\
 -0.22530 \pm 0.00080 			&  0.97462 \pm 0.00018			& 0.04101 \pm 0.00091 \\
 \begin{matrix} (0.00816 \pm 0.00024) \quad \\[-0.6em] \quad -i\, (0.00315 \pm 0.00012) \end{matrix} & -0.04101 \pm 0.00091 & 1
 \end{pmatrix} \,. 
\end{align} 
The lattice studies for the meson decay constants and the bag parameters are summarized in Ref.~\cite{Aoki:2013ldr}, and the recent updates for $f_D$, $f_{D_s}$, and $f_K / f_\pi$ are available in Ref.~\cite{Bazavov:2014wgs}. 
The EM corrections of $\mathcal B (K\to\mu\nu) / \mathcal B (\pi\to\mu\nu)$ and $\mathcal B (\tau \to K\nu) / \mathcal B (\tau \to \pi\nu)$ are given in Refs.~\cite{Antonelli:2010yf,Banerjee:2008hg}. 
The values are listed in Table~\ref{Tab:DecayConstant}. 
As for the parameter $\hat B_{B_q}^{ST} \eta_{B_q}^{ST}$ in (\ref{Eq:deltaM}), the scale dependent expression defined as 
\begin{align}
 \hat B_{B_q}^{ST} \eta_{B_q}^{ST} \equiv 
  B_3^{(q)} (\mu_b) \, \eta_{21} (\mu_b) - B_2^{(q)}(\mu_b) \left ( \frac{5}{8} \eta_{11} (\mu_b) + \frac{5}{2} \eta_{21} (\mu_b) \right ) \,,
\end{align} 
are only evaluated. 
The bag parameters at the $\mu_b = m_b$ scale are given as~\cite{Carrasco:2014nda,Carrasco:2013zta,Bouchard:2011xj}  
\begin{align}
 & f_{B_s} \sqrt{B_2^{(s)} (\mu_b) } = (225 \pm 28) \,\text{MeV} \,, \quad  f_{B_s} \sqrt{B_3^{(s)} (\mu_b) } = (231 \pm 38) \,\text{MeV} \,, \label{Eq:SRRs} \\
 & f_{B_d} \sqrt{B_2^{(d)} (\mu_b) } = (183 \pm 11) \,\text{MeV} \,, \quad  f_{B_d} \sqrt{B_3^{(d)} (\mu_b) } = (190 \pm 36) \,\text{MeV} \,, \label{Eq:SRRd}
\end{align} 
and the QCD corrections are $\eta_{11} (\mu_b)=1.654$ and $\eta_{21} (\mu_b)=-0.007$~\cite{Buras:2001ra}. 
Quark masses that appear in the formulae are the running masses evaluated at the proper scale in the $\overline{\text{MS}}$ scheme, $m_q \equiv \overline{m}_q(\mu)$. 
The matching scale for the Wilson coefficient $\mathcal C_X$ is chosen as $\mu_t = 160\,\text{GeV}$.  
The low energy scales are set as $\mu_B = 5\,\text{GeV}$, $\mu_D = 2\,\text{GeV}$, and $\mu_K = 1\,\text{GeV}$ for the $B_{(s)}$, $D_{(s)}$, and $K$ mesons, respectively. 
To evaluate the RGE running of the quark masses, we utilize the {\tt Mathematica} package {\tt RunDec}~\cite{Chetyrkin:2000yt}, in which QCD RGEs up to the four-loop level are implemented. 
The input values of the initial condition for $\overline{m}_q(\mu)$ are listed in Table~\ref{Tab:DecayConstant}, where $M_t$ indicates the pole mass of the top quark.  
%
%%%%%%Table%%%%%%
\begin{table}
\begin{center}
\scalebox{0.7}{
\begin{tabular}{cc}
 \hline
 \hline  Decay constant 							& Value    \\ \hline
 $f_B$ 										& $(190.5 \pm 4.2)\, \text{MeV}$~\cite{Aoki:2013ldr} \\
 $f_{B_s}$										& $(227.7 \pm 4.5)\, \text{MeV}$~\cite{Aoki:2013ldr} \\
 $f_D$ 										& $(212.6 \pm 1.2)\, \text{MeV}$~\cite{Bazavov:2014wgs} \\
 $f_{D_s}$										& $(249.0 \pm 1.3) \text{MeV}$~\cite{Bazavov:2014wgs} \\
 $f_K$										& $(156.3 \pm 0.9)\, \text{MeV}$~\cite{Aoki:2013ldr} \\
 $f_{K/\pi}$									& $1.1956\pm 0.0024$~\cite{Bazavov:2014wgs} \\
 \hline\hline
\end{tabular}
\begin{tabular}{cc}
 \hline 
 \hline  Bag parameter 							& Value    \\ \hline
 $\hat B_{B_d}$ 								& $1.27 \pm 0.10$~\cite{Aoki:2013ldr} \\
 $\hat B_{B_s}$ 								& $1.33 \pm 0.06$~\cite{Aoki:2013ldr} \\
 $\hat B_{K}$ 									& $0.7661 \pm 0.0099$~\cite{Aoki:2013ldr} \\
 \hline 
 \hline  EM correction 							& Value    \\ \hline
 $\delta_\text{EM}^{K/\pi}$							& $-0.0070 \pm 0.0018$~\cite{Antonelli:2010yf} \\
 $\delta_\text{EM}^{K/\pi,\,\tau}$					& $0.0003 \pm 0.0044$~\cite{Banerjee:2008hg} \\
 \hline\hline
\end{tabular}
\begin{tabular}{cc}
 \hline
 \hline  Quark mass 						& Value~\cite{Agashe:2014kda}    \\ \hline
 $\overline{m}_u (2\,\text{GeV})$ 			& $(2.3 \pm 0.6)\, \text{MeV}$ \\
 $\overline{m}_d (2\,\text{GeV})$			& $(4.8 \pm 0.4)\, \text{MeV}$ \\
 $\overline{m}_s (2\,\text{GeV})$ 			& $(95 \pm 5)\, \text{MeV}$ \\
 $\overline{m}_c (m_c)$					& $(1.275 \pm 0.025) \text{GeV}$ \\
 $\overline{m}_b (m_b)$					& $(4.18 \pm 0.03)\, \text{GeV}$ \\
 $M_t$								& $(174.6 \pm 1.9)\, \text{GeV}$ \\
 \hline\hline
\end{tabular}
}
\end{center}
\caption{
Lattice results of the meson decay constants, the bag parameters and the electromagnetic correction evaluated in Refs.~\cite{Aoki:2013ldr,Bazavov:2014wgs,Antonelli:2010yf,Banerjee:2008hg}, 
and input values of the initial conditions for the evaluation of the running quark masses~\cite{Agashe:2014kda}. 
}
\label{Tab:DecayConstant}
\end{table}
%%%%%%Table%%%%%%
%
For input parameters obtained from the experimental data of the neutral meson mixings, we refer to the HFAG summary in Ref.~\cite{Amhis:2014hma}, 
\begin{align}
 & \Delta \Gamma_s = (0.081 \pm 0.006) \,\text{ps}^{-1}\,, \quad  \Delta M_K = 3.484 \times 10^{-12}\,\text{MeV}\,, \label{Eq:DGs} \\
 & \tau_{B_s^0}^H = 1.607 \,\text{ps} \,, \quad \tau_{B_s^0}^L = 1.422 \,\text{ps} \,, \quad \tau_{B_d^0}^H \simeq \tau_{B_d^0}^L \simeq \tau_{B_d^0}= 1.519 \,\text{ps} \,,
\end{align} 
assuming $\Delta \Gamma_d / \Gamma_d \simeq 0$~\cite{Lenz:2011ti}, where uncertainties less than $1\%$ are neglected for these parameters.  
The other numerical input for our numerical analysis are shown in the Appendix~\ref{App:uncertainty}. 
In addition, we summarize the experimental data for the relevant observables in Table~\ref{Tab:ExpResult}, along with the SM {\it contributions} which we evaluated with use of the input values shown above. 
%
%%%%%%Table%%%%%%
\begin{table}
\begin{center}
\begin{tabular}{ccc}
 \hline 
 \hline Observable 												& \hspace{4em} Experimental result \hspace{4em}						& SM contribution    \\
 \hline $\mathcal B (B\to\tau\nu)$									& $(1.14 \pm 0.22)\times 10^{-4}$~\cite{Amhis:2014hma}					& $(0.78\pm 0.07 )\times 10^{-4}$ \\
 \hline $\mathcal B (D\to\mu\nu)$ 									& $(3.74 \pm 0.17)\times 10^{-4}$~\cite{Agashe:2014kda,Amhis:2014hma}	& $(3.94\pm 0.13) \times 10^{-4}$  \\
 \hline $\mathcal B (D_s\to\tau\nu)$ 									& $(5.55 \pm 0.24)\times 10^{-2}$~\cite{Agashe:2014kda,Amhis:2014hma}	& $(5.17\pm 0.11) \times 10^{-2}$  \\
 \hline $\mathcal B (D_s\to\mu\nu)$									& $(5.57 \pm 0.24)\times 10^{-3}$~\cite{Agashe:2014kda,Amhis:2014hma}	& $(5.28\pm 0.11) \times 10^{-3}$ \\
 \hline $\mathcal B (K\to\mu\nu) / \mathcal B (\pi\to\mu\nu)$	 			& $ 0.6357 \pm 0.0011$~\cite{Agashe:2014kda} 						& $ 0.6231\pm 0.0071 $  \\
 \hline $\mathcal B (\tau \to K\nu) / \mathcal B (\tau \to \pi\nu)$      			& $ 0.0646 \pm 0.0009$~\cite{Agashe:2014kda} 						& $ 0.0655\pm0.0008 $   \\
 \hline $\overline{\mathcal B} (B^0_s \to \mu^+ \mu^-)$ 						& $(2.8\pm 0.7)\times 10^{-9}$~\cite{Archilli:2014cla}						& $(3.66\pm 0.28) \times 10^{-9}$ \\
 \hline $\overline{\mathcal B} (B^0_d \to \mu^+ \mu^-)$						& $(3.9\pm 1.5)\times 10^{-10}$~\cite{Archilli:2014cla}					& $(1.08\pm 0.13) \times 10^{-10}$ \\
 \hline $\overline{\mathcal B}(b\to s\gamma)_{E_\gamma>1.6\,\text{GeV}}$ 	& $(3.43 \pm 0.22)\times 10^{-4}$~\cite{Amhis:2014hma}					& $(3.36\pm 0.24) \times 10^{-4}$  \\
 \hline $\Delta M_s$ 												& $(17.757\pm 0.021) \text{ps}^{-1}$~\cite{Agashe:2014kda,Amhis:2014hma}	& $(18.257\pm 1.505) \text{ps}^{-1}$ \\
 \hline $\Delta M_d$ 												& $(0.510\pm 0.003) \text{ps}^{-1}$~\cite{Agashe:2014kda,Amhis:2014hma}	& $(0.548\pm 0.075) \text{ps}^{-1}$ \\
 \hline $| \epsilon_K|$ 											& $(2.228\pm 0.011)\times 10^{-3}$~\cite{Agashe:2014kda}				& $(1.662\pm 0.354) \times 10^{-3}$ \\
 \hline 
 \hline
\end{tabular}\end{center}
\caption{
Experimental results of the observables combined by the PDG and/or HFAG collaborations in Refs.~\cite{Agashe:2014kda,Amhis:2014hma}. 
As for $\overline{\mathcal B} (B^0_q \to \mu^+ \mu^-)$, the combined results from the LHCb and CMS collaborations are shown as in Ref.~\cite{Archilli:2014cla}.
}\label{Tab:ExpResult}
\end{table}
%%%%%%Table%%%%%%

\subsection{Setup of model parameters}
%%%%%%%%%%%%%%%%%%%%%%%%
%%%%%%%%%%%%%%%%%%%%%%%%
Here, we summarize setup for the parameters of the 2HDMs in our numerical analysis. 
We assume the same masses for the extra Higgs bosons, $m_H=m_A=m_{H^+}$.  
This is favored by the truth that this relation satisfies a theoretical bound from perturbativity~\cite{Broggio:2014mna,Chang:2015goa}, and it is also allowed by the EW precision tests~\cite{Gunion:1989we}. 
In this case, constraints on $m_A$ given by the ATLAS and CMS collaborations~\cite{Khachatryan:2014wca,Aad:2014vgg} are notable for the type~II model. 
This is particularly relevant for the bound from $B^0_q \to \mu^+ \mu^-$ since the CP-odd Higgs boson contributes to the process.

For the mixing angle of $h$ and $H$, we take the SM-like limit $\sin (\beta - \alpha)=1$ in which the heavier CP-even Higgs boson $H$ can not decay into $W^+ W^-$ and $ZZ$. 
This is justified by current Higgs boson searches at the Large Hadron Collider (LHC). 
The current combined fit of  $\sin (\beta - \alpha)$ to the LHC results has been studied in Refs.~\cite{Chakrabarty:2014aya,Chowdhury:2015yja,Chang:2015goa}.

The case that $\sin (\beta - \alpha)$ is close to, but not exactly, one is interesting for collider searches. 
From the viewpoint of flavor physics, $\mathcal B (B^0_q \to \mu^+ \mu^-)$ can be affected as varying $\sin (\beta - \alpha)$. 
But, the small difference of $\sin (\beta - \alpha)$ from one changes only a few \% of $\mathcal B (B^0_q \to \mu^+ \mu^-)$, much smaller than the current experimental and theoretical uncertainties. 
For example, one finds $1.5\%$ reduction of $\mathcal B (B^0_s \to \mu^+ \mu^-)$ in the type~II model for $\sin (\beta - \alpha)=0.9$, $\tan\beta=30$, and $m_H=m_A=m_{H^+}=500\,\text{GeV}$ 
from the $\sin (\beta - \alpha)=1$ case. 
Changing $\sin (\beta - \alpha)=1$ and $m_H=m_A=m_{H^+}$ also affect $\Delta a_\mu^\text{2HDM}$. 
Later, we will loosen these assumption and see the effect.

\subsection{Constraint on the $Z_2$ symmetric models}
\label{SubSec:Z2bound}
%%%%%%%%%%%%%%%%%%%%%%%%
%%%%%%%%%%%%%%%%%%%%%%%%
%
%%%%%%%%%%%%FIG%%%%%%%%%%%%
\begin{figure}[t]
\begin{center}
\includegraphics[viewport=0 0 360 367, width=9.5em]{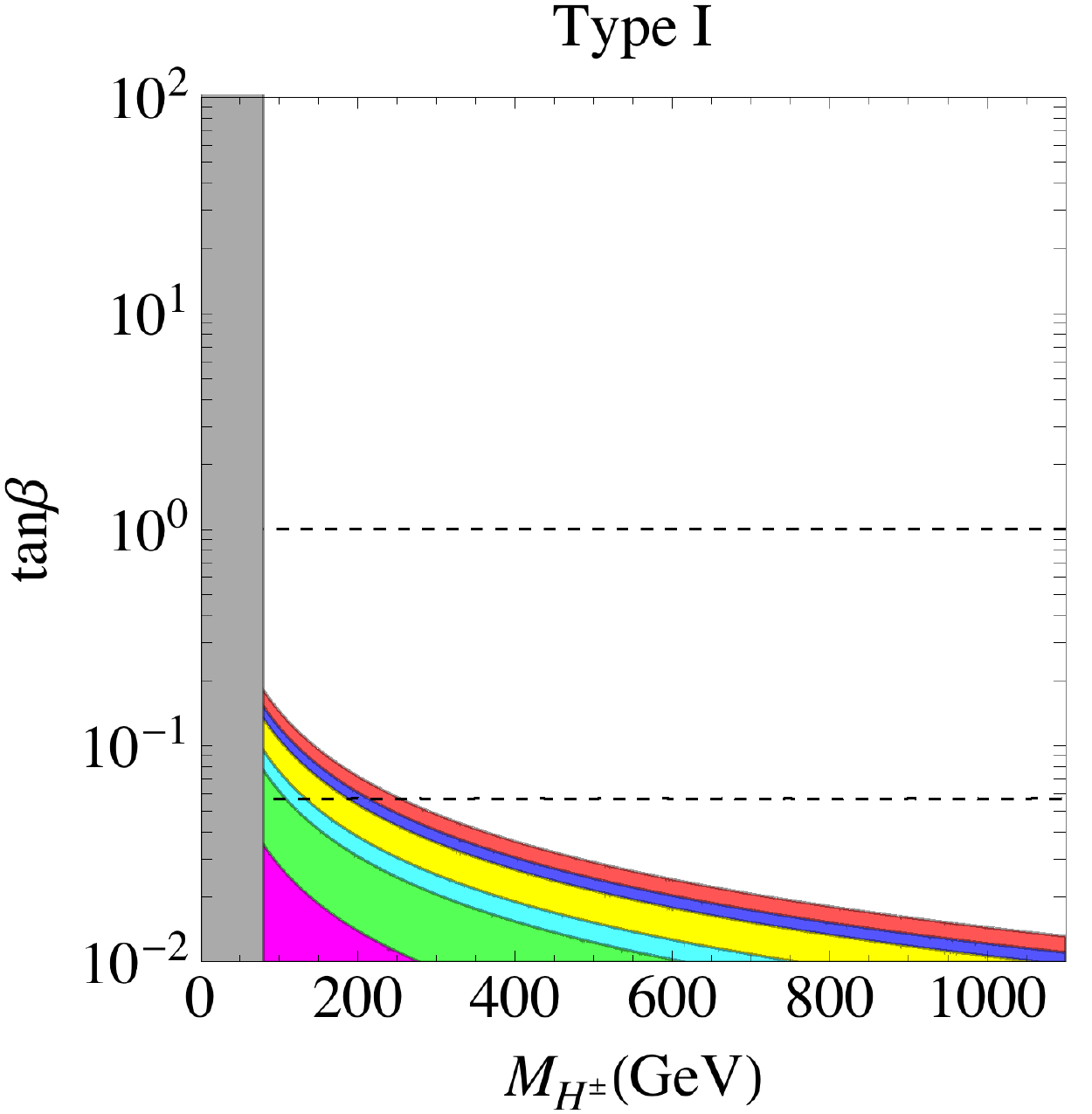}
\includegraphics[viewport=0 0 360 367, width=9.5em]{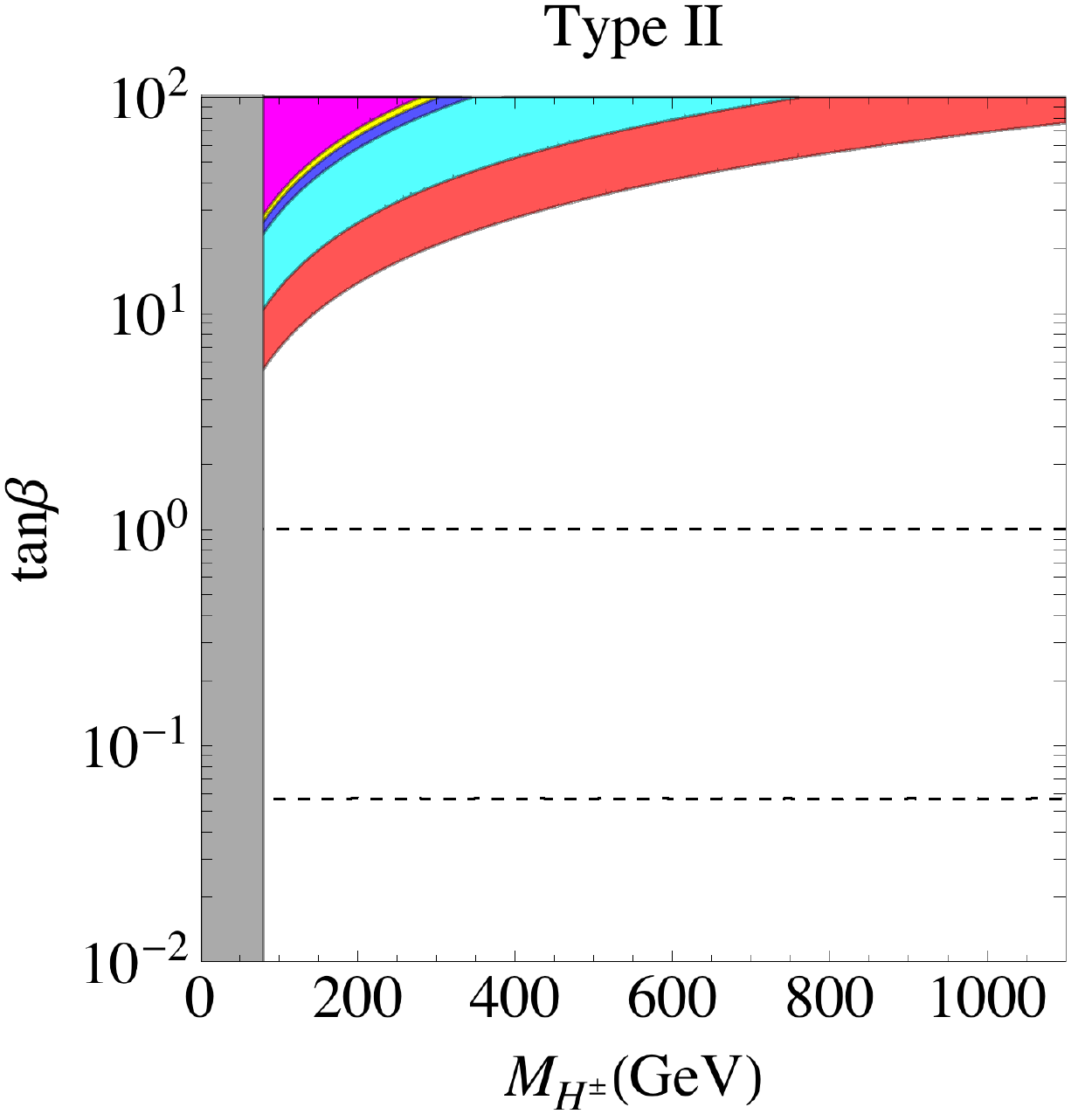}
\includegraphics[viewport=0 0 360 367, width=9.5em]{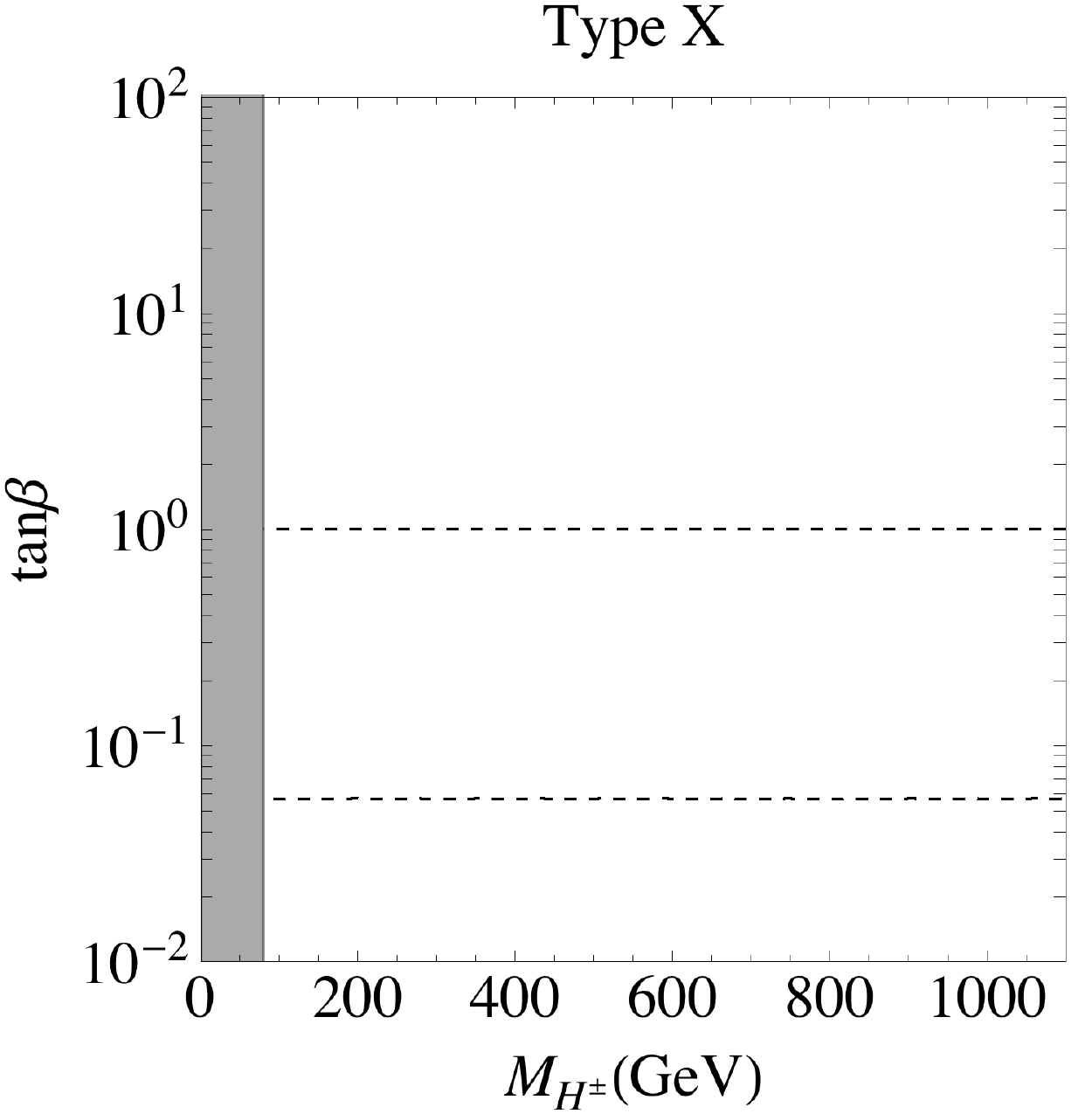}
\includegraphics[viewport=0 0 360 367, width=9.5em]{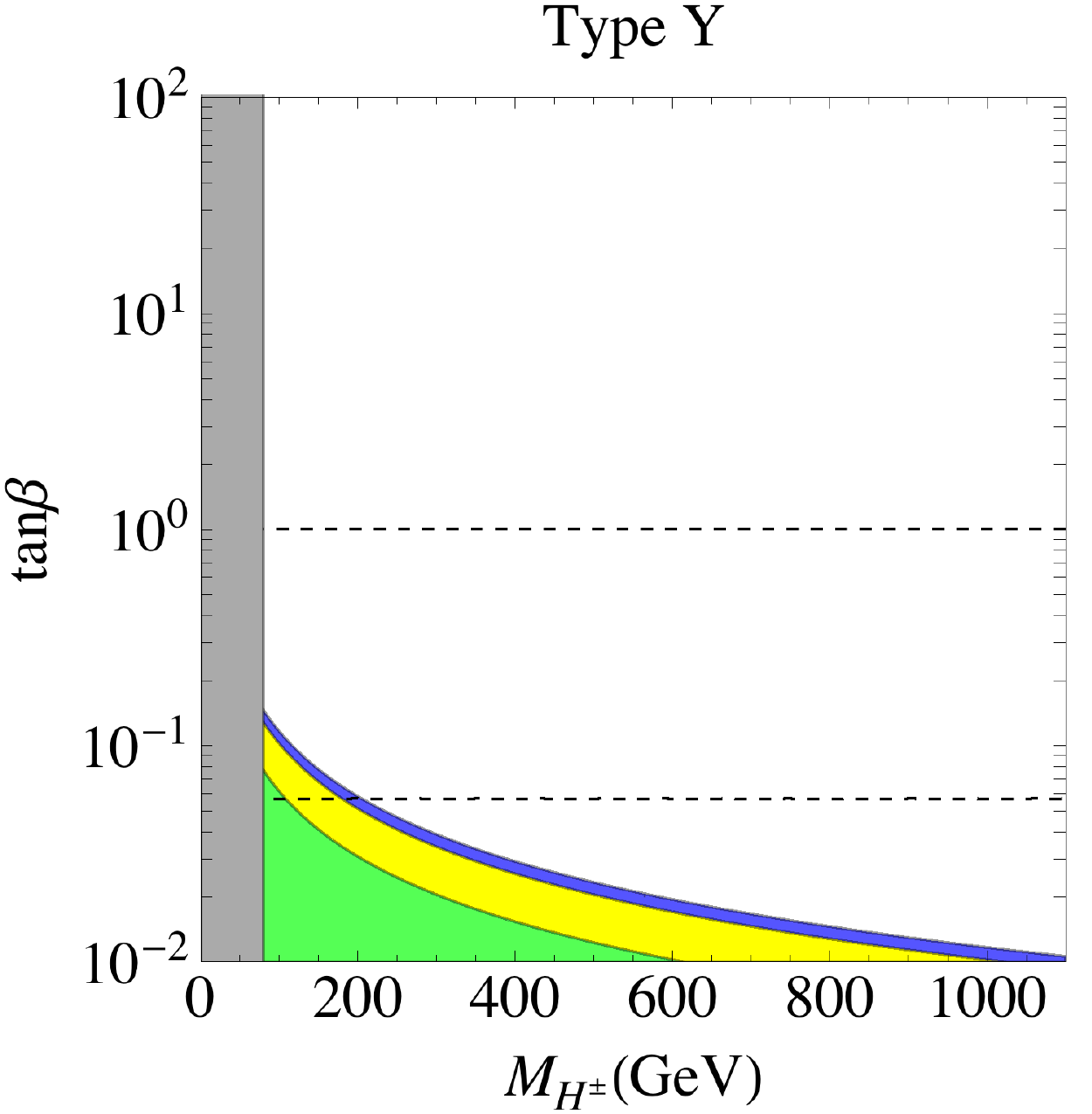} \\[0.5em]
\includegraphics[viewport=0 0 360 367, width=9.5em]{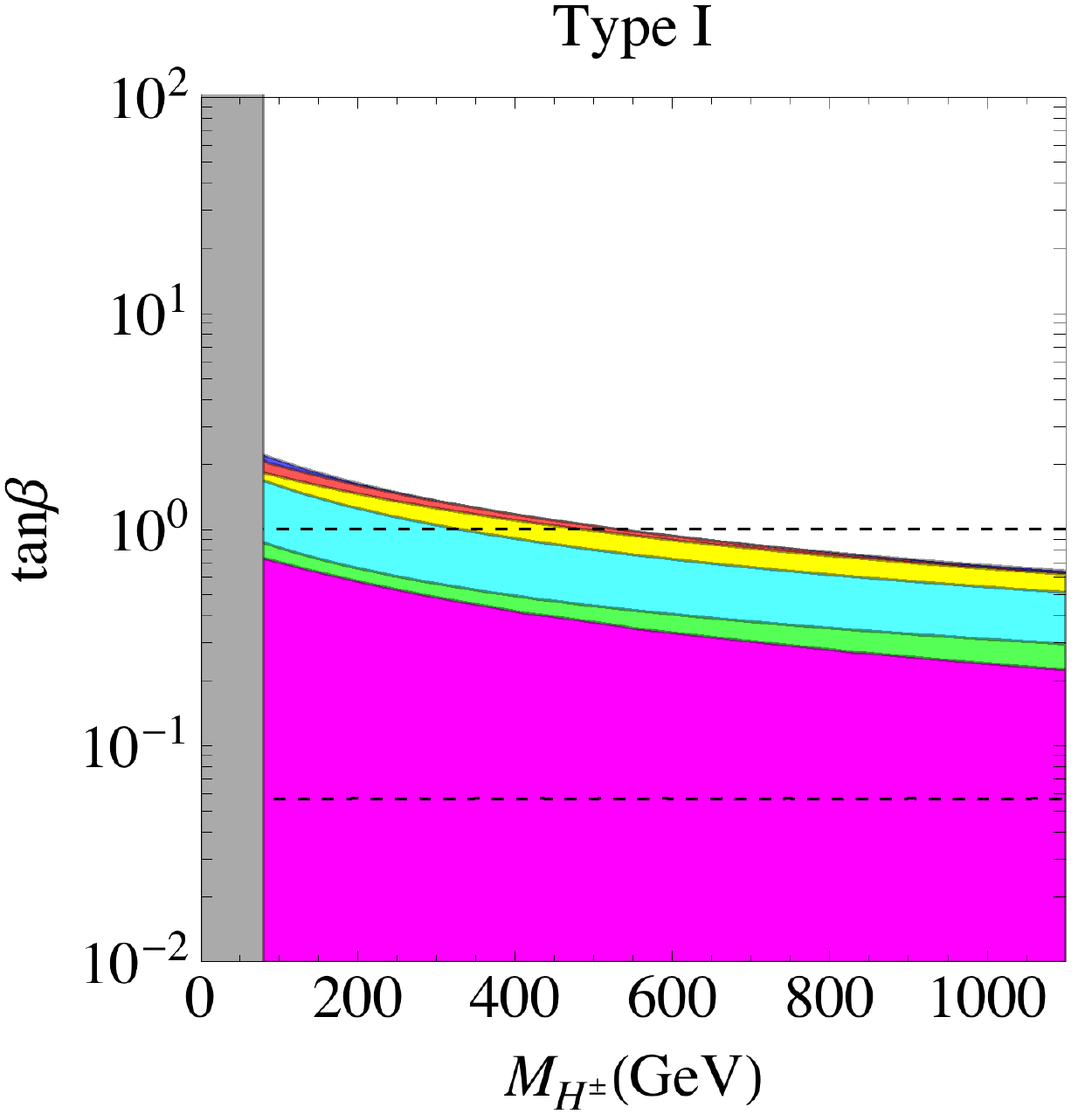}
\includegraphics[viewport=0 0 360 367, width=9.5em]{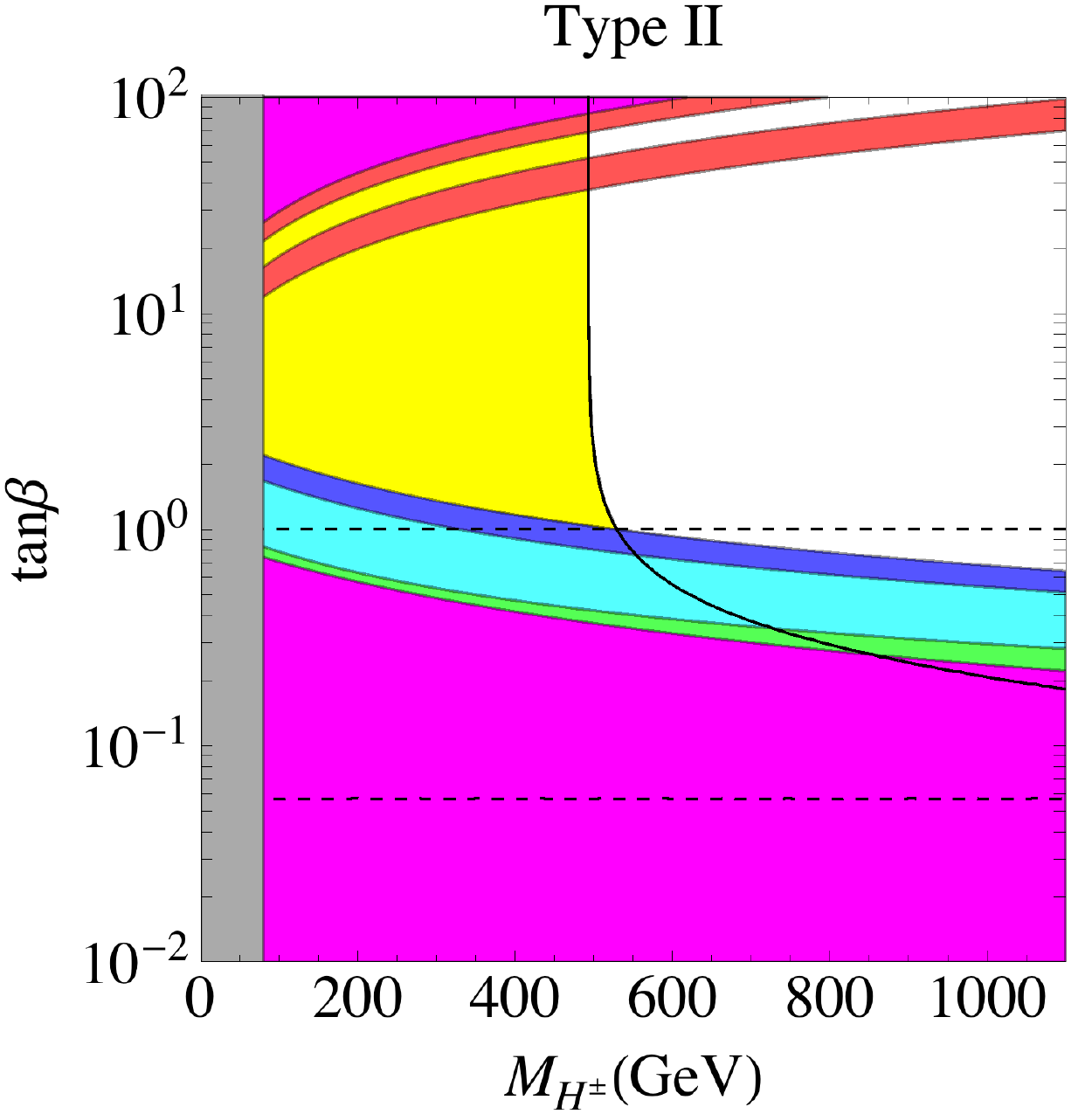}
\includegraphics[viewport=0 0 360 367, width=9.5em]{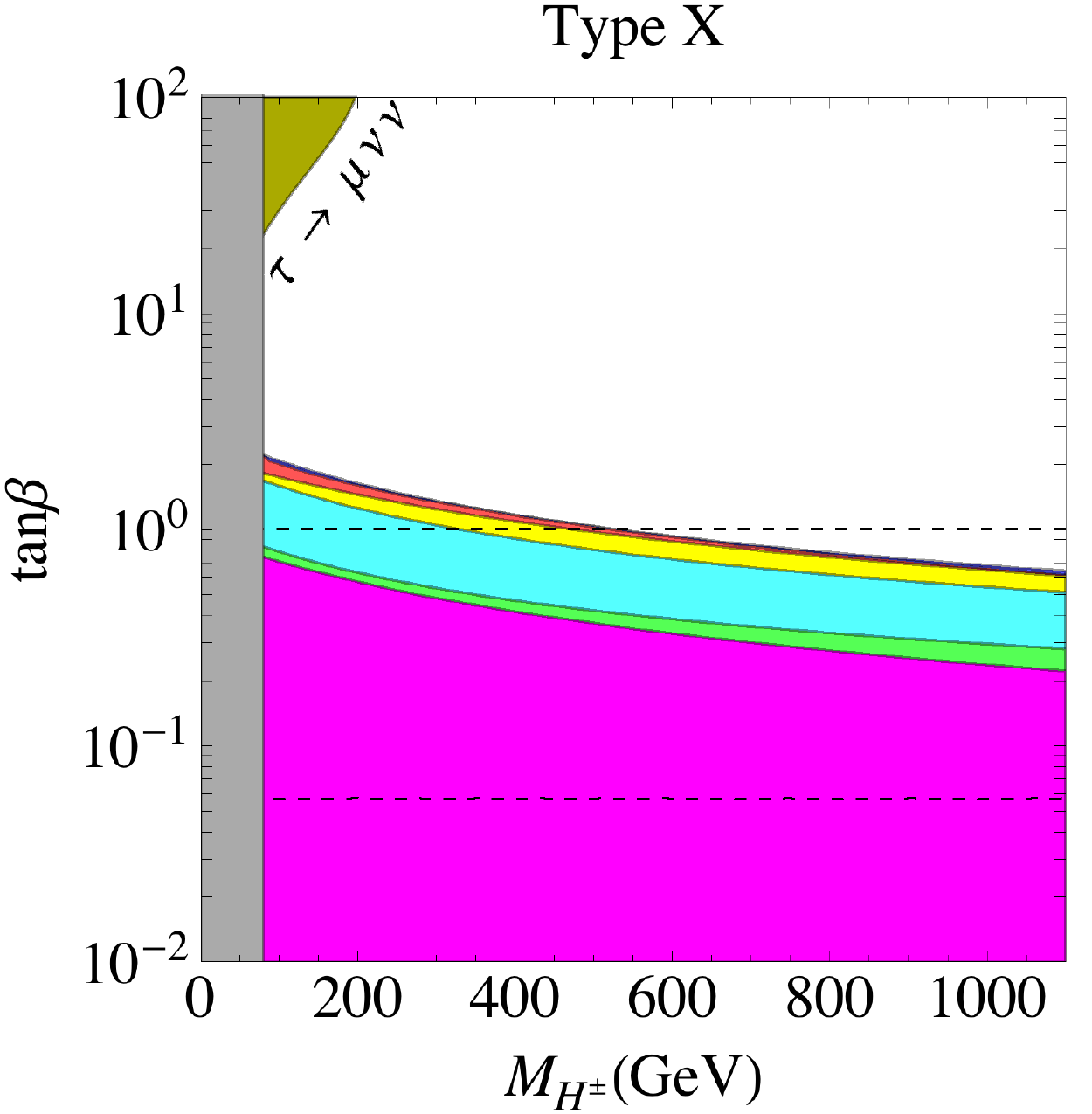}
\includegraphics[viewport=0 0 360 367, width=9.5em]{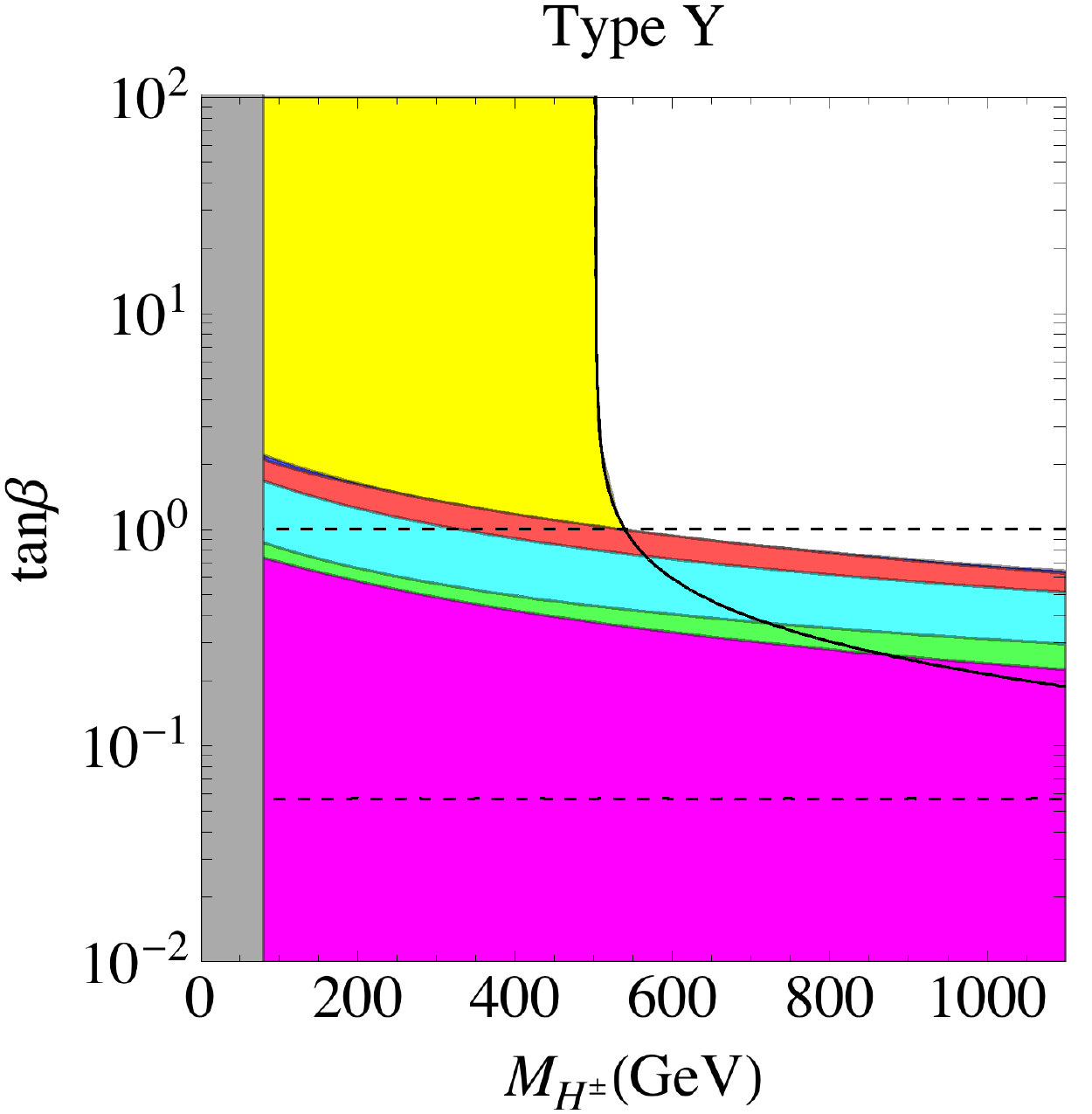}
\caption{
Excluded regions in the $Z_2$ symmetric models on the $(m_{H^+}, \tan\beta)$ plane at $95\%$ CL individually from 
the tree level processes $B\to\tau\nu$ (red), $D\to\mu\nu$ (green), $D_s\to\tau\nu$ (blue), $D_s\to\mu\nu$ (yellow), $K\to\mu\nu / \pi\to\mu\nu$ (cyan), $\tau \to K\nu / \tau \to \pi\nu$ (magenta) in the upper panels, 
and  
the loop induced processes $B^0_s \to \mu^+ \mu^-$ (red), $B^0_d \to \mu^+ \mu^-$ (magenta), $\bar B\to X_s\gamma$ (yellow), $\Delta M_s$ (blue), $\Delta M_d$ (cyan), $| \epsilon_K|$ (green) in the lower panels. 
The black line contour in the type~II and Y is the boundary of $95\%$ CL exclusion from $\bar B\to X_s\gamma$. 
The dashed horizontal lines are ones for $\tan\beta=1$ and $0.057$, corresponding to the top Yukawa coupling to be $1$ and $4\pi$, respectively. 
The gray region is the minimal exclusion from LEP searches~\cite{Abbiendi:2013hk}. 
The exclusion from $\tau\to\mu\nu\nu$ is also shown in the type~X~\cite{Abe:2015oca}. 
}
\label{Fig:Z2bound}
\end{center}
\end{figure}
%%%%%%%%%%%%FIG%%%%%%%%%%%%
 
In Fig.~\ref{Fig:Z2bound}, we show constraints on the plane of $(m_{H^+}, \tan\beta)$ at $95\%$ CL from the individual observables in the type~I, II, X, and Y models. 
These constraints are given by evaluating $\chi^2$ for each observable. 
Comments on the results for each model are as follows. 

\subsubsection*{Type~I:}
The region $\tan\beta \lesssim 1$ is strongly constrained by $\overline{\mathcal B} (B^0_s \to \mu^+ \mu^-)$ and $\Delta M_s$ in the type~I model. 
The other observables also provide the constraints for small $\tan\beta$ in this type. 
We can see that the extra Higgs boson mass is not constrained by the flavor observables in the large $\tan\beta$ range. 

\subsubsection*{Type~II:}
In the type~II model, the dominant constraint comes from $\mathcal B (B\to\tau\nu)$ and $\overline{\mathcal B} (B^0_q \to \mu^+ \mu^-)$ for large $\tan\beta$. 
The branching ratio $\overline{\mathcal B}(b\to s\gamma)_{E_\gamma>1.6\,\text{GeV}}$ gives the lower limit on the mass. 
Our result shows that $m_{H^+} < 493\,\text{GeV}$ is ruled out at $95\%$ CL and close to what was reported in Ref.~\cite{Misiak:2015xwa}. 
Moreover, $m_{H^+} < 408\,\text{GeV}$ is excluded at $99\%$ CL. 
The loop induced processes such as the neutral meson mixings exclude the region for small $\tan\beta$ as well as the type~I model. 

\subsubsection*{Type~X:}
As for the type~X model, the processes $M\to\ell\nu$ and $\tau\to M\nu$ provide no constraint on the $(m_{H^+}, \tan\beta)$ plane from the current data, 
whereas the loop induced processes exclude the range for small $\tan\beta$ as well as for the type~I case. 
Indeed, as the $\tan\beta$ enhancement can be seen only in the lepton sector, we can put a constraint for large $\tan\beta$ region from the measurement of the Fermi constant $G_F$ from $\tau\to\mu\nu\nu$~\cite{Abe:2015oca}. 
In the figure, we show the result from $\tau\to\mu\nu\nu$, where we obtain the theoretical formula (at the one loop level) and the experimental data based on Ref.~\cite{Abe:2015oca}. 
A similar bound is obtained in the type~II model but we have checked that it is smaller than the one from $\mathcal B (B\to\tau\nu)$ and $\mathcal B (D_s\to\tau\nu)$. 
To conclude, the type~X model does not have a significant exclusion for the mass.

\subsubsection*{Type~Y:}
The type~Y model is constrained by $\overline{\mathcal B} (B^0_s \to \mu^+ \mu^-)$ and $\Delta M_s$ for small $\tan\beta$ as is the same with the other models.  
The lower mass limit is obtained by $\overline{\mathcal B}(b\to s\gamma)_{E_\gamma>1.6\,\text{GeV}}$, similarly to the type~II model, as $m_{H^+} < 493\,\text{GeV}$ ($408\,\text{GeV}$) at 95\% CL (99\% CL),  
because of the same couplings, $\xi_u^A = 1/\tan\beta$ and $\xi_d^A = \tan\beta$. 

\subsection{Constraint on the aligned model}
\begin{figure}[t]
\begin{center}
\includegraphics[viewport=0 0 360 363, width=9.5em]{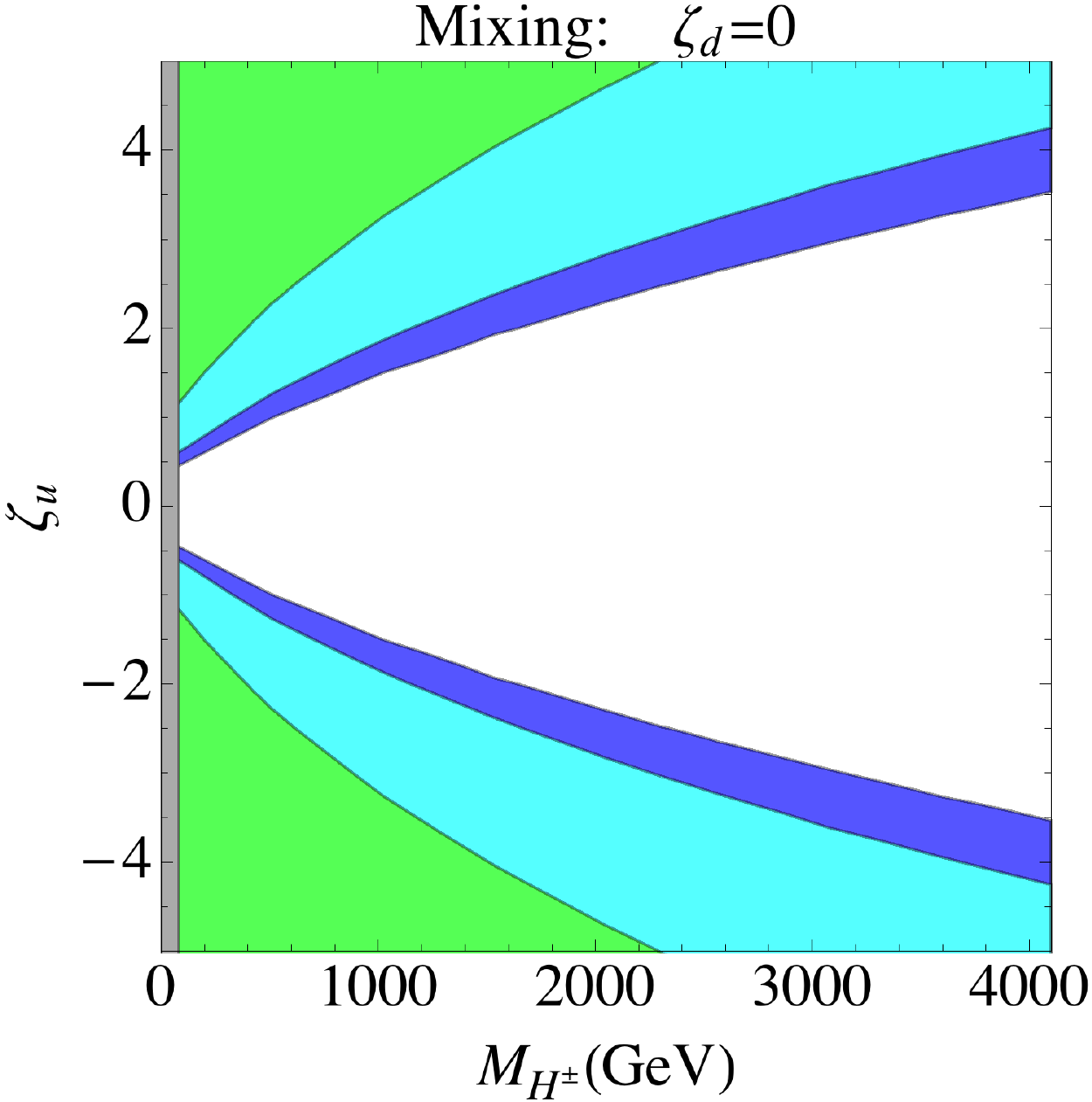}
\includegraphics[viewport=0 0 360 363, width=9.5em]{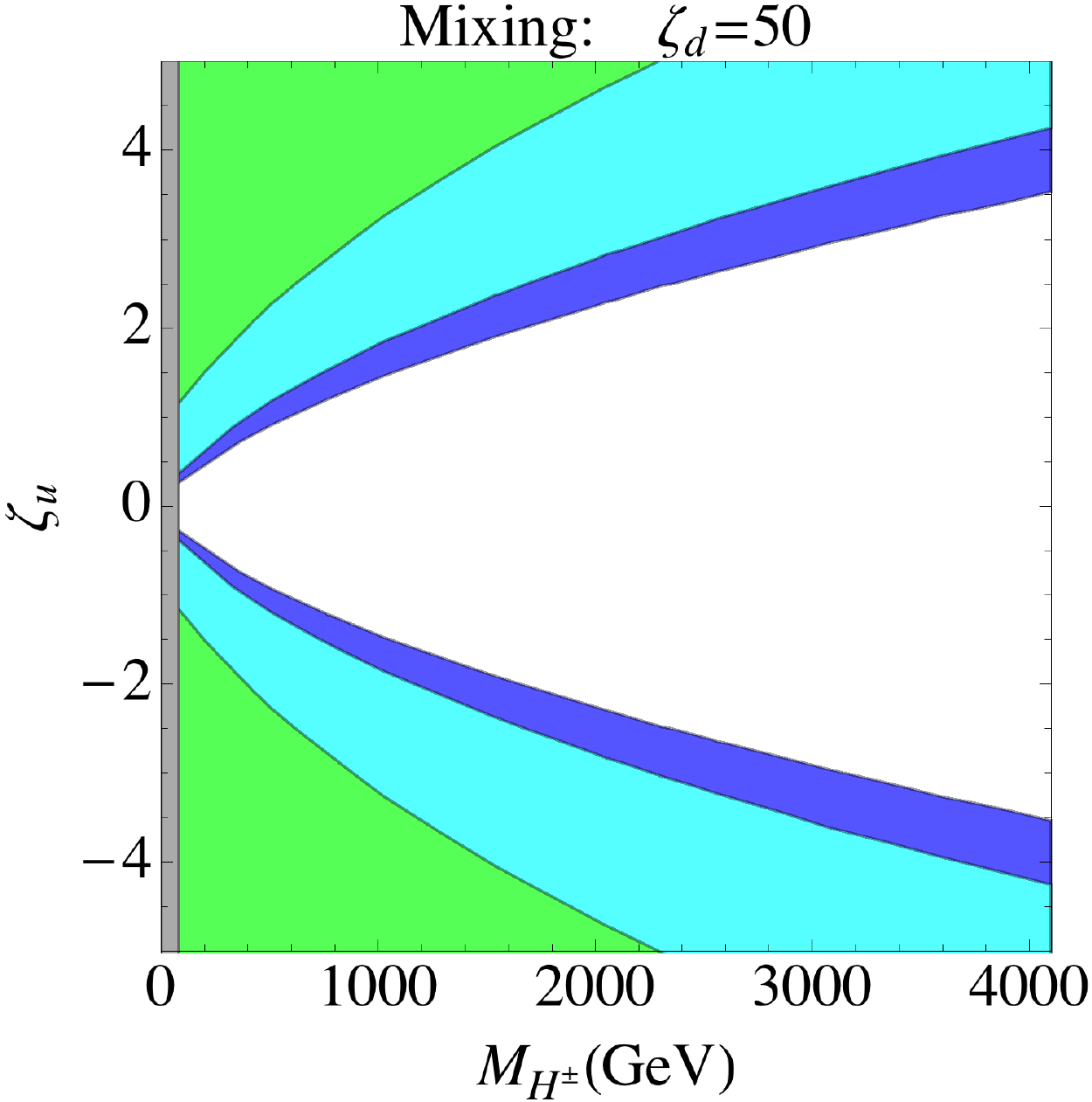}
\includegraphics[viewport=0 0 360 363, width=9.5em]{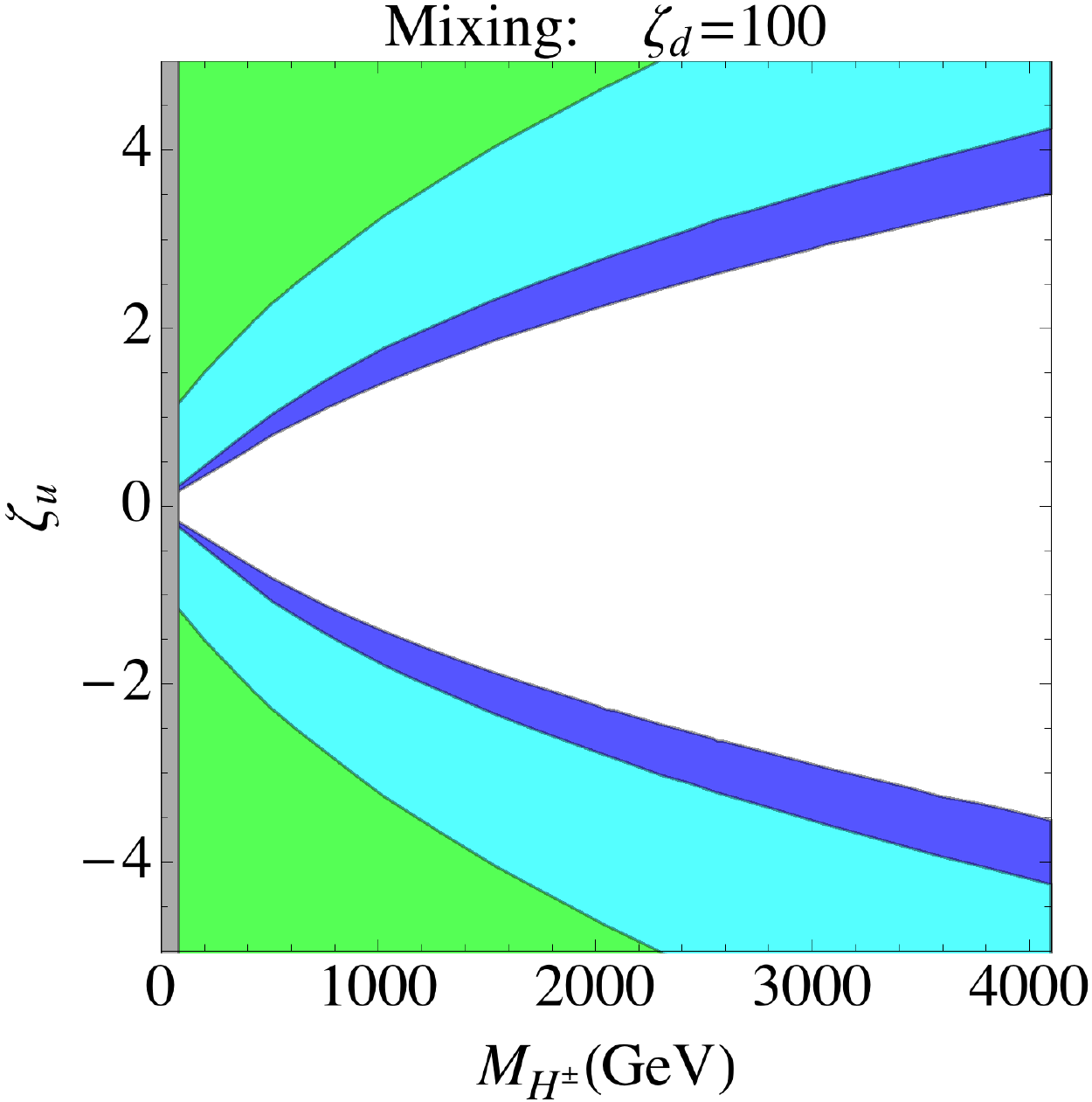}
\includegraphics[viewport=0 0 360 363, width=9.5em]{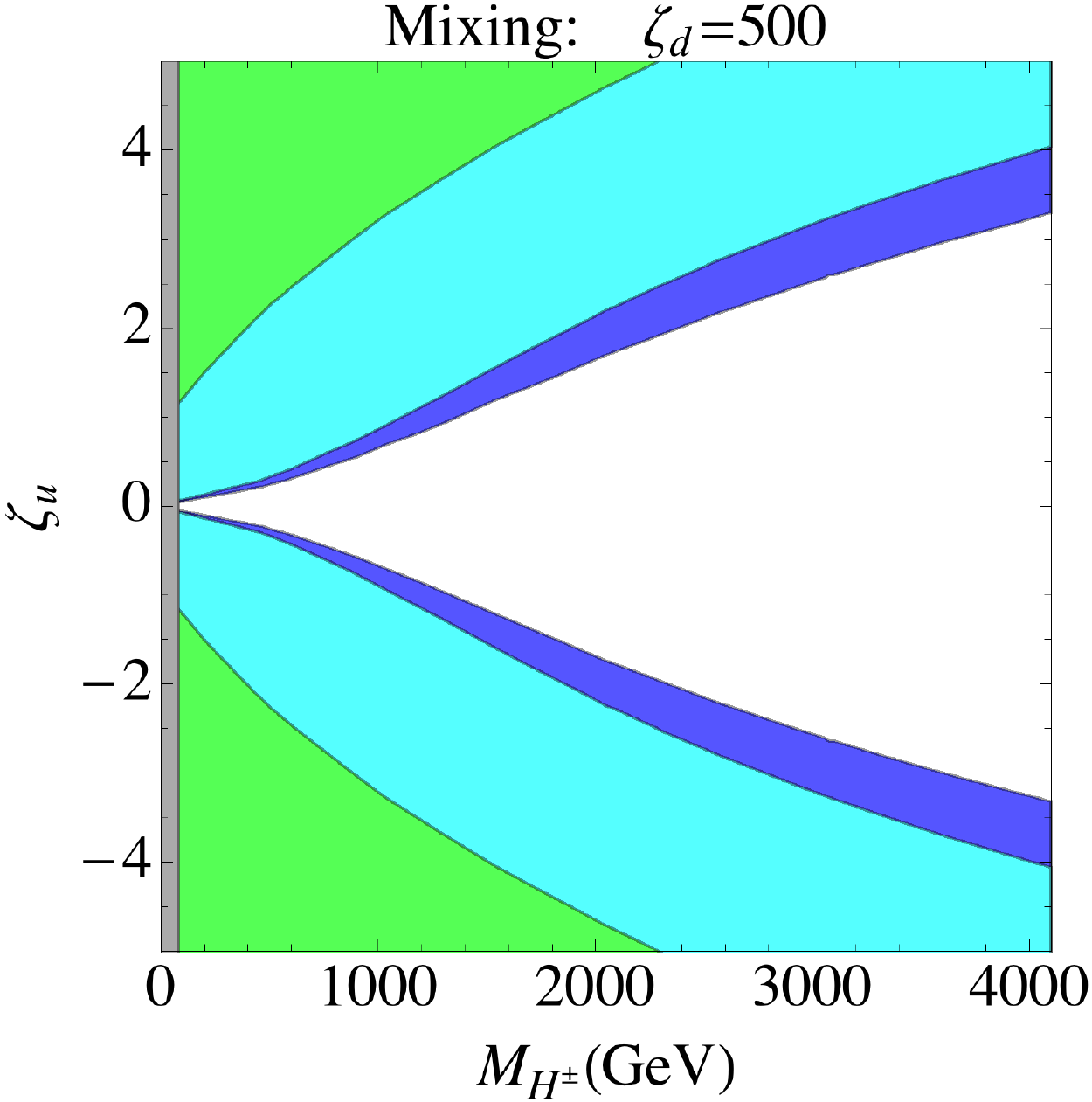}
\caption{
Excluded regions on the $(m_{H^+}, \zeta_u)$ plane at $95\%$ CL from $\Delta M_s$ (blue), $\Delta M_d$ (cyan), and $| \epsilon_K|$ (green) in the aligned model. 
The excluded regions are obtained as varying $\zeta_d=0$, $50$, $100$, and $500$ to see the dependence as denoted in the figure. 
Note that $\zeta_\ell$ is irrelevant for the neutral meson mixings. 
}
\label{Fig:AlignedMixing}
\end{center}
\end{figure}

%%%%%%%%%%%%%%%%%%%%%%%%
%%%%%%%%%%%%%%%%%%%%%%%%
Next, we see constraints on the aligned model. 
The tree processes $M\to\ell\nu$ are insensitive to the aligned model unless the products of $\zeta_u \zeta_\ell$ and/or $\zeta_d \zeta_\ell$ are very large. 
However, the large value of $\zeta_u$ is constrained by the neutral meson mixings as shown in Fig.~\ref{Fig:AlignedMixing}. 
The results do not much depend on $\zeta_d$ since the term including $\zeta_d$ is proportional to $x_b$ in $\Delta M_s$ and $\Delta M_d$ and there is no dependence in $| \epsilon_K|$. 
We can see that the small value of $\zeta_u$ is only allowed, {\it e.g.}, $|\zeta_u| < 1.5$ for $m_{H^+} = 1000\,\text{GeV}$ and $|\zeta_u| < 3.5$ for $m_{H^+} = 4000\,\text{GeV}$, 
as has been pointed out in Ref.~\cite{Jung:2010ik}.  
%
%%%%%%%%%%%%FIG%%%%%%%%%%%%
%%%%%%%%%%%%FIG%%%%%%%%%%%%

\begin{figure}[h]
\begin{center}
\includegraphics[viewport=0 0 360 355, width=9.5em]{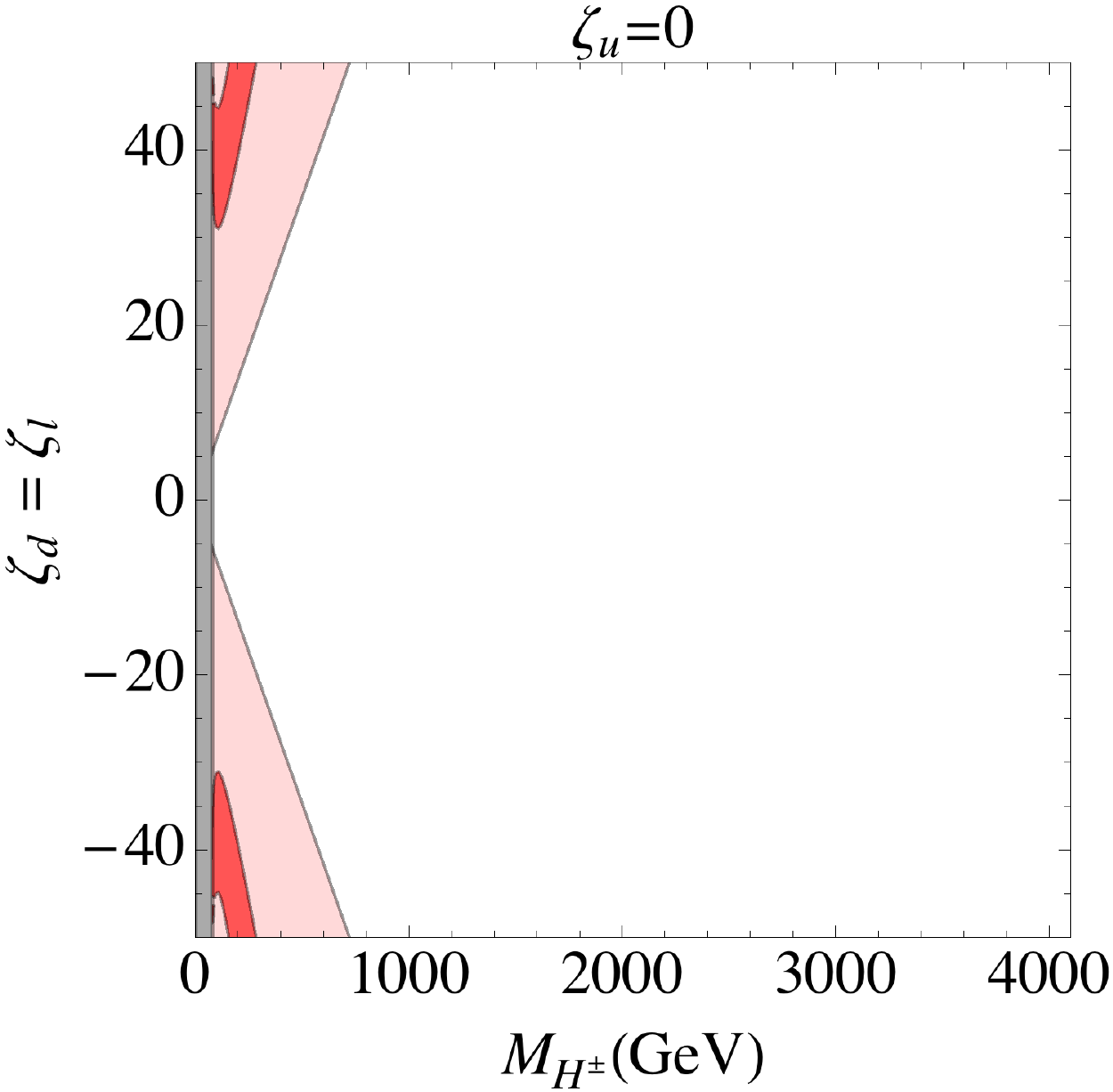}
\includegraphics[viewport=0 0 360 355, width=9.5em]{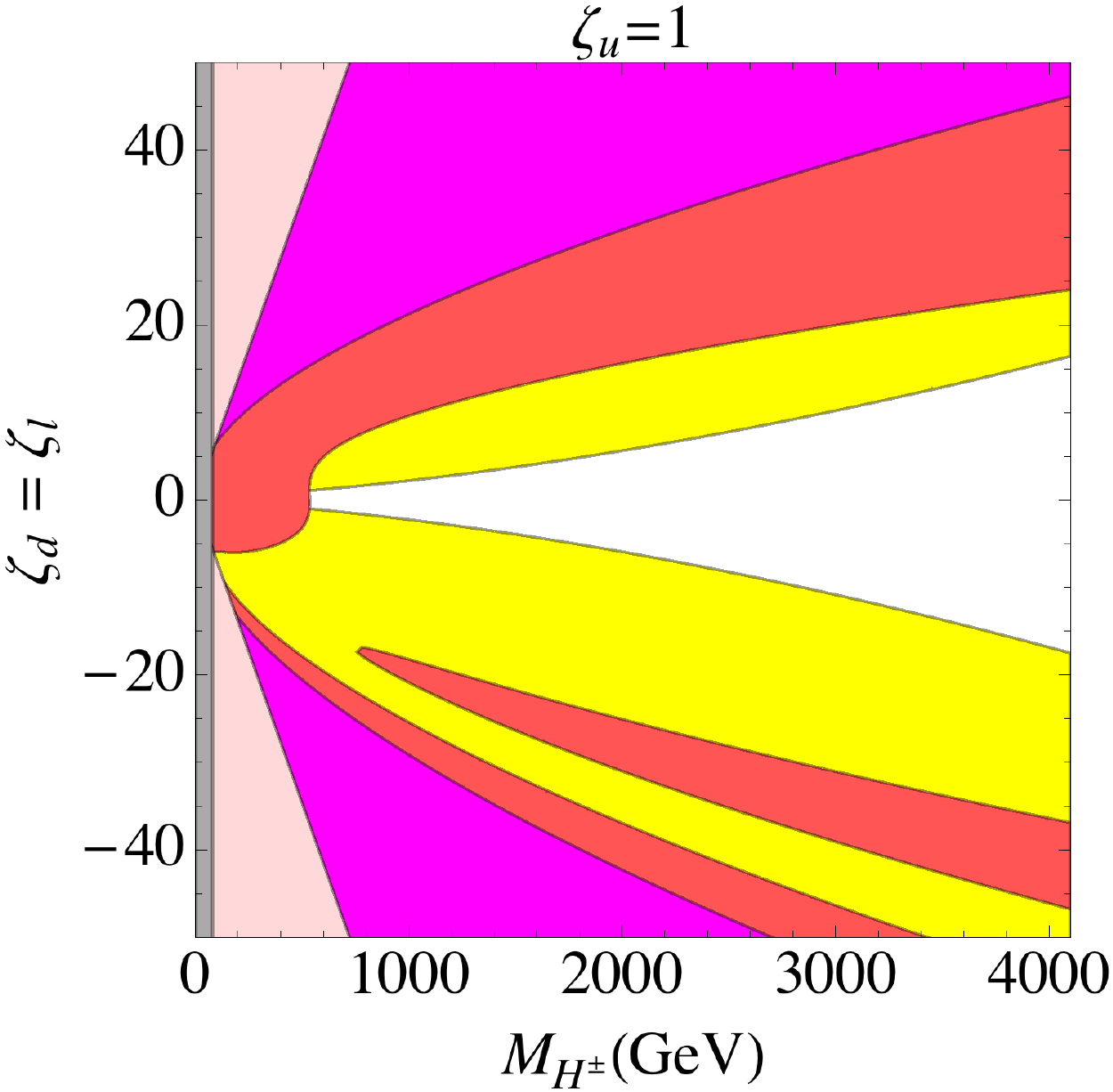}
\includegraphics[viewport=0 0 360 355, width=9.5em]{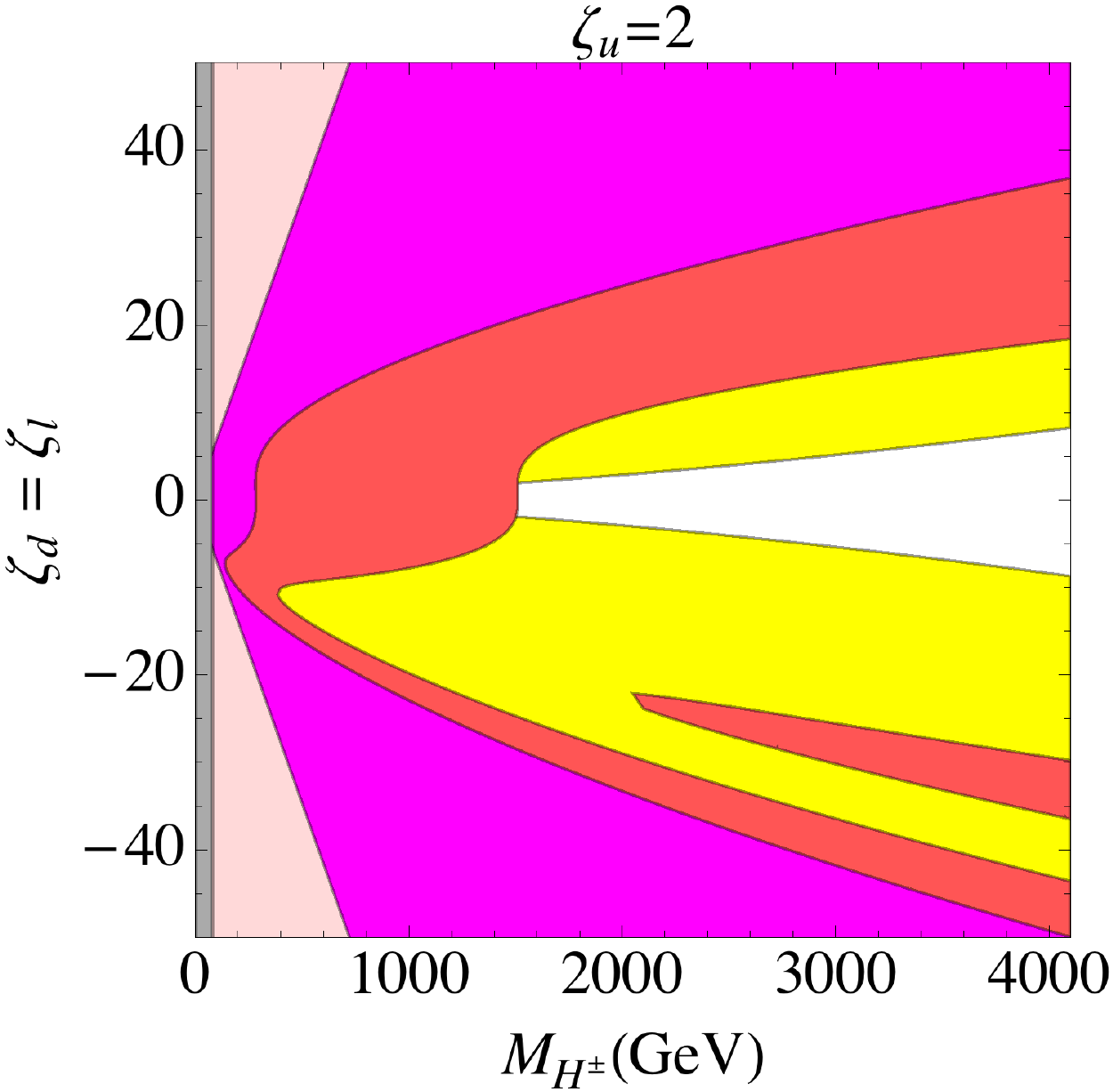}
\includegraphics[viewport=0 0 360 355, width=9.5em]{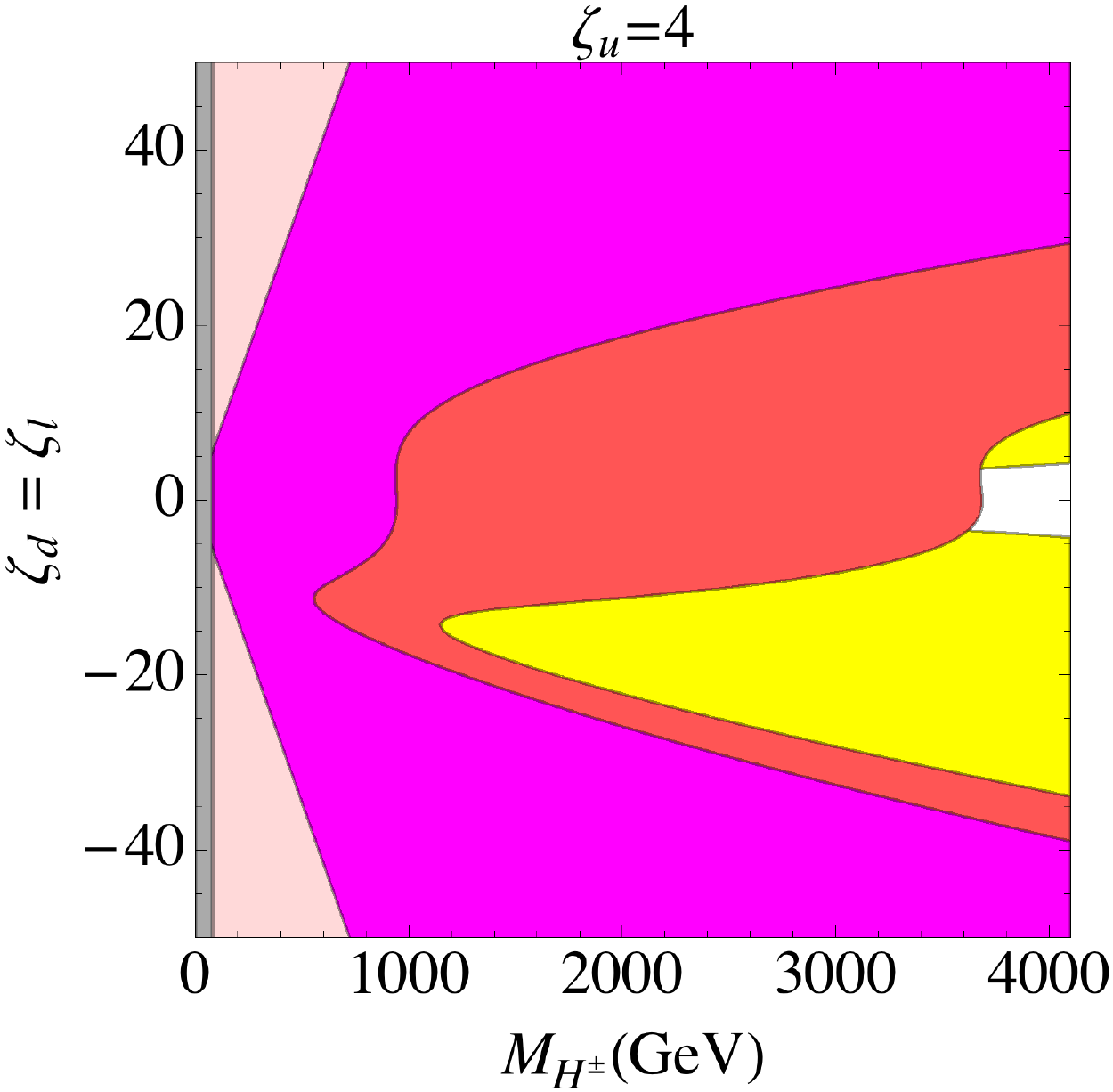} \\[0.5em]
\includegraphics[viewport=0 0 360 355, width=9.5em]{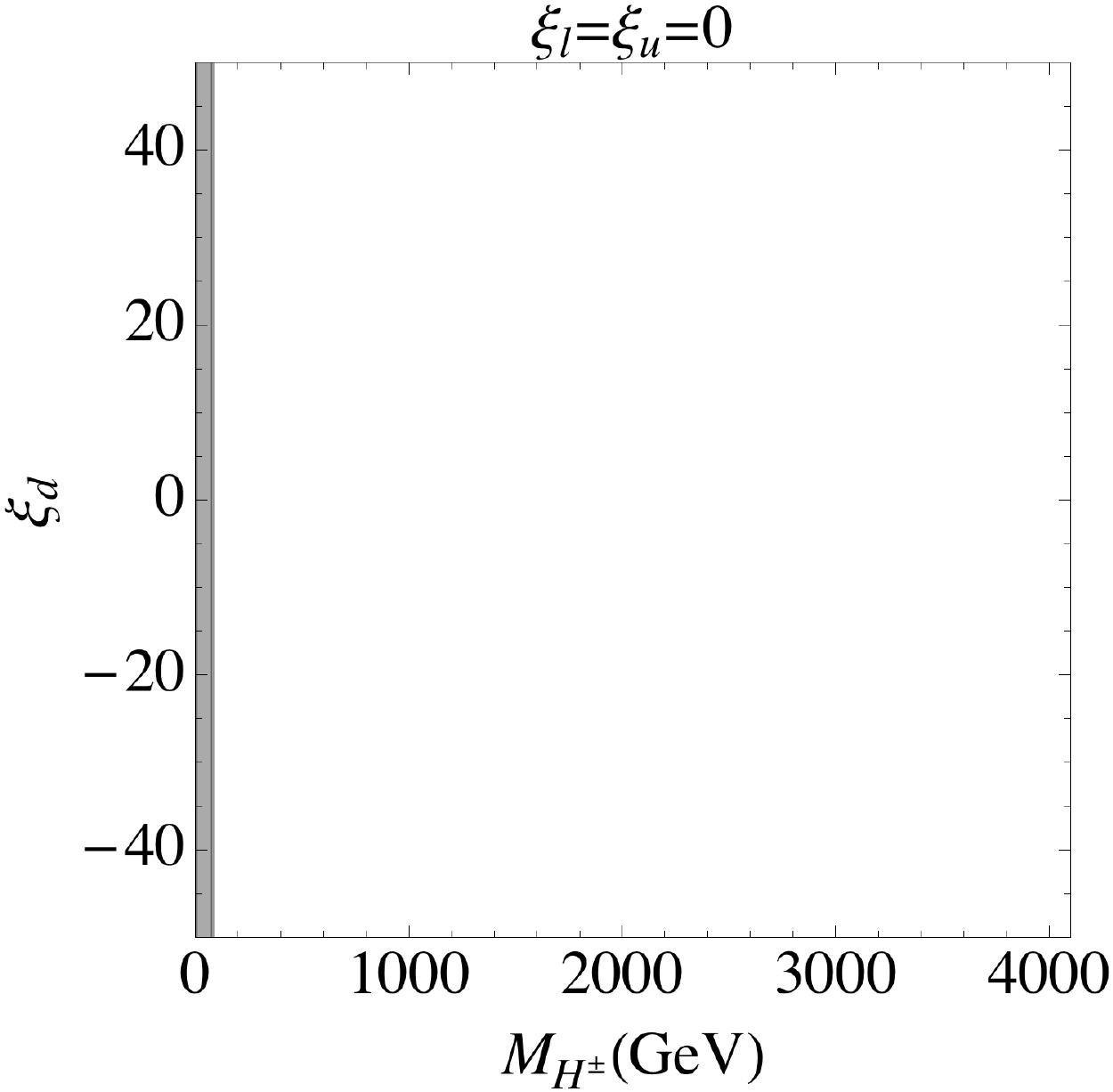}
\includegraphics[viewport=0 0 360 355, width=9.5em]{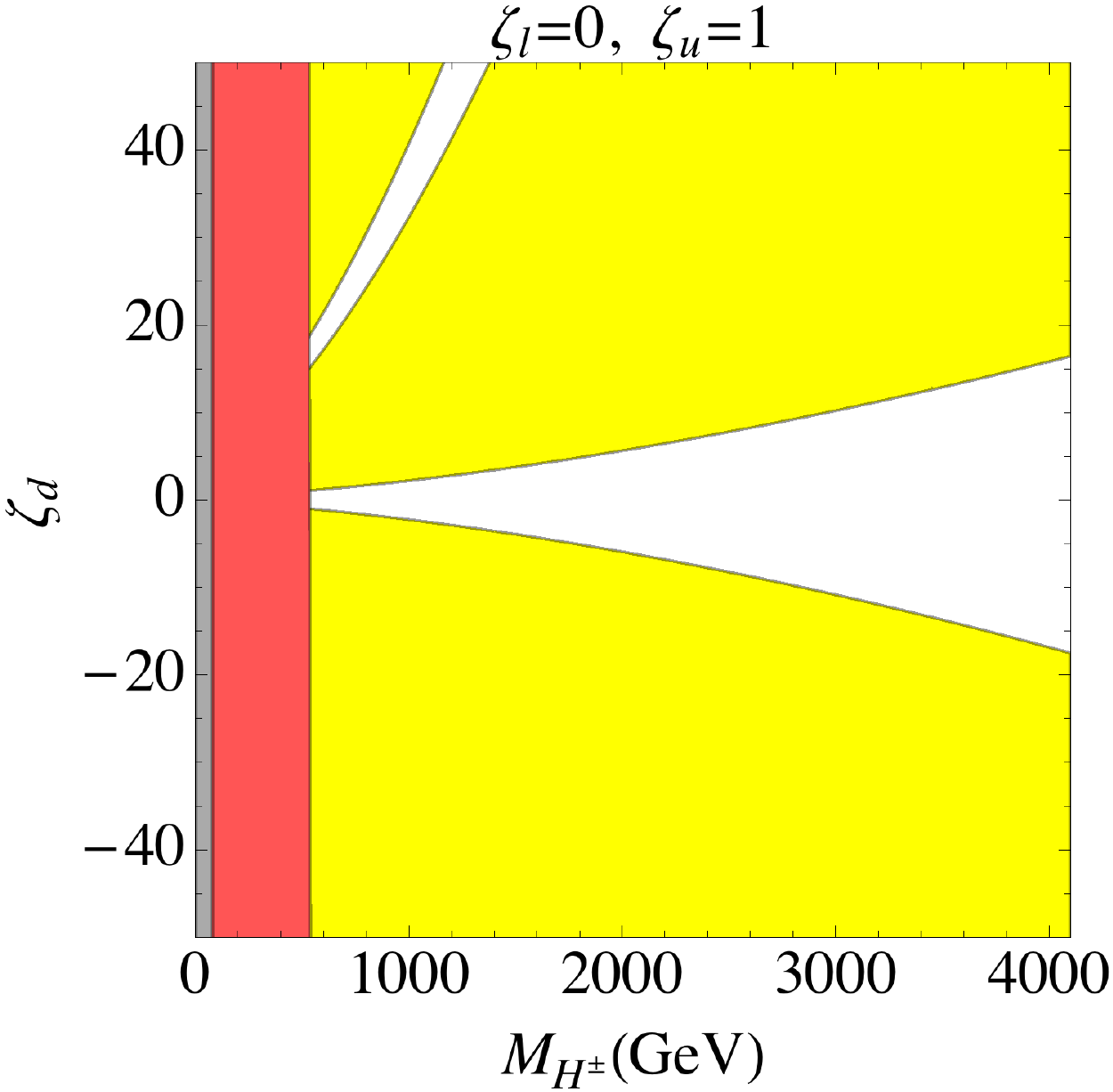}
\includegraphics[viewport=0 0 360 355, width=9.5em]{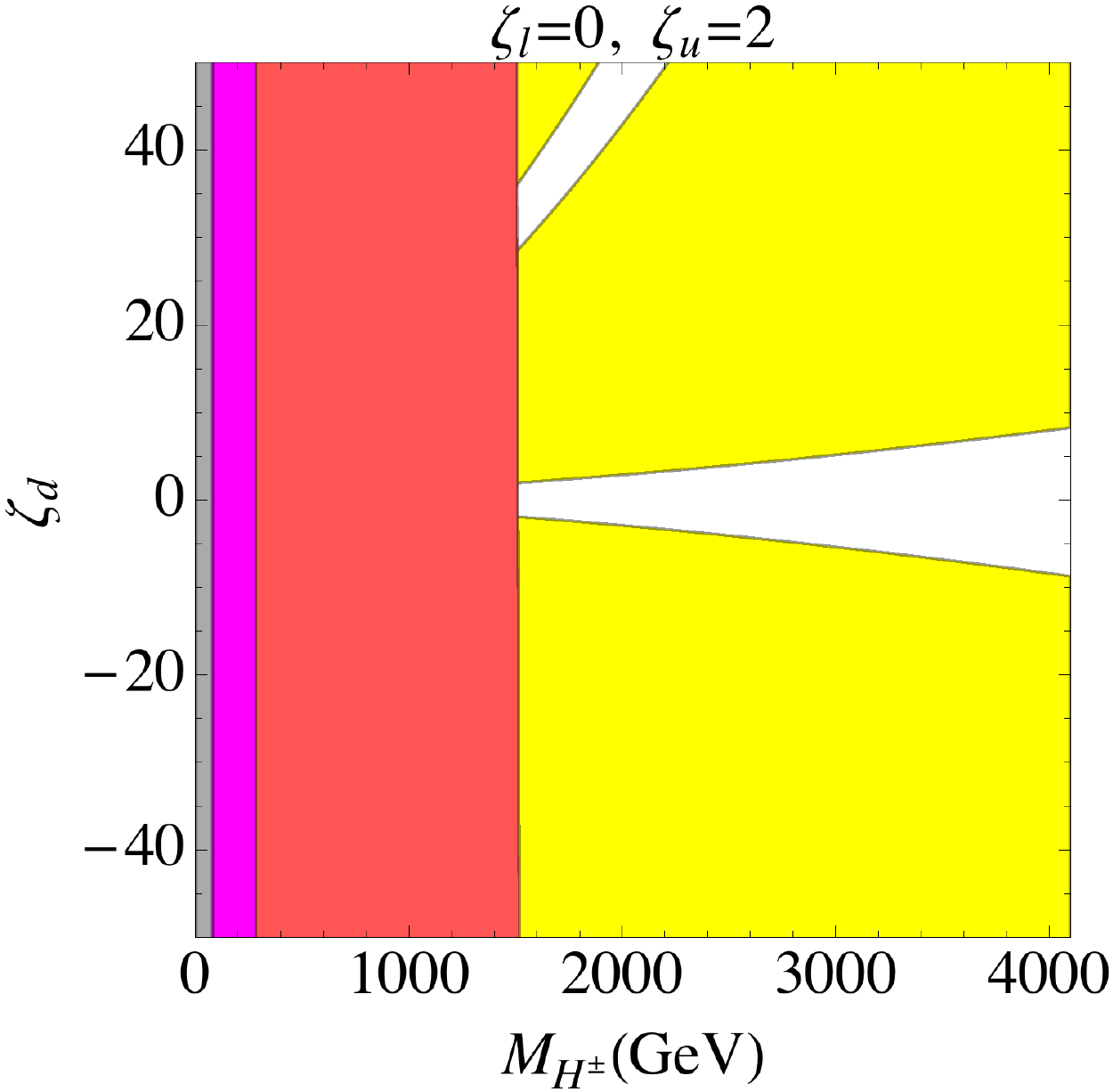}
\includegraphics[viewport=0 0 360 355, width=9.5em]{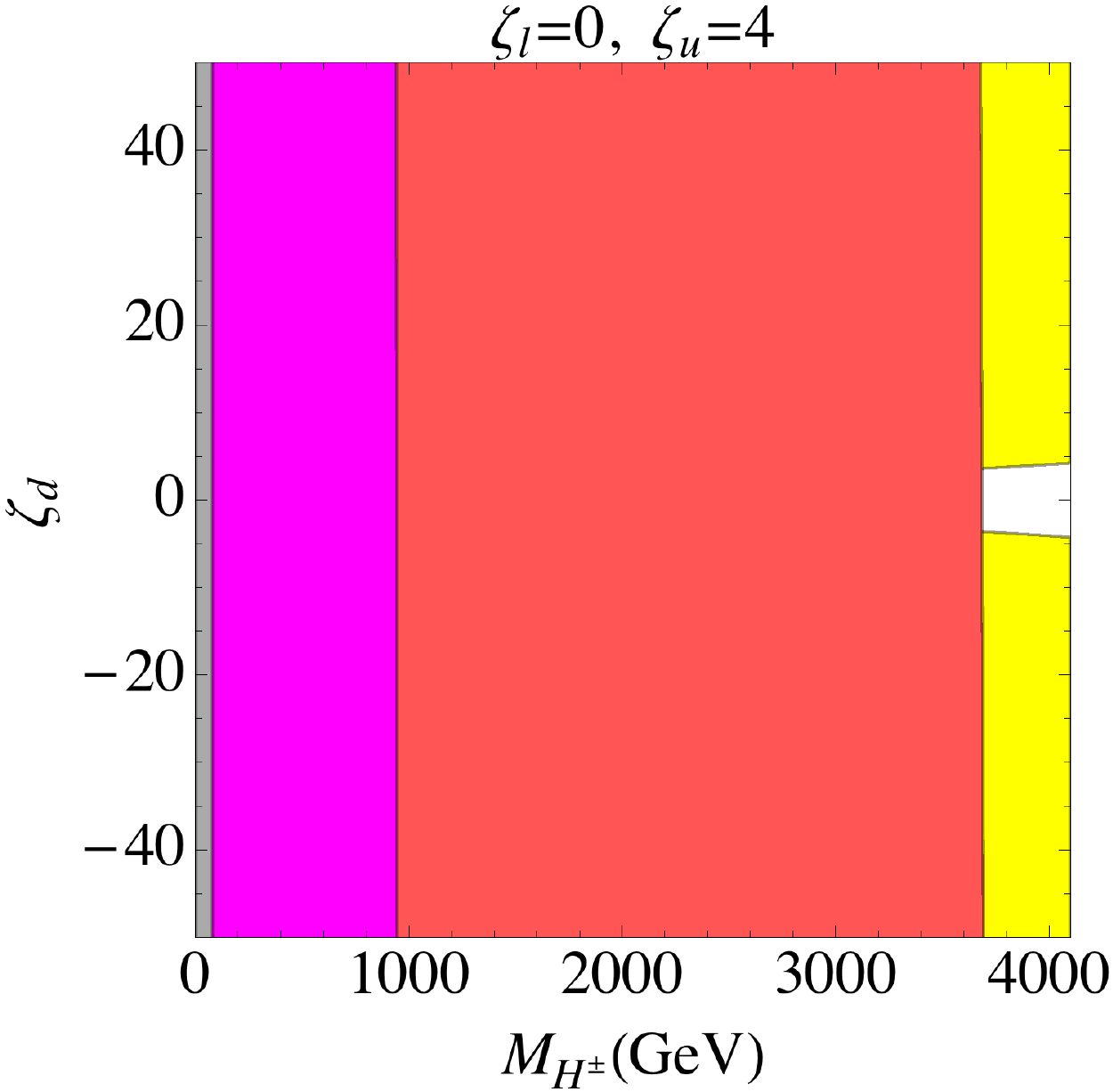}
\caption{
Excluded regions on $(m_{H^+}, \zeta_d)$ at $95\%$ CL from $\overline{\mathcal B}(b\to s\gamma)_{E_\gamma>1.6\,\text{GeV}}$, $\overline{\mathcal B} (B^0_q \to \mu^+ \mu^-)$, and $\mathcal B (B\to \tau\nu)$ 
with varying $\zeta_u$ and $\zeta_\ell$, where the results for $\zeta_\ell= \zeta_d$ and $\zeta_\ell= 0$ are shown in the upper and lower panels, respectively. 
The yellow, red, magenta, and light red regions are excluded by $\bar B\to X_s\gamma$, $B^0_s \to \mu^+ \mu^-$, $B^0_d \to \mu^+ \mu^-$, and $B\to \tau\nu$, respectively. 
}
\label{Fig:AlignedBsgmumu}
\end{center}
\end{figure}

On the other hand, $\zeta_d$ can be limited by $\overline{\mathcal B} (B^0_q \to \mu^+ \mu^-)$ and $\overline{\mathcal B}(b\to s\gamma)_{E_\gamma>1.6\,\text{GeV}}$. 
In Fig.~\ref{Fig:AlignedBsgmumu}, we show the constraints on $(m_{H^+}, \zeta_d)$ from these observables as varying $\zeta_u$ and $\zeta_\ell$. 
The upper and lower figures are the results for $\zeta_\ell=\zeta_d$ and $\zeta_\ell=0$, respectively. 
Constraints in the case of the negative value of $\zeta_u$ are obtained by replacing $\zeta_d$ to $-\zeta_d$ in the vertical axis of these plots. 
The parameter $\zeta_\ell$ is irrelevant for $b\to s\gamma$. 
For $\zeta_u =0$ and $\zeta_\ell \neq 0$, the constraint from $\mathcal B (B\to \tau\nu)$ becomes dominant but is insensitive for large mass. 
There is (trivially) no significant constraint for $\zeta_u =\zeta_\ell =0$. 
We can see that, for $\zeta_u\neq 0$, the combination of the constraints from $\overline{\mathcal B} (B^0_q \to \mu^+ \mu^-)$ and $\overline{\mathcal B}(b\to s\gamma)_{E_\gamma>1.6\,\text{GeV}}$ provides the lower mass limit. 
For example, we obtain the exclusions such as $m_{H^+} < 1500\,\text{GeV}$ for $|\zeta_u|=2$ and $m_{H^+} < 3700\,\text{GeV}$ for $|\zeta_u|=4$ at $95\%$ CL. 
This is the updated result of Refs.~\cite{Jung:2010ik,Jung:2010ab,Li:2014fea}.  
%
%%%%%%%%%%%%FIG%%%%%%%%%%%%
%%%%%%%%%%%%FIG%%%%%%%%%%%%

\subsection{Analysis of anomalies}
%%%%%%%%%%%%%%%%%%%%%%%%
%%%%%%%%%%%%%%%%%%%%%%%%
In this subsection, we study the 2HDM effect on $a_\mu$ and $R(D^{(*)})$, in which the deviations between the SM predictions and the experimental results have been reported.

The muon anomalous magnetic moment has been measured by the Muon G-2 collaboration as in Ref.~\cite{Bennett:2006fi}. 
Since this is a high precision test of the EW corrections, higher oder contributions in the SM are important and have been evaluated. 
Discrepancies between the experimental result reported in Ref.~\cite{Bennett:2006fi} and the SM predictions in Refs.~\cite{Jegerlehner:2009ry,Davier:2010nc,Hagiwara:2011af} are presented as 
\begin{align}
 a_\mu^\text{exp.} -a_\mu^\text{SM} = 
 \begin{cases}
  (282 \pm 91) \times 10^{-11} &  \quad\text{from Ref.~\cite{Jegerlehner:2009ry}}  \\
  (287 \pm 85) \times 10^{-11} &  \quad\text{from Ref.~\cite{Davier:2010nc}}  \\
  (261 \pm 80) \times 10^{-11} &  \quad\text{from Ref.~\cite{Hagiwara:2011af}} 
 \end{cases} \,\,\,.
 \label{Eq:g2_SM}
\end{align}
Even though there has been only one experimental measurement up to now, the deviation between the SM prediction and the experimental value is around $3\sigma$ as shown in (\ref{Eq:g2_SM}). 
As a reference value, we take $a_\mu^\text{exp.} -a_\mu^\text{SM} = (261 \pm 80) \times 10^{-11}$ from Ref.~\cite{Hagiwara:2011af} in the following study.

Excess of the observables $R(D^{(*)})$ in the semi-tauonic $B$ meson decays has been reported by the BaBar, Belle, and LHCb collaborations in Refs.~\cite{Lees:2012xj,Lees:2013uzd,Huschle:2015rga,Aaij:2015yra}. 
The latest combined result suggests that the deviations from the SM predictions are described as
\begin{align}
 & R(D)^\text{exp.} - R(D)^\text{SM} = 0.089 \pm 0.051 \,, \\
 & R(D^*)^\text{exp.} - R(D^*)^\text{SM} = 0.070 \pm 0.022  \,, 
\end{align}
where the discrepancy reaches around $4 \sigma$ taking experimental correlations into account. 
We note that as $\bar B \to D^{(*)} \tau\bar\nu$ occur at the tree level in the SM, these deviations have an impact on limiting new physics.

In the 2HDM these three observables are affected by the Yukawa interactions of the extra Higgs bosons in (\ref{Eq:Z2interaction}). 
The formulae for $a_\mu$ and $R(D^{(*)})$ are shown in Sec.~\ref{Sec:FO}. 
As for the input parameters in $R(D^{(*)})$, we use~\cite{Amhis:2014hma} $\rho_1^2 = 1.186 \pm 0.054$, $\rho_{A_1}^2 = 1.207 \pm 0.026$, $R_1 = 1.403 \pm 0.033$, and $R_2 = 0.854 \pm 0.020$. 
One can easily see that the type~I and Y models cannot explain the present anomalies of $a_\mu$ and $R(D^{(*)})$ at all, since no large contributions to these processes are available.

%%%%%%%%%%%%FIG%%%%%%%%%%%%
\begin{figure}[t]
\begin{center}
\includegraphics[viewport=0 0 840 280, width=40em]{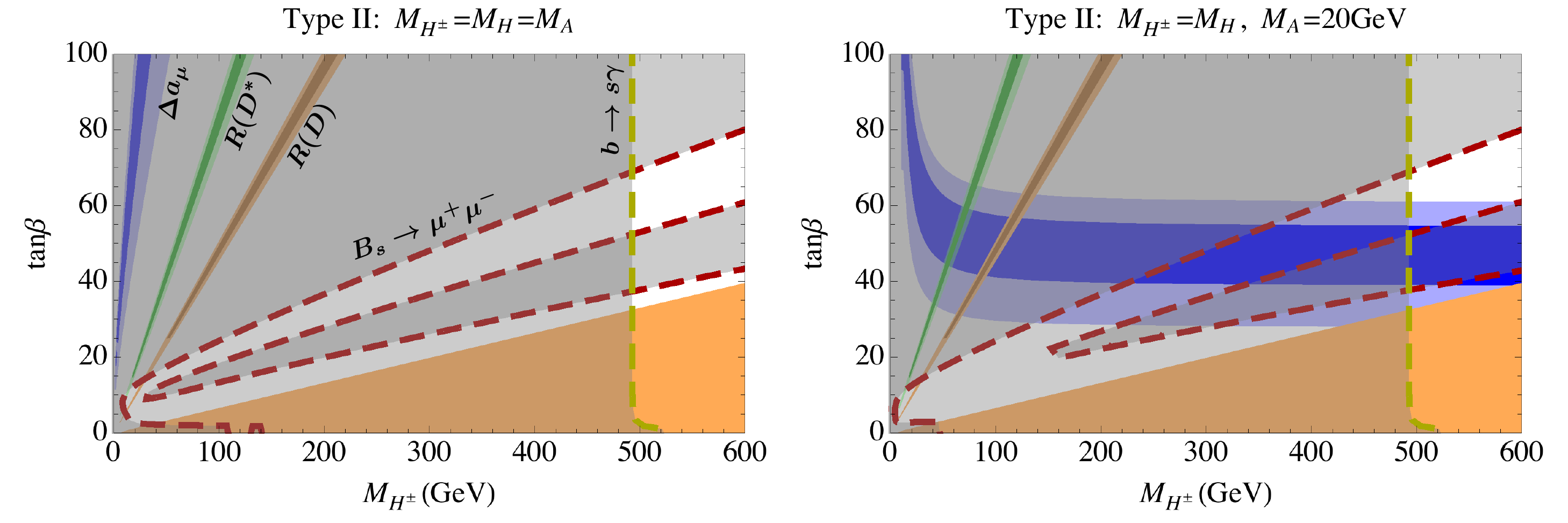}
\caption{
Allowed region plots on the $(m_{H^+}, \tan\beta)$ plane from $a_\mu$, $R(D)$, and $R(D^*)$ in the type~II model. 
The darker (lighter) blue, orange, and green regions are the results for $a_\mu$, $R(D)$, and $R(D^*)$ within $1\sigma$ ($2\sigma$).  
The excluded regions from $B^0_s \to \mu^+ \mu^-$ and $B\to\tau\nu$ at $95\%$ CL are shaded with red and yellow dashed boundaries, respectively.  
}
\label{Fig:Anomaly_II}
\end{center}
\end{figure}
%%%%%%%%%%%%FIG%%%%%%%%%%%%
%
The type~II model can explain these anomalies individually, however, it is inconsistent with each other and also contradictory to the other constraints obtained in Sec.~\ref{SubSec:Z2bound}. 
In Fig.~\ref{Fig:Anomaly_II}, we exhibit the allowed regions from $a_\mu$ and $R(D^{(*)})$ 
along with the excluded regions from $\overline{\mathcal B} (B^0_s \to \mu^+ \mu^-)$ and $\overline{\mathcal B}(b\to s\gamma)_{E_\gamma>1.6\,\text{GeV}}$ as indicated in the legend. 
We can see that the small value of $m_A$ is required to explain the anomaly in $a_\mu$. 
%The CP odd Higgs boson is irrelevant for $R(D^{(*)})$ and so their allowed regions in the upper and lower left figures are the same. 
But, in any case for the relation among the extra Higgs boson masses, the three anomalies cannot be explained at the same time, and are not consistent with the present constraints.

\begin{figure}[t]
\begin{center}
\includegraphics[viewport=0 0 800 540, width=40em]{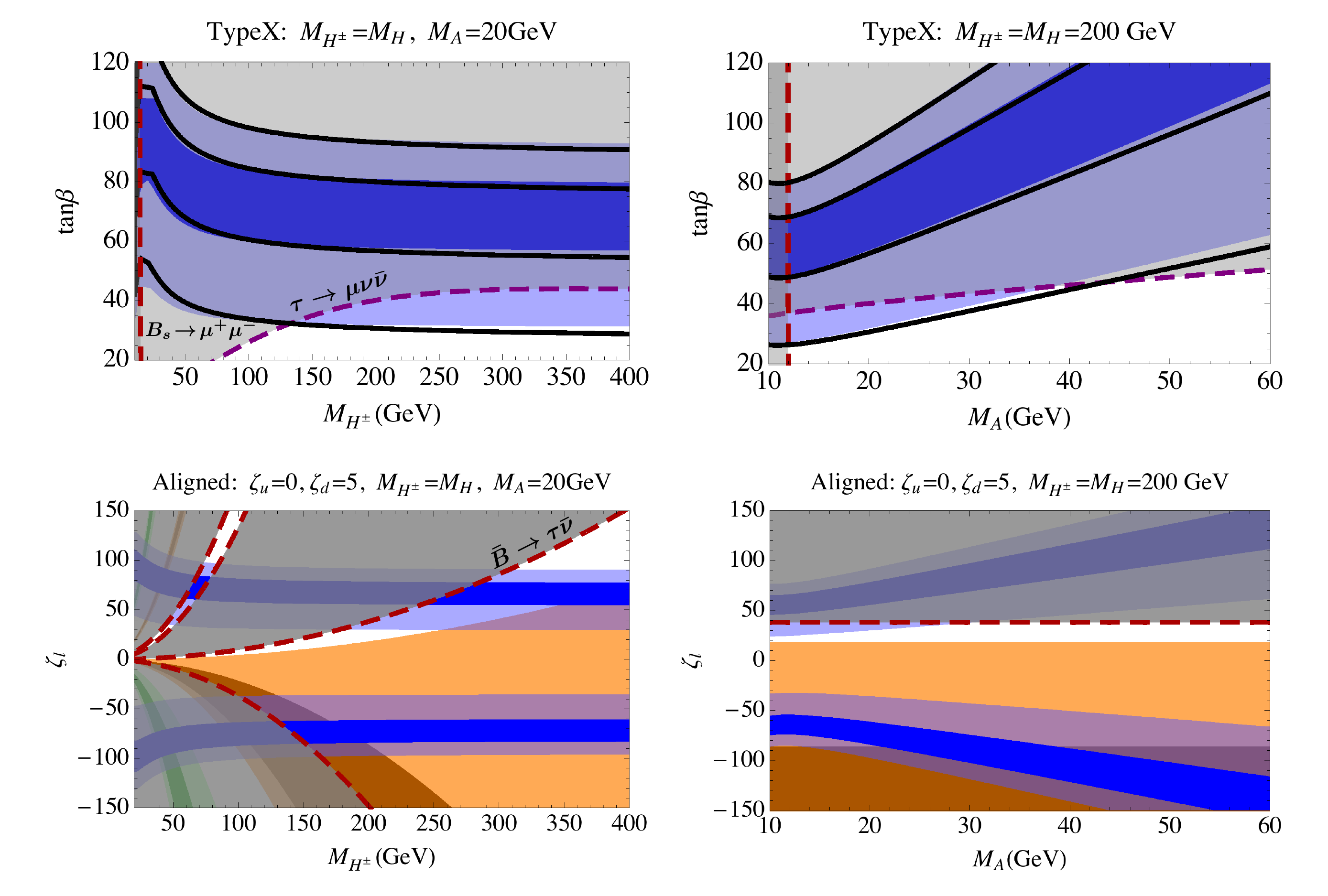}
\caption{
Allowed region plots from the $a_\mu$, $R(D)$, and $R(D^*)$ anomalies in the type~X (upper) and aligned (lower) models are shown with blue, orange, and green colors, as is similar with Fig.~\ref{Fig:Anomaly_II}. 
For the alined model, we take $\zeta_u=0$ and $\zeta_d=5$. 
For the type~X model, we show the allowed boundary from the $a_\mu$ anomaly with black thick curves in the case of $\sin(\beta - \alpha)=0.9$. 
In the left and right panels, $M_A = 20\,\text{GeV}$ and $m_H=m_{H^+}= 200\,\text{GeV}$ are taken, respectively. 
The excluded regions from $B^0_s \to \mu^+ \mu^-$, $B\to\tau\nu$, and $\tau \to \mu \nu \nu$ at $95\%$ CL are shaded with dashed boundaries as denoted in the figure.  
}
\label{Fig:Anomaly_AX}
\end{center}
\end{figure}

The type~X is often discussed as one of the solutions for the anomaly in $a_\mu$. 
The recent studies for the type~X model aiming at this anomaly are given in Refs.~\cite{Broggio:2014mna,Wang:2014sda,Abe:2015oca,Hektor:2015zba,Crivellin:2015hha,Chun:2015hsa}. 
In the upper panels of Fig.~\ref{Fig:Anomaly_AX}, we review the allowed region plot for $a_\mu$ in the type~X model. 
With small $m_A$, it can explain the $a_\mu$ anomaly. 
According to the study in Ref.~\cite{Abe:2015oca}, however, the constraint from $\tau \to \mu \nu \nu$ has turned out to be severe. 
In the figure, we also show the $95\%$ CL exclusion dashed line from $\tau \to \mu \nu \nu$. 
As can be seen, the explanation of the $a_\mu$ anomaly at the $1\sigma$ level is not possible, but that at the $2\sigma$ level is accessible in the type~X model. 
The result for the case of $\sin(\beta - \alpha)=0.9$ is also shown with the black lines and one find it does not change the above conclusion. 
Remind that the constraints from the meson observables are negligible. 
Then, in any case, this model cannot accommodate the excesses in $R(D)$ and $R(D^*)$.

%%%%%%%%%%%%FIG%%%%%%%%%%%%
%%%%%%%%%%%%FIG%%%%%%%%%%%%
%
In the aligned model, the parameters $\zeta_u$, $\zeta_d$, and $\zeta_\ell$ are independent and thus there is a larger parameter space than that in the $Z_2$ symmetric models. 
Nevertheless, $\zeta_u$ is severely limited by $\Delta M_s$ and then 
the constraints on $\zeta_d$ and $\zeta_\ell$ come from $\overline{\mathcal B} (B^0_s \to \mu^+ \mu^-)$ and $\overline{\mathcal B}(b\to s\gamma)_{E_\gamma>1.6\,\text{GeV}}$, which are correlated to the value of $\zeta_u$. 
They are less bounded if $\zeta_u=0$ as we have seen in Figs.~\ref{Fig:AlignedMixing} and \ref{Fig:AlignedBsgmumu}. 
To see how the aligned model affect $a_\mu$ and $R(D^{(*)})$, we simply take $\zeta_u=0$ for clarity. 
The possible significant constraint comes from $\mathcal B (B\to\tau\nu)$ in this model as we can see in the leftmost panels of Fig.~\ref{Fig:AlignedBsgmumu}. 
We surveyed several parameter set and found that the anomalies in $R(D)$ and $a_\mu$ can be explained simultaneously in a specific region, 
such as $\zeta_u=0$, $\zeta_d=5$, $\zeta_\ell\sim -70$, $m_H=m_{H^+}\sim 200\,\text{GeV}$, and $m_A=20\,\text{GeV}$. 
The results are plotted in the lower panels of Fig.~\ref{Fig:Anomaly_AX}. 
We can also see that it is hard to accommodate the excess in $R(D^*)$ without any contradiction to the other constraints. 
This specific point is not excluded by the other constraints yet, but close to the excluded region from $\mathcal B (B\to\tau\nu)$. 
We note that the sizable value of $\zeta_u \zeta_\ell$ are required to explain both the excesses in $R(D)$ and $R(D^*)$ simultaneously as studied in Ref.~\cite{Tanaka:2012nw}. 
%Thus, the future update by the SuperKEKB/Belle~II can be conclusive about this possibility. 
 %namely $\mathcal C_{S_2}$ in (\ref{Eq:BDtaunu_CS})

%%%%%%%%%%%%%%%%%%%%%%%%%%%%%%%%%%%%%%%%%%%%%%%%%%
\section{Summary}
\label{Sec:Summary}
%%%%%%%%%%%%%%%%%%%%%%%%%%%%%%%%%%%%%%%%%%%%%%%%%%
We have given the comprehensive study from the observables of $\pi$, $K$, $D_{(s)}$, and $B_{(s)}$ for limiting the 2HDMs with the hypothesis of {\it natural flavor conservation}, namely, the $Z_2$ symmetric and aligned models. 
Then we have obtained the significant constraints on the masses and couplings of the extra Higgs bosons, 
and shown the possibilities to accommodate the anomalies in the muon anomalous magnetic moment and the tauonic $B$ meson decays.

We have considered the following flavor observables; $\mathcal B (B\to\tau\nu)$, $\mathcal B (D\to\mu\nu)$, $\mathcal B (D_s\to\tau\nu)$, $\mathcal B (D_s\to\mu\nu)$, $\mathcal B (K\to\mu\nu) / \mathcal B (\pi\to\mu\nu)$, $\mathcal B (\tau \to K\nu) / \mathcal B (\tau \to \pi\nu)$, $\overline{\mathcal B} (B^0_s \to \mu^+ \mu^-)$, $\overline{\mathcal B} (B^0_d \to \mu^+ \mu^-)$, $\overline{\mathcal B}(b\to s\gamma)_{E_\gamma>1.6\,\text{GeV}}$, $\Delta M_s$, $\Delta M_d$, and $| \epsilon_K|$, and collected the formulae of them, in some of which we have taken the updated calculations into account. 
In addition, we have also obtained the updated formula of $\Delta M_q$.

We have re-fitted the CKM matrix elements to the experimental data to which the extra Higgs bosons do not give large contributions, and evaluated the effect on the determination of $|V_{cb}|$. 
As a result we have seen no significant difference between the re-fitted values and the global SM fit values. 
With the use of the re-fitted CKM matrix elements and the latest combined experimental results summarized by the PDG and HFAG collaborations, 
we have evaluated the excluded regions on the model parameters of the 2HDMs, with carefully considering the uncertainties from the input parameters.  
To obtain the results, we have assumed the same masses among the extra Higgs bosons and the SM-like limit $\sin(\beta-\alpha)=1$, favored by the theoretical bounds, the EW precision tests, and the collider searches.

As a consequence of our work, in the $Z_2$ symmetric models, it has been found that $B^0_s \to \mu^+ \mu^-$ plays a significant role to constraint $\tan\beta$ as well as the $B^0_s$-$\bar{B}^0_s$ mixing. 
The charged Higgs boson mass is constrained by the process $\bar B \to X_s \gamma$ in the type~II and Y models. 
The updated theoretical evaluation and the experimental result of $\overline{\mathcal B}(b\to s\gamma)_{E_\gamma>1.6\,\text{GeV}}$ suggest that 
$m_{H^+} < 493\, (408) \,\text{GeV}$ is excluded at $95\%$ ($99\%$) CL in these two models. 
There is no severe constraint on the mass in the type~I and X models.

In the aligned model, there are three free parameters in the Yukawa interaction term of the charged Higgs, $\zeta_f$ for $f = u,d,\ell$. 
The neutral meson mixings constrain $\zeta_u$ and we have obtained severe bound as $|\zeta_u| < 1.5$ for $m_{H^+} = 1000\,\text{GeV}$ and $|\zeta_u| < 3.5$ for $m_{H^+} = 4000\,\text{GeV}$, 
which are mostly independent on the other couplings, $\zeta_d$ and $\zeta_\ell$. 
With a non-zero value of $\zeta_u$, the parameters $\zeta_d$ and $\zeta_\ell$ are limited by $\overline{\mathcal B}(b\to s\gamma)_{E_\gamma>1.6\,\text{GeV}}$ and $\overline{\mathcal B} (B^0_s \to \mu^+ \mu^-)$. 
For example, $|\zeta_d| \lesssim 5$ is excluded for $m_{H^+} = 1000\,\text{GeV}$ and $\zeta_u=1$. 
We have also shown that the combination of these two observables gives the lower mass limit  as $m_{H^+} < 1500\,\text{GeV}$ for $|\zeta_u|=2$ and $m_{H^+} < 3700\,\text{GeV}$ for $|\zeta_u|=4$.

In addition, we have summarized the current status of the anomalies in the muon anomalous magnetic moment $a_\mu$ and the tauonic $B$ meson decays $R(D^{(*)})$, in the context of the 2HDMs. 
We have shown that the type~II model can explain each anomaly, however, the allowed regions are not only inconsistent with each other but also contradictory to the other constraints. 
The type~X model is often considered as one of the good candidates to accommodate the excess of $a_\mu$. 
We have reconfirmed that this model can solve the excess of $a_\mu$ individually, but it is not consistent with the constraint from $\tau\to\mu\nu\nu$ at the $1 \sigma$ level. 
Note that this model cannot resolve the excesses in $R(D^{(*)})$, in any case. 
We also surveyed the possibility to explain the anomalies in the aligned model. 
We have pointed out that these three anomalies cannot be explained simultaneously, 
whereas the excesses of $a_\mu$ and $R(D)$ can be explained for $\zeta_u=0$, $\zeta_d=5$, $\zeta_\ell\sim -70$, $m_H=m_{H^+}\sim 200\,\text{GeV}$, and $m_A=20\,\text{GeV}$. 
This parameter set is allowed by the other flavor constraints yet, but close to the excluded region from $\mathcal B (B\to\tau\nu)$.

We have not considered semi-leptonic meson decays such as $\bar B \to \pi \tau\bar\nu$, $B\to (K^{(*)},\phi) \mu^+\mu^-$, and others to obtain the constraints. 
Although form factors in $B \to \pi, K^{(*)}$ transitions still include large uncertainties in fit parameters, these decays can provide constraints on new physics and will become more significant at the future experiments. 
The recent studies for $\bar B \to \pi \tau\bar\nu$ are given in Ref.~\cite{Bernlochner:2015mya} and for $B\to (K^{(*)},\phi) \mu^+\mu^-$ in Refs.~\cite{Altmannshofer:2014rta,Descotes-Genon:2015uva}. 
Exclusive radiative $B$ meson decays are also important, see {\it e.g.}, Ref.~\cite{Jung:2012vu}.

The bounds obtained in this work are expected to be the last updated status before starting the future flavor experiments such as the SuperKEKB/Belle~II~\cite{Aushev:2010bq} and the LHCb run~II~\cite{Koppenburg:2015wxa}. 
Future searches at the Belle~II and the LHCb run~II will improve sensitivities to the 2HDMs and may reveal the source of the excesses in the semi-tauonic $B$ meson decays. 
Future muon $g-2$ searches at the J-PARC~\cite{Mao:2011bz} and the Fermilab~\cite{Grange:2015fou} will also change the present situation on the anomaly.  
The requirement for explaining the excess of $a_\mu$ implies the mass of the CP odd Higgs boson should be small. 
Therefore, collider signatures from $h \to AA \to 4\tau, 4b$ can be important.

%%%%%%%%%%%%%%%%%%%%%%%%%%%%%%
%%%%%%%%%%%%%%%%%%%%%%%%%%%%%%
%%%%%%%%%%%%%%%%%%%%%%%%%%%%%%

%%%%%%%%%%%%%%%%%%%%%%%%%%%%%%%%%%%%%%%%%%%%%%%%%%
\begin{acknowledgments}
We are grateful to Mikolaj Misiak for his helpful and crusial remarks on the charged Higgs contribution to $\bar B \to X_s \gamma$.  
We are grateful to Xin-Qiang Li for an useful comment on the importance of the external momentum of the $b$-quark in the neutral meson mixings. 
We thank Minoru Tanaka for a helpful comment on the running quark masses, and the CKMfitter group for a suggestion on the uncertainties in the determination of $|V_{cb}|$. 
This work was supported by IBS under the project code, IBS-R018-D1 for RW. 
\end{acknowledgments}
%%%%%%%%%%%%%%%%%%%%%%%%%%%%%%%%%%%%%%%%%%%%%%%%%%

%\newpage
\appendix
%%%%%%%%%%%%%%%%%%%%%%%%%%%%%%%%%%%%%%%%%%%%%%%%%%
\section{Evaluation of the uncertainties and input parameters}
\label{App:uncertainty}
%%%%%%%%%%%%%%%%%%%%%%%%%%%%%%%%%%%%%%%%%%%%%%%%%%
Here, we explain the way to evaluate the uncertainty in the observable coming from the one in the input parameters. 
Suppose the observable is expressed as $F(y; \{ x_i \})$, where $x_i$ is an input parameter measured (or calculated) as $x_i = x_i^0 \pm \delta x_i$, $\{ x_i \}$ shows a set of parameters for $i=1,2,\cdots, k$, 
and $y$ indicates model parameters. 
We define the uncertainty of $F(y; \{ x_i \})$ as 
\begin{align}
 \delta F_\text{th.}(y) = \sqrt{ \sum_{i=1}^k \left| \left. \frac{\partial F(y; \{ x_i \})}{\partial x_i} \right|_{x_i^0} \delta x_i \right|^2 }  \,, 
\end{align}
and the central value is shown as $F_\text{th.}(y) = F(y; \{ x_i^0 \})$. 
To obtain excluded and allowed regions of a parameter space $y$, we evaluate the $\chi^2$ function. 
In our analysis it is defined as 
\begin{align}
 \chi^2 (y) = \frac{(F_\text{th.}(y) - F_\text{exp.})^2}{\delta F_\text{th.}(y)^2 +\delta F_\text{exp.}^2} \,,
\end{align}
where the experimental result is shown as $F_\text{exp.} \pm \delta F_\text{exp.}$. 
The parameters taken as $\{ x_i \}$ in our analysis are listed in Table~\ref{Tab:DecayConstant}, and (\ref{Eq:SRRs})--(\ref{Eq:DGs}). 
The other input values used in our numerical analysis are shown in Table~\ref{Tab:BasicInput}. 
%%%%%%Table%%%%%%
\begin{table}
\begin{center}
\scalebox{0.9}{
\begin{tabular}{cc}
 \hline 
 \hline  Input 							& Value    \\ \hline
 $\alpha_s (m_Z)$						& $0.1185$ \\
 $\alpha$ 								& $1/137$  \\
 $G_F$								& $1.16637 \times 10^{-5}\,\text{GeV}^{-2}$ \\
 $v = \sqrt{v_1^2+v_2^2} $				& $246\,\text{GeV}$ \\
 $m_h$								& $125\,\text{GeV}$ \\
 $m_W$								& $80.40\,\text{GeV}$ \\
 $m_Z$								& $91.19\,\text{GeV}$ \\
 $m_e$								& $0.5101 \times 10^{-3}\,\text{GeV}$ \\
 $m_\mu$								& $0.1057\,\text{GeV}$ \\
 $m_\tau$								& $1.7768\,\text{GeV}$ \\
 \hline\hline
\end{tabular}
\quad %
\begin{tabular}{cc}
 \hline 
 \hline  Input 							& Value    \\ \hline
 $m_{\pi^\pm}$							& $0.140\,\text{GeV}$ \\
 $m_{\pi^0}$							& $0.135\,\text{GeV}$ \\
 $m_{K^\pm}$							& $0.494\,\text{GeV}$ \\
 $m_{K_L}$							& $0.498\,\text{GeV}$ \\
 $m_{D^\pm}$							& $1.870\,\text{GeV}$ \\
 $m_{D^0}$							& $1.865\,\text{GeV}$ \\
 $m_{D_s}$							& $1.969\,\text{GeV}$ \\
 $m_{B^\pm}$							& $5.279\,\text{GeV}$ \\
 $m_{B^0}$							& $5.279\,\text{GeV}$ \\
 $m_{B_s}$							& $5.367\,\text{GeV}$ \\
 \hline\hline
\end{tabular}
\quad %
\begin{tabular}{cc}
 \hline 
 \hline  Input 							& Value    \\ \hline
 $\tau_{\pi^\pm}$						& $2.6033 \times 10^{-8}\,\text{s}$   \\
 $\tau_{\pi^0}$							& $8.5200 \times 10^{-17}\,\text{s}$   \\
 $\tau_{K^\pm}$						& $1.2380 \times 10^{-8}\,\text{s}$   \\
 $\tau_{K_L}$							& $5.116 \times 10^{-8}\,\text{s}$   \\
 $\tau_{D^\pm}$						& $1.040 \times 10^{-12}\,\text{s}$ \\
 $\tau_{D^0}$							& $0.410 \times 10^{-12}\,\text{s}$ \\
 $\tau_{D_s}$							& $0.500 \times 10^{-12}\,\text{s}$ \\
 $\tau_{B^\pm}$						& $1.638 \times 10^{-12}\,\text{s}$ \\
 $\tau_{B^0}$							& $1.519 \times 10^{-12}\,\text{s}$ \\
 $\tau_{B_s}$							& $1.512 \times 10^{-12}\,\text{s}$ \\
 \hline\hline
\end{tabular}
}
\end{center}
\caption{
Input values for fundamental parameters.  
}
\label{Tab:BasicInput}
\end{table}
%%%%%%Table%%%%%%

%%%%%%%%%%%%%%%%%%%%%%%%%%%%%%%%%%%%%%%%%%%%%%%%%%
\section{Analytic formulae for flavor observables}
\label{App:formulae}
%%%%%%%%%%%%%%%%%%%%%%%%%%%%%%%%%%%%%%%%%%%%%%%%%%
Here we give the functions of analytic formulae for the flavor observables, which are used to obtain the constraints in this paper. 

\subsection{$B^0_{q}\to\ell^+\ell^-$}
\label{App:form_Bsmumu}
%%%%%%%%%%%%%%%%%%%%%%%%
%%%%%%%%%%%%%%%%%%%%%%%%

\subsubsection{Functions}
The analytic formula for the averaged time-integrated branching ratio is given in terms of the Wilson coefficients $\mathcal C_{10}$, $\mathcal C_P$, and $\mathcal C_S$. 
The SM contributions in (\ref{Eq:CSc}) and (\ref{Eq:CPc}) are written as  
\begin{align}
 \mathcal C_{S}^\text{c,\,SM} = - \frac{x_t(x_t-2)}{12(x_t-1)^2} + \frac{(x_t-2)(3x_t-1)}{24(x_t-1)^3}\, \ln x_t \,, 
\end{align}
\begin{align}
\hspace{-2em}
 \mathcal C_{P}^\text{c,\,SM} = 
 &\frac{1}{24}\,\left[\frac{x_t(36 x_t^3-203 x_t^2+352 x_t-209)}{6 (x_t-1)^3} + \frac{17 x_t^4-34 x_t^3+ 4 x_t^2+23 x_t-6}{(x_t-1)^4} \,\ln x_t \right] \notag \\
 & - \frac{s_W^2}{36}\,\left[\frac{x_t(18 x_t^3-139 x_t^2+274 x_t-129)}{2 (x_t-1)^3} + \frac{24 x_t^4-33 x_t^3-45 x_t^2+50 x_t-8}{(x_t-1)^4}\, \ln x_t\right] \,. 
\end{align}
The functions $G_i$ and $F_i$ written in (\ref{Eq:CSn}) and (\ref{Eq:CPn}) are described as 
\begin{align}
 G_1 (\xi_u^{A},\xi_d^{A},x_{H^+},x_t) = -\frac{3}{4} + \xi_d^A \xi_u^{A} F_4 + ( \xi_u^A )^2 F_5 \,, 
\end{align}
\begin{align}
 &G_2 (\xi_u^{A},\xi_d^{A},x_{H^+},x_t) = \xi_d^A (\xi_d^A \xi_u^{A} +1) F_6 -\xi_d^A (\xi_u^{A})^2 F_7  \notag \\
 &\hspace{10em} + ( \xi_u^{A} )^2 (\xi_d^A F_8 + \xi_u^A F_9 - \xi_u^{A} F_{10} ) +\xi_u^{A} F_{11} - \xi_u^{A} F_{12} \,, 
\end{align}
\begin{align}
 &G_3 (\xi_u^{A},\xi_d^{A},x_{H^+},x_t) =  \xi_d^A (\xi_d^A \xi_u^{A} +1) F_6 +\xi_d^A (\xi_u^{A})^2 F_7  \notag \\
 &\hspace{10em} + ( \xi_u^{A} )^2 (\xi_d^A F_8 + \xi_u^A F_9 + \xi_u^{A} F_{10} ) +\xi_u^{A} F_{11} + \xi_u^{A} F_{12} \,, 
\end{align}
\begin{align}
 F_0 = \frac{1}{8(x_t-1)^2} \left[ \frac{x_t-3}{2} -\frac{x_t(x_t-2)}{x_t-1} \ln x_t \right] \,, 
\end{align}
\begin{align}
 F_1 = \frac{1}{4(x_{H^+}-x_t)} \left[\frac{x_t\ln x_t}{x_t-1}-\frac{x_{H^+}\ln x_{H^+}}{x_{H^+}-1}\right] \,, 
\end{align}
\begin{align}
 \hspace{-2.5em}  F_2 = \frac{1}{8(x_{H^+}-x_t)}\left[\frac{x_{H^+}}{x_{H^+}-1} + \frac{x_t^2\ln x_t}{(x_t-1)(x_{H^+}-x_t)} - \frac{x_{H^+}(x_{H^+}x_t + x_{H^+}-2 x_t)}{(x_{H^+}-1)^2(x_{H^+}-x_t)}\ln x_{H^+}\right] \,, 
\end{align}
\begin{align}
 F_3 = \frac{1}{8(x_{H^+}-x_t)}\left[\frac{x_{H^+}-x_t}{(x_{H^+}-1)(x_t-1)} + \frac{x_t(x_t-2)}{(x_t-1)^2}\ln x_t - \frac{x_{H^+}(x_{H^+}-2)}{(x_{H^+}-1)^2}\ln x_{H^+}\right] \,,
\end{align}
\begin{align}
 F_4 = \frac{x_t}{x_{H^+}-x_t} \left[1- \frac{x_{H^+}}{x_{H^+}-x_t} \ln \frac{x_{H^+}}{x_t} \right] \,, 
\end{align}
\begin{align}
 F_5 = \frac{x_t}{2(x_{H^+}-x_t)^2} \left[\frac{x_{H^+}+x_t}{2} - \frac{x_{H^+} x_t}{x_{H^+}-x_t} \ln \frac{x_{H^+}}{x_t} \right] \,,  
\end{align}
\begin{align}
 F_6 = \frac{1}{2 (x_{H^+}- x_t)}\left[-x_{H^+}+x_t +x_{H^+} \ln x_{H^+}-x_t \ln x_t\right]\,, 
\end{align}
\begin{align}
 F_7 = \frac{1}{2(x_{H^+}-x_t)} \left[x_t - \frac{x_{H^+} x_t}{x_{H^+}-x_t}(\ln x_{H^+}-\ln x_t)\right]\,, 
\end{align}
\begin{align}
 F_8 = \frac{1}{2(x_{H^+}-x_t)} \left[x_{H^+}-\frac{x_{H^+}^2 \ln x_{H^+}}{x_{H^+}-x_t}+\frac{x_t (2x_{H^+}- x_t) \ln x_t}{x_{H^+}-x_t}\right]\,, 
\end{align}
\begin{align}
 F_9 = \frac{1}{4(x_{H^+}-x_t)^2} \left[\frac{x_t \left(3 x_{H^+}-x_t\right)}{2}-\frac{x_{H^+}^2 x_t}{x_{H^+}- x_t}(\ln x_{H^+}-\ln  x_t)\right]\,, 
\end{align}
\begin{align}
 F_{10} = \frac{1}{4(x_{H^+}-x_t)^2} \left[\frac{x_t (x_{H^+}-3 x_t)}{2}-\frac{x_{H^+} x_t (x_{H^+}-2 x_t)}{x_{H^+}- x_t}(\ln x_{H^+}-\ln   x_t)\right]\,, 
\end{align}
\begin{align}
 &\hspace{-2.5em} 
 F_{11} = \frac{1}{2(x_{H^+}-x_t)} \Bigg[\frac{x_t\left(x_t^2-3x_{H^+}x_t+9x_{H^+}-5 x_t-2\right)}{4 (x_t-1)^2} +\frac{x_{H^+} \left(x_{H^+} x_t-3x_{H^+} + 2 x_t\right) }{2 (x_{H^+}-1)(x_{H^+}- x_t)}\ln x_{H^+} \notag \\
 &\hspace{-2em} +\frac{x_{H^+}^2 \left(-2 x_t^3+6 x_t^2-9 x_t+2\right)+3 x_{H^+} x_t^2 (x_t^2-2 x_t+3)-x_t^2 \left(2 x_t^3-3 x_t^2+3 x_t+1\right)}{2 (x_t-1)^3 (x_{H^+}- x_t)}\ln x_t\Bigg]\,, 
\end{align}
\begin{align}
 &F_{12} =\frac{1}{2(x_{H^+}-x_t)}\left[\frac{\left(x_t^2+x_t-8\right) (x_{H^+}- x_t)}{4 (x_t-1)^2}-\frac{x_{H^+}   (x_{H^+}+2) }{2 (x_{H^+}-1)}\ln x_{H^+}\right.\notag \\
 & \hspace{8em} \left.+\frac{x_{H^+} \left(x_t^3-3 x_t^2+3 x_t+2\right)+3 \left(x_t-2\right) x_t^2}{2 (x_t-1)^3}\ln x_t\right]\,.
\end{align}
The SM Higgs contributions for $\mathcal C_{S,P}^\text{n}$ are included in (\ref{Eq:CSn}) and (\ref{Eq:CPn}), which can be extracted as a SM-like limit. 
Taking $\cos (\beta- \alpha)\to 0$, $\sin (\beta- \alpha) \to 1$, $\xi_f^A \to 0$, and $m_{\phi} \to \infty$ for $\phi = H^+,H,A$, the SM-like limit is obtained as   
\begin{align}
 \mathcal C_{S}^\text{n,\,SM} = -\frac{3x_t}{8 x_h} + x_t F_0 \,, 
 \quad\quad \mathcal C_{P}^\text{n,\,SM} = 0 \,.
\end{align}

\subsubsection{Assumptions} 
In the formulae of $\mathcal C_{S,P}^\text{n}$, we have ignored FCNC contributions induced by a running effect of the Yukawa interaction term at the low energy scale. 
In Ref.~\cite{Li:2014fea} such contributions in $\mathcal C_{S,P}^\text{n}$ are estimated as 
\begin{align}
 & \mathcal R_S = \frac{x_t}{2x_H} \zeta_\ell (\zeta_u - \zeta_d) (1 + \zeta_u \zeta_d) \mathcal C_{\mathcal R} (\mu_t) \,, \label{Eq:RSform} \\
 & \mathcal R_P = -\frac{x_t}{2x_A} \zeta_\ell (\zeta_u - \zeta_d) (1 + \zeta_u \zeta_d) \mathcal C_{\mathcal R} (\mu_t) \,, \label{Eq:RPform}
\end{align}
in the SM-like limit, where $\mathcal C_{\mathcal R} (\mu_t)$ shows the renormalized coupling of FCNC term in the Yukawa Lagrangian ((2.15) in Ref.~\cite{Li:2014fea}). 
We can see from (\ref{Eq:RSform}) and (\ref{Eq:RPform}) that $\mathcal R_{S,P}=0$ in the $Z_2$ symmetric models. 
This is because that the $Z_2$ symmetry can protect the alignment condition at any scale. 
On the other hand, in the aligned model the condition is guaranteed only at the scale where the model is set and thus the non-zero contribution can appear at the low energy scale. 
In this paper, we simply ignore this effect in all types of 2HDM. 
We also neglect contributions proportional to light quark mass $m_q$ and Higgs self couplings $\lambda_{3,7}$~\cite{Li:2014fea}. 
As for the Higgs self couplings, we have confirmed that the effect is negligible.

\subsubsection{Averaged time-integrated branching ratio} 
The averaged time-integrated branching ratio $\overline{\mathcal B} (B^0_q \to \ell^+\ell^-)$ can be understood as follows. 
The ``untagged'' decay rate for $P \to f$ is defined and described as 
\begin{align} 
 \langle \Gamma(P(t) \to f) \rangle 
 &\equiv \Gamma(P^0(t) \to f) +\Gamma(\bar P^0(t) \to f) \\
 &= A_H e^{-\Gamma_H t} + A_L e^{-\Gamma_L t}  \\
 &=\left( A_H + A_L \right) e^{-\Gamma_P t} \times \left[ \cosh \frac{\Delta \Gamma_P t}{2} + \mathcal A_f \sinh \frac{\Delta \Gamma_P t}{2} \right] \,,
\end{align} 
with $\Gamma_P = (\Gamma_P^L + \Gamma_P^H)/2 = 1/\tau_P$, $\Delta \Gamma_P = \Gamma_P^L - \Gamma_P^H$, and $\mathcal A_f = (A_H - A_L)/(A_H + A_L)$, 
where ``H'' and ``L'' denote two mass eigenstates with difference lifetimes, $1/\Gamma_P^L$ and $1/\Gamma_P^H$. 
In experiment, a branching ratio is usually extracted from the total event yield. 
It means that the lifetime of neutral mesons is nothing to do with the measurement of branching ratio. 
Thus, the {\it experimentally measurable} branching ratio can be defined as 
\begin{align} 
 \mathcal B(P \to f)_\text{exp} 
 \equiv \frac{1}{2} \int_0^\infty \langle \Gamma(P(t) \to f) \rangle dt = \frac{1}{2} \left( \frac{A_H}{\Gamma_P^H} + \frac{A_L}{\Gamma_P^L} \right) \,.
\end{align} 
On the other hand, the {\it theoretical} branching ratio is considered as 
\begin{align} 
 \mathcal B(P \to f)_\text{theo} 
 \equiv \frac{\tau_P}{2} \langle \Gamma(P(t=0) \to f) \rangle = \frac{\tau_P}{2} \left( A_H + A_L \right) \,. 
\end{align} 
Therefore, $\mathcal B(P \to f)_\text{exp}$ is represented as
\begin{align} 
 \label{Eq:barBr}
 \overline{\mathcal B} (P \to f) \equiv \mathcal B(P \to f)_\text{exp} = \frac{1 + \mathcal A_f \, y_P}{1-y_P^2} \mathcal B(P \to f)_\text{theo} \,,
\end{align} 
where $y_P = \Delta \Gamma_P/(2 \Gamma_P)$. 
For $B^0_q \to \ell^+\ell^-$, the SM predicts $\mathcal A_{\ell^+\ell^-} =+1$ since there is only one contribution from $\mathcal O_{10}$ to the process.   
Thus one finds 
\begin{align} 
 \label{Eq:barBrSM}
 \overline{\mathcal B} (B^0_q \to \ell^+\ell^-)_\text{SM} = \frac{1}{1-y_{B^0_q}} \mathcal B (B^0_q \to \ell^+\ell^-)_\text{SM} = \frac{\Gamma_{B^0_q}}{\Gamma_{B^0_q}^H} \mathcal B (B^0_q \to \ell^+\ell^-)_\text{SM}  \,.
\end{align} 
Hence (\ref{Eq:barBrStandardModel}) is obtained. 
If we consider new physics, it is possible to have two different CP-violating phases, a relative phase difference. 
This can be described in the amplitude as 
\begin{align} 
 \label{Eq:CPamp}
 A(\bar B^0_q \to \ell^+\ell^-) = \mathcal P + \mathcal S \,, \quad A( B^0_q \to \ell^+\ell^-) = - \mathcal P^* + \mathcal S^* \,,
\end{align} 
where $\mathcal P = e^{i \phi_{\mathcal P}} |\mathcal P|$ and $\mathcal S =e^{i \phi_{\mathcal S}} |\mathcal S|$ denote the contributions from the effective operators $\mathcal O_{10},\mathcal O_{P}$ and $\mathcal O_{S}$, respectively. 
The CP asymmetry $\mathcal A_{\ell^+\ell^-}$ is represented in terms of (\ref{Eq:CPamp}) as 
\begin{align} 
 \label{Eq:Aellell}
 \mathcal A_{\ell^+\ell^-} 
 \equiv \frac{2\text{Re} (\lambda_{\ell^+\ell^-})}{1 + |\lambda_{\ell^+\ell^-}|^2} 
 = \frac{ |\mathcal P|^2 \cos 2 \phi_{\mathcal P} -|\mathcal S|^2 \cos 2 \phi_{\mathcal S} }{|\mathcal P|^2 + |\mathcal S|^2} \,, 
\end{align} 
where
\begin{align} 
 \lambda_{\ell^+\ell^-}
 \equiv \frac{A(\bar B^0_q \to \ell^+\ell^-)}{A( B^0_q \to \ell^+\ell^-)}
 =\frac{\mathcal P + \mathcal S}{- \mathcal P^* + \mathcal S^*} \,.
\end{align} 
Therefore, when we define (the normalization is adjusted as appropriate), 
\begin{align} 
 \frac{ \mathcal B(B^0_q \to \ell^+\ell^-)_\text{NP} }{ \mathcal B(B^0_q \to \ell^+\ell^-)_\text{SM} } =  |\mathcal P|^2 + |\mathcal S|^2 \,,
\end{align}
we obtain 
\begin{align} 
 \hspace{-2em} \overline{\mathcal B} (B^0_q \to \ell^+\ell^-)_\text{NP} 
 & = \mathcal B(B^0_q \to \ell^+\ell^-)_\text{NP} \frac{1 + \mathcal A_{\ell^+\ell^-} \, y_{B^0_q}}{1-y_{B^0}^2}  \\
 & = \mathcal B(B^0_q \to \ell^+\ell^-)_\text{SM} \left[ \frac{1 + y_{B^0_q} \cos 2 \phi_{\mathcal P} }{1-y_{B^0_q}^2} |\mathcal P|^2 + \frac{1 - y_{B^0_q} \cos 2 \phi_{\mathcal S} }{1-y_{B^0}^2} |\mathcal S|^2 \right]  \\
 & = \overline{\mathcal B}(B^0_q \to \ell^+\ell^-)_\text{SM} \left[ \frac{1 + y_{B^0_q} \cos 2 \phi_{\mathcal P} }{1+y_{B^0_q}} |\mathcal P|^2 + \frac{1 - y_{B^0_q} \cos 2 \phi_{\mathcal S} }{1+y_{B^0_q}} |\mathcal S|^2 \right]  \,,
\end{align}
where it can be derived with use of (\ref{Eq:barBr}), (\ref{Eq:barBrSM}), and (\ref{Eq:Aellell}). 
In the 2HDMs of $Z_2$ symmetric types and of aligned type with real $\zeta_f$, we trivially see $\phi_{\mathcal P} = \phi_{\mathcal S} =0$. 
In this case, finally we can derive (\ref{Eq:barBrNewPhysics}) in terms of $y_{B^0_q} =\Delta \Gamma_{B^0_q}/(2 \Gamma_{B^0_q}) = (\Gamma_{B^0_q}^L - \Gamma_{B^0_q}^H)/(\Gamma_{B^0_q}^L + \Gamma_{B^0_q}^H)$.

\subsection{Neutral meson mixings}
\label{App:form_mixing}
%%%%%%%%%%%%%%%%%%%%%%%%
%%%%%%%%%%%%%%%%%%%%%%%%

%\subsubsection{Functions} 
The SM and 2HDM contributions derived from one-loop diagrams are involved in the forms $A_{VV'}^{(ST)}$, which are described as 
\begin{align}
 A_{WW}(x_t) = 1+\frac{9}{1-x_{t}} -\frac{6}{(1-x_{t})^2} -\frac{6x_{t}^2 \ln x_{t}}{(1-x_{t})^3} \,,  
\end{align}
\begin{align}
 &\hspace{-4em} A_{WH}(x_t,x_b) = ( \xi_u^A )^2 \Bigg[ \frac{4 - x_t}{(x_t -1)(x_{H^+} -x_t)} 
 +\frac{(x_{H^+} -4) x_{H^+} \ln x_{H^+} }{(x_{H^+} -1)(x_{H^+} -x_t)^2} +\frac{(3x_t^2 - (x_t^2-2x_t+4) x_{H^+} ) \ln x_t}{(x_t -1)^2(x_{H^+} -x_t)^2} \Bigg] \notag  \\[0.2cm]
 & \hspace{-2em} + \frac{2x_b}{x_t} \Bigg[ 2 \xi_d^A \xi_u^{A} \bigg( -\frac{ 1 }{ (x_{H^+} - x_t) (x_t -1 ) } 
 +\frac{ x_{H^+} \ln x_{H^+} }{(x_{H^+} -1 ) (x_{H^+} - x_t)^2 } - \frac{ (-x_{H^+}+x_t^2) \ln x_t }{ (x_{H^+}-x_t)^2 (-1+x_t)^2 } \bigg) \notag \\[0.2cm] 
 & \hspace{-2em} + ( \xi_u^A )^2 
 \bigg( 
 -\frac{1}{36(-1+x_{H^+})^2 (x_{H^+}-x_t)^3 (-1+x_t)^3} \Big [ -x_{H^+}^4 (-12+65 x_t+2 x_t^2+5 x_t^3)  \notag \\[0.2cm]
 & \hspace{-2em} +2 x_{H^+}^3 (-12+47 x_t+85 x_t^2-11 x_t^3+11 x_t^4) +x_{H^+}^2 (12+91 x_t-574 x_t^2+246 x_t^3-130 x_t^4-5 x_t^5) \notag \\[0.2cm]
 & \hspace{-2em} +2 x_{H^+} x_t (-30+95 x_t+49 x_t^2-11 x_t^3+17 x_t^4) +x_t^2 (-24+43 x_t-110 x_t^2+31 x_t^3) \Big ]  \notag \\[0.2cm]
 & \hspace{-2em} - \frac{ x_t ( (x_{H^+}^3 -3 x_{H^+}^2 x_t) (1+9 x_t) -x_t^4 (12-3 x_t+x_t^2) +3 x_{H^+} x_t (4-12 x_t+24 x_t^2-7 x_t^3+x_t^4) ) \ln x_t}{6 (x_{H^+}-x_t)^4 (-1+x_t)^4} \notag \\[0.2cm] 
 & \hspace{-2em} - \frac{ x_tx_{H^+} (-3 x_{H^+}^3 (-3+x_t) +12 x_t (1+x_t) -3 x_{H^+} x_t (13+x_t) +x_{H^+}^2 (1+10x_t+x_t^2) ) \ln x_{H^+}}{6 (-1+x_{H^+})^3 (x_{H^+}-x_t)^4} 
 \bigg)
 \Bigg] \,, \label{Eq:WH} 
\end{align}
\begin{align}
 &A_{HH}(x_t,x_b) = ( \xi_u^A )^4 \Bigg[ \frac{x_t +x_{H^+}}{(x_t -x_{H^+})^2} -\frac{2x_t x_{H^+}}{(x_t -x_{H^+})^3} \ln \frac{x_t}{x_{H^+}} \notag \\[0.2cm]
 & \hspace{4.5em} -x_b \left( \frac{5 x_{H^+}^2-22 x_{H^+} x_t+5 x_t^2}{9 \left(x_{H^+} - x_t\right)^4} + \frac{ x_{H^+}^3 - 3 x_{H^+}^2 x_t - 3 x_{H^+} x_t^2+x_t^3}{3 \left(x_{H^+} - x_t\right)^5} \ln \frac{x_t}{x_{H^+}} \right)
 \Bigg]  \,, \label{Eq:HH} 
\end{align}
\begin{align}
 & \hspace{-4em} A_{WH}^{ST}(x_t) = ( \xi_u^A )^2 \Bigg[ \frac{(x_t^2 + x_{H^+}^4)(-11+7x_t-2x_t^2)+x_{H^+} x_t(7+53x_t-55x_t^2+19x_t^3)}{9(1-x_{H^+})^2(x_{H^+}-x_t)^3(1-x_t)^3} \notag \\[0.2cm] 
 & \hspace{2em} +\frac{x_{H^+}^2 (-2-55x_t+15x_t^2+17x_t^3-11x_t^4)+x_{H^+}^3 (19+17x_t-19x_t^2+7x_t^3)}{9(1-x_{H^+})^2(x_{H^+}-x_t)^3(1-x_t)^3} \notag \\[0.2cm]
 & \hspace{2em} +\frac{2x_{H^+} (x_{H^+}^2+(-3+x_{H^+})x_{H^+} x_t+(3+(x_{H^+}-3) x_{H^+})x_t^2)\ln{x_{H^+}}}{3(1-x_{H^+})^3(x_{H^+}-x_t)^4} \notag \\[0.2cm]
 & \hspace{2em} -\frac{2 (x_{H^+}^3-3x_{H^+}^2x_t+3x_{H^+} x_t^2-3x_t^4+3x_t^5-x_t^6)\ln{x_t}}{3(x_{H^+} - x_t)^4(1-x_t)^4} \Bigg] \notag \\[0.2cm]
 & \hspace{2.5em} +\xi_d^A \xi_u^{A} \Bigg[ \frac{ (x_{H^+}^2+x_t)(-3+x_t)+x_{H^+}(1+6x_t-3x_t^2) }{2(1-x_{H^+})(x_{H^+}-x_t)^2(1-x_t)^2} \notag \\[0.2cm] 
 & \hspace{2em} +\frac{ (x_{H^+}^2-2x_{H^+} x_t-(-2+x_t)x_t^3)\ln{x_t}}{(x_{H^+} - x_t)^3(1-x_t)^3} -\frac{x_{H^+} (x_{H^+} -2x_t+x_{H^+} x_t)\ln{x_{H^+}}}{(1-x_{H^+})^2(x_{H^+}-x_t)^3} \Bigg] \,, \label{Eq:WHst} 
\end{align}
\begin{align}
 &A_{HH}^{ST}(x_t) = ( \xi_d^A \xi_u^{A} )^2 \, \left[ \frac{2}{(x_{H^+} - x_t)^2} +\frac{x_t +x_{H^+}}{(x_{H^+} - x_t)^3} \ln \frac{x_t}{x_{H^+}} \right] \notag \\[0.2cm] 
 &\hspace{4em} + ( \xi_u^A )^4 \left[ \frac{ 5x_{H^+}^2 -22 x_{H^+} x_t +5x_t^2}{18(x_{H^+} - x_t)^4} +\frac{ x_{H^+}^3-3x_{H^+}^2x_t-3x_{H^+}x_t^2+x_t^3}{6(x_{H^+}-x_t)^5} \ln \frac{x_t}{x_{H^+}} \right] \notag \\[0.2cm] 
 &\hspace{4em} +\xi_d^A (\xi_u^{A})^3 \left[ \frac{2}{(x_{H^+}-x_t)^2} +\frac{x_{H^+}+x_t}{(x_{H^+}-x_t)^3} \ln \frac{x_t}{x_{H^+}} \right] \,. \label{Eq:HHst}
\end{align}
The formulae of $A_{WH}^{ST}$ and $A_{HH}^{ST}$ can be obtained by taking non-zero external momenta into account. 
The non-zero external momenta also leads to the $x_b$ terms in $A_{WH}$ and $A_{HH}$, which are not described in Ref.~\cite{Chang:2015rva}. 
In the $K^0$-$\bar K^0$ mixing, additional loop functions $B_{VV'}$ are derived due to two non-zero masses of $t$ and $c$ quarks. 
They are described as 
\begin{align}
 B_{WW}(a,b) = -\frac{3}{(a-1)(b-1)} +\frac{a^2-8a+4}{(a-b)(b-1)^2}\ln a -\frac{b^2-8b+4}{(a-b)(a-1)^2}\ln b \,, 
\end{align}
\begin{align}
 & B_{WH}(a,b) = ( \xi_u^A )^2 \left[ \frac{b^2 \ln b}{(1-b)(b-a)(b-x_{H^+})} +\frac{a^2 \ln a}{(1-a)(a-b)(a-x_{H^+})} \right. \notag \\[0.2cm]
 &\hspace{8.5em} \left. +\frac{x_{H^+}^2 \ln x_{H^+}}{(1-x_{H^+})(x_{H^+}-a)(x_{H^+}-b)} \right] \,, 
\end{align}
\begin{align}
 & B_{HH}(a,b) = ( \xi_u^A )^4 \left[ \frac{x_{H^+}}{(a-x_{H^+})(b-x_{H^+})} +\frac{b^2 \ln b}{(b-a)(b-x_{H^+})^2} -\frac{a^2 \ln a}{(b-a)(a-x_{H^+})^2} \right. \notag \\[0.2cm]
 &\hspace{8.5em}  \left. -\frac{x_{H^+}(a\, x_{H^+}+b\, x_{H^+} -2a\,b) \ln x_{H^+}}{(a-x_{H^+})^2(b-x_{H^+})^2}  \right] \,.  
\end{align}

%\subsubsection{Long distance effect} 
%AA

\subsection{$B \to X_q \gamma$}
\label{App:form_bsgamma}
%%%%%%%%%%%%%%%%%%%%%%%%
%%%%%%%%%%%%%%%%%%%%%%%%
The loop functions $G_a^i$, $C_a^i$, and $D_a^i$ in (\ref{Eq:bsgLO}) and (\ref{Eq:bsgNLO}) are given as  
\begin{align}
  G_1^7(y) = \frac{y (7-5y-8y^2)}{24(y-1)^3} + \frac{y^2(3y-2)}{4(y-1)^4}\ln y \,, \quad  G_2^7(y) = \frac{y (3-5y)}{12(y-1)^2} + \frac{y(3y-2)}{6(y-1)^3}\ln y \,, 
\end{align} 
\begin{align}
  G_1^8(y) = \frac{y (2+5y-y^2)}{8(y-1)^3} - \frac{3y^2}{4(y-1)^4}\ln y \,, \quad  G_2^8(y) = \frac{y (3-y)}{4(y-1)^2} - \frac{y}{2(y-1)^3}\ln y \,, 
\end{align} 
\begin{align}
 & C_1^7(y) = \frac{2}{9} y \left[ \frac{y(18-37y+8y^2)}{(y-1)^4}{\rm Li}_2 \left( 1 - \frac{1}{y} \right)+ \frac{y(-14+23y+3y^2)}{(y-1)^5}\ln^2y \right. \notag \\
 & \hspace{6em} +\frac{-50+251y-174y^2-192y^3+21y^4}{9(y-1)^5}\ln y -\frac{3y-2}{3(y-1)^4} \ln y    \notag \\
 & \hspace{6em} \left. +\frac{797-5436y+7569y^2-1202y^3}{108(y-1)^4} -\frac{16-29y+7y^2}{18(y-1)^3} \right] \,,
\end{align} 
\begin{align}
 & C_2^7(y) = - \frac{4}{3} y \left[ \frac{4(-3+7y-2y^2)}{3(y-1)^3}{\rm Li}_2  \left( 1 - \frac{1}{y} \right) + \frac{8-14y-3y^2}{3(y-1)^4}\ln^2y \right. \notag \\
 & \hspace{6em} \left. +\frac{2(-3-y+12y^2-2y^3)}{3(y-1)^4}\ln y +\frac{7-13y+2y^2}{(y-1)^3}\right] \,, 
\end{align} 
\begin{align}
 & C_1^8(y) = \frac{1}{6} y \left[ \frac{y(30-17y+13y^2)}{(y-1)^4}{\rm Li}_2 \left( 1 - \frac{1}{y} \right) - \frac{y(31+17y)}{(y-1)^5}\ln^2y \right. \notag \\ 
 & \hspace{6em}  -\frac{226-817y-1353y^2-318y^3-42y^4}{36(y-1)^5}\ln y -\frac{3y-2}{6(y-1)^4} \ln y   \notag \\
 & \hspace{6em} \left. +\frac{1130-18153y+7650y^2-4451y^3}{216(y-1)^4} -\frac{16-29y+7y^2}{36(y-1)^3}  \right] \,,
\end{align} 
\begin{align}
 & C_2^8(y) = - \frac{1}{3} y \left[ \frac{-36+25y-17y^2}{2(y-1)^3}{\rm Li}_2 \left( 1 - \frac{1}{y} \right) + \frac{19+17y}{(y-1)^4}\ln^2y \right. \notag \\ 
 & \hspace{6em} \left. +\frac{-3-187y+12y^2-14y^3}{4(y-1)^4}\ln y +\frac{3(143-44y+29y^2)}{8(y-1)^3}\right]  \,,
\end{align} 
\begin{align}
 D_1^7(y) = \frac{2}{9} y \left[ \frac{-31-18y+135y^2-14y^3}{ 6(y-1)^4}+\frac{y(14-23y-3y^2)}{(y-1)^5}\ln y \right] \,, 
\end{align} 
\begin{align}
 D_2^7(y) = -\frac{2}{9}y\left[ \frac{21-47y+8y^2}{(y-1)^3}+\frac{2(-8+14y+3y^2)}{(y-1)^4}\ln y \right] \,, 
\end{align} 
\begin{align}
 D_1^8 (y) = \frac{1}{6} y \left[ \frac{-38-261y+18y^2-7y^3}{6(y-1)^4}+\frac{y(31+17y)}{(y-1)^5}\ln y \right] \,, 
\end{align} 
\begin{align}
 D_2^8 (y) = -\frac{1}{3} y \left[ \frac{81-16y+7y^2}{2(y-1)^3}-\frac{19+17y}{(y-1)^4}\ln y \right] \,.
\end{align}

\subsection{$\bar B \to D^{(*)} \tau\bar\nu$}
\label{App:form_BDtaunu}
%%%%%%%%%%%%%%%%%%%%%%%%
%%%%%%%%%%%%%%%%%%%%%%%%
The coefficients $\Gamma_i^{D\ell}$ and $\Gamma_i^{D^{*}\ell}$ for the light leptonic mode are 
\begin{align}
&\Gamma_1^{D\ell} =8.788 \,, \\ 
&\Gamma_2^{D\ell} =-5.230\,, \\ 
&\Gamma_3^{D\ell} =0.819\,, \\
&\Gamma_1^{D^*\ell} =32.87 -0.05\,R_1 +0.81\,R_1^2 -17.68\,R_2 +3.35\,R_2^2\,, \\ 
&\Gamma_2^{D^*\ell} =-15.34 +0.02\,R_1 -0.35\,R_1^2 +9.52\,R_2 -1.88\,R_2^2\,, \\
&\Gamma_3^{D^*\ell} =1.99 +0.04\,R_1^2 -1.34\,R_2 +0.27\,R_2^2\,.
\end{align}
The coefficients $\Gamma_i^{D\tau}$ and $\Gamma_i^{D^{*}\tau}$ are written as 
\begin{align}
&\Gamma_1^{D\tau} =1.845\,, \\ 
&\Gamma_2^{D\tau} =-0.676\,, \\
&\Gamma_3^{D\tau} =0.069\,, \\
&\Gamma_4^{D\tau} =2.493\,, \\ 
&\Gamma_5^{D\tau} =-0.790\,, \\ 
&\Gamma_6^{D\tau} =0.074 \,, \\
&\Gamma_7^{D\tau} =1.578\,, \\ 
&\Gamma_8^{D\tau} =-0.447\,, \\ 
&\Gamma_9^{D\tau} =0.039 \,, 
\end{align}
\begin{align}
&\Gamma_1^{D^*\tau} =5.593 -0.005\,R_1 +0.134\,R_1^2 -2.051\,R_2 +0.352\,R_2^2 \,, \\ 
&\Gamma_2^{D^*\tau} =-1.581 +0.002\,R_1 -0.040\,R_1^2 +0.692\,R_2 -0.123\,R_2^2 \,, \\
&\Gamma_3^{D^*\tau} =0.131 -0.003\,R_1^2 -0.063\,R_2 +0.011\,R_2^2 \,, \\
&\Gamma_4^{D^*\tau} =1.289 -1.137\,R_2 +0.251\,R_2^2 \,, \\
&\Gamma_5^{D^*\tau} =-0.408 +0.361\,R_2 -0.080\,R_2^2\,, \\
&\Gamma_6^{D^*\tau} =0.036 -0.032\,R_2 +0.007\,R_2^2\,, \\
&\Gamma_7^{D^*\tau} =0.384 -0.338\,R_2 +0.074\,R_2^2\,, \\
&\Gamma_8^{D^*\tau} =-0.113 +0.100\,R_2 -0.022\,R_2^2\,, \\
&\Gamma_9^{D^*\tau} =0.009 -0.008\,R_2 +0.002\,R_2^2\,.
\end{align}

%%%%%%%%%%%%%%%%%%%%%%%%%%%%%%%%%%%%%%%%%%%%%%%%%%
\bibliographystyle{JHEP}
\bibliography{reference_EW}
%%%%%%%%%%%%%%%%%%%%%%%%%%%%%%%%%%%%%%%%%%%%%%%%%%

\end{document}